\documentclass[aps,pre,amsmath,amsfonts,floatfix,superscriptaddress,color]{revtex4}
\usepackage{mathrsfs} 
\usepackage{graphicx} 
\usepackage{bbm}
\usepackage{color}

\let\G=\Gamma
 \def\la{\left\langle} \def\ra{\right\rangle}
\newcommand{\rme}{{\rm e}}
\newcommand{\rmd}{{\rm d}}
\newcommand{\eff}{{\rm eff}}
\newcommand{\ie}{{\it i.e.}}
\newcommand{\eg}{{\it e.g.}}
\newcommand{\reff}[1]{(\ref{#1})}
\newcommand{\Y}{\Upsilon}
\newcommand{\beq}{\begin{equation}} \newcommand{\eeq}{\end{equation}}
\newcommand{\bea}{\begin{eqnarray}} \newcommand{\eea}{\end{eqnarray}}
\newcommand{\ket}[1]{\left\vert#1\right\rangle}
\newcommand{\bra}[1]{\left\langle#1\right\vert}

\newcommand{\Dt}{\Delta}
\newcommand{\IM}{\text{Im}}
\newcommand{\RE}{\text{Re}}
\newcommand{\OO}{{\cal O}}

\begin{document}

\title{Dynamic correlations, fluctuation-dissipation relations, and effective temperatures after a quantum quench of the transverse field Ising chain}
  \author{Laura Foini} 
\affiliation{Universit\'e Pierre et Marie Curie -- Paris VI, Laboratoire de Physique Th\'eorique
et Hautes Energies, UMR
  7589, Tour 15 5\`eme \'etage, 4 Place Jussieu, 75252 Paris Cedex 05, France} 
  \author{Leticia F. Cugliandolo}
\affiliation{Universit\'e Pierre et Marie Curie -- Paris VI, Laboratoire de Physique Th\'eorique
et Hautes Energies, UMR
  7589, Tour 15 5\`eme \'etage, 4 Place Jussieu, 75252 Paris Cedex 05, France} 
  \author{Andrea  Gambassi} 
\affiliation{SISSA -- International School for Advanced
  Studies and INFN, via Bonomea 265, 34136 Trieste, Italia}

\date{\today}

\begin{abstract}
Fluctuation-dissipation relations, {\it i.e.}, the relation between two-time correlation and linear response functions, 
were successfully used to search for signs of equilibration and 
to identify effective temperatures in the non-equilibrium behavior of a number of  macroscopic 
classical and quantum systems in contact with thermal baths. Among the most relevant 
cases in which the effective temperatures thus defined were shown to have a 
thermodynamic meaning one finds the stationary dynamics of driven super-cooled 
liquids and vortex glasses, and the relaxation of glasses.
Whether and under which conditions an effective thermal behavior can be found in 
quantum isolated many-body systems after a global quench is a question of 
current interest. We propose to study the possible emergence of thermal behavior long after the quench 
by studying fluctuation-dissipation relations 
in which (possibly time- or frequency-dependent) parameters replace 
the equilibrium temperature.  If thermalization within the Gibbs ensemble 
eventually occurs these parameters should be constant and equal for all pairs of observables in "partial" or "mutual" 
equilibrium. We analyze these relations in the paradigmatic quantum system, \ie, the  
quantum Ising chain, in the stationary regime after a quench of the transverse field.
The lack of thermalization to a Gibbs ensemble becomes apparent within this approach. 

\end{abstract}

\maketitle

\tableofcontents

\section{Introduction}
 
Motivated by recent experimental advances in the field of cold atoms, 
the theoretical study of the non-equilibrium dynamics of isolated interacting many-body 
quantum systems is currently receiving increasing attention~\cite{polkovnikov2010,GMHB02,TCFMSEB-11,CBPESFGBK-12}.
Among the several questions that have been addressed,  a central one 
concerns the way in which a macroscopically large isolated system evolving with unitary quantum dynamics
from a generic initial state approaches equilibrium. 
In this work we discuss how such a problem can be effectively addressed from a novel 
perspective, inspired by analogous studies of non-equilibrium classical systems. 
A brief account of this study can be found in Ref.~\cite{Foini11}.

In classical systems thermalization is usually justified by advocating
a chaotic dynamics that should ensure ergodicity in phase space and thermalization
in terms of a microcanonical ensemble~\cite{gallavotti1999}. 
This implies (for systems with short-range interactions) 
that a large subpart of a much larger system thermalizes to a canonical ensemble. 
This condition is typically satisfied for  generic Hamiltonians, 
even though the time required to reach such an ergodic regime might increase with the volume of the sample and
thus be extremely large. 
Ergodicity implies that the time average of an observable for a single realization of the system 
coincides with its ensemble average. Within  the statistical ensemble one derives exact relations that govern the 
dynamics of the system, such as the fluctuation-dissipation theorem. 
The way in which equilibration and ergodicity are understood for quantum systems 
evolving in a unitary (isolated) manner, together with the 
more appropriate way to rationalize the coarse grained description of 
their properties in terms of statistical ensembles, has been debated for the 
last 80 years~\cite{neumann1929,GolLeb10,tasaki,Srednicki,Deutsch}.

Our aim is to revisit here the problem of thermalization in isolated many-body quantum systems
with tools developed for the study of the non-equilibrium behavior
of classical glassy systems~\cite{Cugliandolo97,Teff-reviews,Cugliandolo-review}. 
Our study builds upon a large number of papers published in recent years. Basic questions as to
whether a stationary state is reached and how this state could be characterized  have been 
addressed in a
number of simple models, including the one-dimensional systems reviewed
in Refs.~\cite{Karevski,polkovnikov2010}. The first picture which emerged was the following: 
non-integrable systems are expected to reach a thermal stationary state 
characterized by a Gibbs distribution with a single temperature. 
Integrable systems, instead, are not expected to thermalize.
However, their asymptotic stationary state should nonetheless
be described by the so-called generalized Gibbs
ensemble (GGE) in which each conserved quantity is characterized by 
a generally different effective 
temperature~\cite{Rigol,Cazalilla_Iucci09,CIC-12,Fioretto,Calabrese11,CEF12,BRI-12}. 
On the other hand, other 
works~\cite{KLA07,CCRSS-11,BKL10,RS2012,BCH11,BPV-11,Carleo-Schiro,Rossini10_short,Rossini10_long,SC-10,EKW-09,schiro} 
have shown (or at least argued) that this scenario could actually be 
significantly richer. 

Indeed, it was suggested in Refs.~\cite{Rossini10_short} that depending on the system's parameters
and the specific quantity under study a conventional Gibbs ensemble might effectively 
capture some relevant features of the non-equilibrium dynamics even in integrable systems.
In particular, a combined analytic and numeric study of the transverse field Ising chain~\cite{Rossini10_short,Rossini10_long}
suggested that observables that are non-local with respect to the 
excitation of the Hamiltonian, display the same relaxation 
scales as if they were in equilibrium at a finite temperature, 
at least for small quenches~\cite{Rossini10_short,Calabrese11}. 
Instead, local quantities such as the transverse magnetization 
do not show thermal behavior with the notable exception of 
quenches to the critical point. 
In fact,  compared to non-critical 
quenches, the critical point shows some remarkable 
properties~\cite{Sachdev_book,Calabrese06,Rossini10_long}
that can be attributed to the gapless spectrum and the linearity of the
dispersion relation at low momenta. 
This is clearly seen in the scaling limit of one-dimensional models with these properties, 
for which the results of Conformal Field Theory (CFT)~\cite{Calabrese06}
suggest the emergence of a unique scale which plays the role of a temperature,
at least for certain (non-generic) initial states \cite{CEF12,Cardy-talk}.
Accordingly, in what follows, we will focus specifically on critical quenches. 
Though it has been recently shown that the dynamical properties of the transverse field Ising chain are eventually described by the GGE~\cite{Calabrese11,CEF12},
our aim here is to highlight this non-thermal behavior within a different approach, that can be extended to cases for which analytical solutions are not available. 

Most of the previous studies of thermalization in quantum systems 
compare the correlation lengths, the coherence times, or the expectation values
of particular time-independent quantities with their values in equilibrium and extract 
in this way effective thermodynamic parameters such as an effective temperature of the out of 
equilibrium system~\cite{polkovnikov2010,Rigol,KLA07,Rossini10_short,Rossini10_long,schiro,Calabrese11,Carleo-Schiro,EKW-09}. 
However, these requirements may be too restrictive
and/or insufficient to investigate equilibration issues in systems 
with complex dynamics. The first two criteria are restricted to exponential relaxation~\cite{footnote-supercooled}, whereas the last one ignores the dynamics of the system.

In Gibbs equilibrium the (times-dependent) correlation function between any two observables is 
linked to the 
linear response of one of these observables to a linear perturbation applied to the other one in a 
model-independent way. 
Indeed, while the functional form of the correlation and linear response may 
 depend on the 
pair of observables used, they can be affected by the spectral density of the bath,
and they may of course be model-dependent, the relation between them remains unaltered and 
just determined by the temperature of the environment. This relation is the statement of the 
fluctuation-dissipation theorem (FDT) that involves only one parameter, \ie, the temperature of the system
(for simplicity we assume that the number of particles is fixed). 
Quite naturally, a test of Boltzmann-Gibbs equilibration then consists in determining the correlation 
and linear response of a chosen pair 
of observables and to verify whether FDT holds for them. 

The analysis of fluctuation-dissipation relations (FDRs), \ie, the relation between 
correlation and linear responses, 
in classical  dissipative macroscopic systems out of equilibrium has revealed 
a very rich and somehow unexpected structure. For instance, the spatio-temporal relaxation in 
classical glassy systems
(or even non-disordered coarsening systems \cite{Godreche,Teff-critical,Malte2}) 
is very different from an equilibrium one with, \eg, breakdown of stationarity (aging effects) and other 
peculiar   
features. Still, the FDRs show that the dynamics can be interpreted
as taking place in different temporal regimes each of them in equilibrium at a different 
value of an effective temperature with 
good thermal properties~\cite{Cugliandolo97,Teff-reviews,Cugliandolo-review,footnote-crit}. 
Similar results were found in quantum dissipative glassy models of mean-field type~\cite{Culo}.

The main purpose of this contribution  
is to propose the use of FDRs as possible tests of
(at least partial) equilibration in isolated quantum systems. In particular, this can be done
by "measuring" independently
the two-time symmetric correlation $C^{AB}$ and at the linear response
 $R^{AB}$ (defined in more detail further below) of two generic quantities $A$ and $B$. On the basis of $C^{AB}$ and $R^{AB}$ in the frequency domain one extracts a frequency- 
 and observable-dependent inverse "temperature" $\beta^{AB}_{\rm eff}(\omega)$
 through the fluctuation-dissipation relation:
\beq
\displaystyle \hbar~\mbox{Im} \tilde R^{AB}(\omega) =
\tanh\Big(\frac{\beta^{AB}_{\rm eff}(\omega) \, \hbar\, \omega}{2}\Big)  \,
\tilde C^{AB}_+(\omega).
\eeq
The quantity $\beta^{AB}_{\rm eff}(\omega)$ provides important information on the possible equilibration of the system
and on the various time/energy scales within which partial equilibration might occur. 

Concretely, 
we apply this idea to test equilibration in an integrable quantum system, the quantum Ising chain, 
quenched to its critical point, for which it has been argued that at least some observables could equilibrate 
in the usual sense of having their static and dynamic properties determined  
by a single global temperature.
More precisely, we compute independently the correlation function and linear response of several 
pairs of observables. We note that despite the existence of many studies of the dynamics after a quench of the transverse 
field~\cite{Calabrese06,Rossini10_short,Rossini10_long,Calabrese11,CEF12,Igloi00,Igloi11,rieger2011,JamirSilva,Silva,GambSilva,MDDS-07,BRI-12}
none of them discussed the behavior of the linear response functions. For each FDR we extract 
a parameter (actually a time- or frequency-dependent function) that with a definite abuse of language we call  
"effective temperatures". The analysis of these quantities, especially whether they are constant over 
certain time or frequency regimes and whether they coincide for different observables, 
will inform us about the (non-)thermal 
character of the dynamics.

Let us emphasize that the idea of using FDRs to 
investigate thermalization properties in non-equilibrium system is completely general.
Here, for illustration purposes, we apply it to a specific problem -- the Ising model in a transverse field -- 
in order to demonstrate that equilibration does not occur
in this case, in spite of some evidence for the contrary, mentioned above. 
We expect our approach to provide an efficient test of thermalization also for non-integrable quantum systems,  
even though the characterization of their real-time dynamics is a very hard
problem, often limited to small system sizes and short time intervals.

The paper is organized as follows. In Sec.~\ref{sec:eff-temp} we
review several definitions of effective temperatures proposed in the 
context of quantum quenches~\cite{Rigol,Cazalilla_Iucci09,Fioretto,Calabrese11,Rossini10_short,Rossini10_long,Calabrese06} 
and classical and quantum dissipative 
glassy dynamics~\cite{Cugliandolo97,Teff-reviews,Cugliandolo-review,Culo,Caso}.  Section~\ref{sec:Ising} 
summarizes 
those features of the static and dynamic behavior of a quantum Ising chain
that are relevant to our study. Section~\ref{sec:results} illustrates our results
on the FDRs for several observables: the local and 
global transverse magnetization and the order parameter. Finally, in Sec.~\ref{sec:conclusions}
we summarize our findings, discussing their implications and some 
ideas for future investigations. As already mentioned, a preliminary account of some of our results 
appeared in Ref.~\cite{Foini11}.

\section{Effective temperatures}
\label{sec:eff-temp}

The evolution of a quantum system with Hamiltonian 
$\hat{H}(\Gamma)$ is ruled by the unitary dynamics
\beq
i \hbar \frac{\rmd}{\rmd t} |\psi(t)\rangle = \hat{H}(\Gamma) |\psi(t)\rangle ,
\label{eq:Udyn}
\eeq
where $\Gamma$ is some control parameter and $|\psi(0)\rangle=|\psi_0\rangle$
an arbitrary initial state of the system.
The initial condition is often chosen to be the ground state of the Hamiltonian 
$\hat{H}(\Gamma_0)$ corresponding to a different value of the parameter $\Gamma$. In this case, 
one usually refers to the evolution in Eq.~\reff{eq:Udyn} as resulting from a quantum quench, \ie, from a    
sudden change $\Gamma_0\to\Gamma$ of the parameter of the Hamiltonian. Alternatively, one might 
consider the case in which $|\psi(0)\rangle$ is not a pure state, \eg, it is a mixed state corresponding to 
the canonical distribution with Hamiltonian $\hat{H}(\Gamma_0)$ and inverse temperature 
$\beta_0$~\cite{SCC09}. The initial state is then a generic excited state that does not correspond 
to an equilibrium state of the new Hamiltonian $\hat{H}(\Gamma)$ and right after the quench the system 
is in a non-stationary, non-equilibrium, regime. 

In order to investigate the possible emergence of an effective thermal behavior of the system  
after the quench, one can introduce effective temperatures on the basis of the behavior of 
various quantities. In Sec.~\ref{sec:Teff_static} we discuss some of the possible definitions 
based on the (asymptotic) behavior of one-time quantities, such as those widely investigated 
so far in the literature on quantum quenches. In Sec.~\ref{sec:Teff_dynamic}, instead, we focus on the 
definitions based on two-time (dynamic) quantities that have been used in the context of 
glassy dynamics. We insist upon the fact that we still do not know whether the effective temperatures thus 
introduced can be attributed a  thermodynamic meaning. 

\subsection{Energy and constants of motion}
\label{sec:Teff_static}

Among the various quantities that one can focus on in order to define 
an effective temperature, a special role is expected to be played by the 
energy of the system~\cite{polkovnikov2010,EKW-09,Rossini10_short,Rossini10_long,Carleo-Schiro}: indeed
the average energy
$E(t) \equiv \langle \psi(t)| \hat{H}(\Gamma) | \psi(t)\rangle = \langle \psi_0|\hat{H}(\Gamma)|\psi_0\rangle = E(t=0)$ 
is conserved because the dynamics after the quench is unitary. 
Rather generally, one can define the density matrix $\hat{\rho}$ 
possibly describing the asymptotic state of the system as the one which maximizes the von Neumann 
entropy $S[\hat\rho]=- \text{Tr}[\hat{\rho} \log \hat{\rho}]$, subject to
the constraint of having the correct expectation value of the energy. 
This amounts to assuming 
that the asymptotic state of the system long after the quench is effectively described by a Gibbs canonical distribution 
$\propto \exp\{- \beta_{\rm eff}^E \hat{H}(\Gamma)\}$, in which the value of 
the effective temperature $1/\beta_{\rm eff}^E$ is fixed by the constraint
\beq\label{Teff_quench}
\langle \psi_0 | \hat{H}(\Gamma) | \psi_0 \rangle = \frac{1}{Z} \mbox{Tr}[ \rme^{- \beta_{\rm eff}^E\hat{H}(\Gamma)} \hat{H}(\Gamma) ],
\eeq
where $Z$ is the partition function. 
The average on the l.h.s.~is the energy of the system after the quench
while the one on the r.h.s.~is the average energy of an equilibrium state of $\hat{H}(\Gamma)$
at temperature $T=T_{\rm eff}^E = 1/\beta_{\rm eff}^E$.
(Hereafter we set the Boltzmann constant $k_B=1$.) 
In this specific example, the l.h.s.~of Eq.~\reff{Teff_quench}
is independent of time because the energy is a constant of motion. 

In general, however, one would like to check that the temperature thus identified
also describes the stationary limit of the average value of other observables.
The time dependence of a generic observable $\hat{{\cal O}}$ can be conveniently studied within the Heisenberg 
representation
\beq
\label{eq:Hrep}
\hat{{\cal O}}(t) \equiv \rme^{i \hat{H} t} \hat{{\cal O}} \rme^{- i \hat{H} t},
\eeq
within which 
the time-dependent expectation value on a generic mixed quantum state represented by a density matrix $\hat{\rho}$ 
(assumed to be normalized to one) is 
given by 
\beq
\langle \hat{{\cal O}}(t) \rangle  = \text{Tr}[ \hat{\rho}  ~ \hat{{\cal O}}(t) ].
\label{eq:exvaleq}
\eeq
In canonical equilibrium at temperature $T=\beta^{-1}$, $\hat \rho(T)$ is the Gibbs density matrix
$\hat\rho(T)= \exp(-\beta \hat{H})/Z(\beta)$
and therefore, due to $[\hat{H},\hat{\rho}]=0$, the expectation value 
$\langle \hat{{\cal O}}(t) \rangle $ in Eq.~\reff{eq:exvaleq} is actually independent of time. 
In a generic non-equilibrium case, instead, the density matrix $\hat \rho$ over 
which the expectation value is taken describes the initial state of the system 
and it reduces to $\hat\rho = |\psi_0\rangle\langle\psi_0|$ when the system is initially prepared in a pure state $|\psi_0\rangle$. 
In this case, 
analogously to what has been done above for the energy, one can compare the generic stationary values (if any) of the averages 
after the quench with the (time-independent) expectation value of the same observable $\hat{\cal O}$ taken on an equilibrium canonical 
ensemble at the temperature $T_{\rm eff}^{\cal O}$ and
then determine $T_{\rm eff}^{\cal O}$ in such a way that these two averages coincide, \ie, 
\begin{equation}
\lim_{t\to\infty} \langle \psi_0 | \hat{{\cal O}}(t) |\psi_0  \rangle
= \langle \hat{{\cal O}} \rangle_{T=T_{\rm eff}^{\cal O}}.
\end{equation}
An effective thermal-like behavior of the 
system in the stationary state would require these temperatures $T_{\rm eff}^{\cal O}$ to be 
independent of ${\cal O}$ and, in particular, to coincide with $T_{\rm eff}^E$ defined above;
however, this is not always the case~\cite{Rossini10_short,Rossini10_long,Carleo-Schiro}.

The discussion above implicitly assumes that 
the energy is the only quantity conserved by the dynamics and  that in minimizing $S[\hat\rho]$ one has to account only for one 
constraint, 
which naturally leads to the Gibbs canonical ensemble. However,
if the system is integrable, the situation turns out to be rather subtle 
because of the many "independent" quantities which are conserved by the dynamics in addition to the 
energy~\cite{Rigol,Cazalilla_Iucci09,Fioretto,Calabrese11}. 
Consider, for simplicity, a non-interacting Hamiltonian
 that can be written in the diagonal form
\beq\label{Hdiagonal}
\hat{H}(\Gamma) = \sum_k \epsilon_k(\Gamma) \, \hat{\eta}_k^{\dag}\hat{\eta}_k,
\eeq
where the $\hat{\eta}_k$'s are creation operators for free (bosonic or fermionic) 
excitations of energy $\epsilon_k$, labeled by  a set $k$ of quantum numbers. 
(For free theories $k$ is the momentum.) 
The operators $\hat{\eta}_k$'s satisfy the canonical commutation relations 
and the number $\hat{n}_k$ of excitations  is given by  $\hat{n}_k= \hat{\eta}_k^{\dag}\hat{\eta}_k$.
Clearly, $[\hat{n}_k,\hat{H}]=0$, \ie, the set $\{\hat{n}_k\}$
is a set of constants of motion induced by $\hat{H}(\Gamma)$,
independently of the initial state over which expectation values are calculated,
which merely fixes the values of these constraints. 
Therefore, the dynamics after the quench are constrained by a large number of integrals of motion, the values 
of which have to be conserved. In repeating the minimization of $S$ which leads to 
Eq.~\reff{Teff_quench}, 
it is necessary to introduce a number of Lagrange multipliers (see further below), one for each ``conserved quantity",  which 
eventually turn into a set of  ``effective temperatures" $\{T_{\rm eff}^k\}$ determined by the condition 
$\la  \psi_0 |  \hat{n}_k | \psi_0 \ra = \la \hat{n}_k\ra_{T=T_{\rm eff}^k}$.
These quantities prove to be particularly useful since they naturally appear in the calculation 
of (stationary and non-stationary) expectation values~\cite{Rigol,Cazalilla_Iucci09,Fioretto}.
It was in fact suggested~\cite{Rigol} that the stationary behavior of the system after 
quenches towards Hamiltonians of the form \reff{Hdiagonal} 
can be described in terms of the density matrix $\hat{\rho}_{\rm GGE}$ obtained by maximizing
the von Neumann entropy $S[\hat\rho]$ under 
the constraints on the expectation values of $\la \hat{n}_k\ra$.
This density matrix is of the form
\beq
\label{density_matrix}
\hat{\rho}_{\rm GGE} = 
\frac{1}{Z} \ \rme^{ - \sum_k \lambda_k  \hat{n}_k },
\eeq
where $\lambda_k = \epsilon_k(\Gamma)  / T_{\rm eff}^k$ are the Lagrange 
multipliers enforcing the values of the integrals of motion.

\subsection{Dynamic correlations and response functions}
\label{sec:Teff_dynamic}

As anticipated above, the aim of the present work is to 
introduce a definition of effective temperature which probes the dynamics of the system rather than 
the asymptotic time-independent properties discussed in Sec.~\ref{sec:Teff_static}. In particular, in this context, 
we are naturally led to consider FDRs, which turned out to be particularly 
useful in understanding various instances and features of the non-equilibrium dynamics of classical and quantum glassy 
systems~\cite{Cugliandolo97,Teff-reviews,Cugliandolo-review}.

The basic quantities which intervene in the FDRs are 
the two-time correlation between two generic operators $\hat{A}$ and $\hat{B}$ in the Heisenberg representation 
[see Eq.~\reff{eq:Hrep}], defined by
\beq
C^{AB}(t,t') = \langle \hat{A}(t) \hat{B}(t') \rangle  
= \text{Tr}[ \hat{\rho}   \hat{A}(t) \hat{B}(t')] .
\label{eq:CAB}
\eeq
Clearly, for generic $\hat{A}$ and $\hat{B}$ one has
$\langle \hat{A}(t) \hat{B}(t') \rangle  \neq \langle \hat{B}(t') \hat{A}(t) \rangle$ and it is natural 
to define symmetric and antisymmetric correlations as follows:
\beq
\label{eqC8:Cpm}
C_{\pm}^{AB}(t,t') = \langle [ \hat{A}(t) ,\hat{B}(t') ]_{\pm} \rangle, 
\eeq
where $[X,Y]_{\pm} = ( X Y \pm Y X ) / 2$.
Without loss of generality we will consider either operators with zero average
or we will imply that the average value is subtracted from the definition of the generic operator $\hat{\cal O}$: 
$\hat{\mathcal{O}}(t) \to \hat{\mathcal{O}}(t) - \langle \hat{\mathcal{O}}(t)\rangle$.

In addition to $C^{AB}$, another dynamic quantity of primarily 
importance is the instantaneous linear response function $R^{AB}$ which quantifies, up to the linear term,
the variation of the expectation $\langle \hat{A}(t)\rangle$ due to a perturbation
which couples to the operator $\hat{B}$,
\beq
R^{AB}(t,t') \equiv \left.\frac{\delta \langle \hat{A}(t) \rangle}{\delta h_B(t')} \right|_{h_B = 0}
,
\label{eq:Resp}
\eeq
where $\hat{A}(t)$ is obtained by evolving $\hat A$ -- according to Eq.~\reff{eq:Hrep} -- with the time-dependent perturbed 
Hamiltonian $\hat H_{h_B}(t) \equiv \hat H - h_B(t)\hat B$. 
In and out of equilibrium $R^{AB}(t,t')$ is related to the antisymmetric correlation $C_-^{AB}(t,t')$ defined in 
Eq.~\reff{eqC8:Cpm} by the so-called Kubo formula~\cite{Kubo},
\beq\label{Kubo}
\hbar \, R^{AB}(t,t') = 2 i \theta(t-t') C_-^{AB}(t,t'),
\eeq
where $\theta(t)$ is the step function $\theta(t < 0)=0$ and $\theta(t > 0)=1$ that enforces causality.
In the following we will be concerned primarily with correlation functions of Hermitian operators, for which 
$[C^{AB}(t,t')]^* = \langle \hat B(t') \hat A(t)\rangle$. Their symmetric and antisymmetric correlators 
can be expressed in terms of $C^{AB}$ in Eq.~\reff{eq:CAB} as
\beq
C^{AB}_+(t,t') = \mbox{Re\,}C^{AB}(t,t') \quad \;\;\; \mbox{and} \;\;\; \quad C^{AB}_-(t,t') = i \,\mbox{Im\,}C^{AB}(t,t'),
\label{eq:ReC}
\eeq
so that Eq.~\reff{Kubo} yields
\beq
\hbar\,R^{AB}(t,t') = - 2 \theta(t-t') \, \mbox{Im\,}C^{AB}(t,t').
\label{eq:ImC}
\eeq 

In equilibrium, the dynamics are invariant under time translations and therefore correlation and response functions are
stationary, $C_{\pm}^{AB}(t,t') = C_{\pm}^{AB}(t-t')$, whereas out of equilibrium this is not necessarily the case. 
When dealing with stationary cases  it is convenient to consider the Fourier transform
of these quantities, for which we adopt the following convention:
\beq
\tilde{f}(\omega) = \int_{-\infty}^{\infty}\!\!\rmd t~ \rme^{i \omega t} f(t)
\qquad\mbox{and}\qquad
f(t) = \int_{-\infty}^{\infty}\!\frac{\rmd\omega}{2\pi}~ \rme^{ - i \omega t} \tilde{f}(\omega). 
\label{eq:FT}
\eeq

The canonical fluctuation-dissipation theorem  (FDT) establishes a relation between the linear 
response of a system to  an external perturbation and the 
spontaneous fluctuations occurring within the same system in thermal equilibrium at a temperature $\beta^{-1}$.
Remarkably, this relation does not depend on the particular system under consideration and takes the same functional form 
independently of the quantities which the correlation and the response refer to. 

In the case of canonical (Gibbs) equilibrium, the quantum ``bosonic"  
FDT can be expressed in the time domain as
\beq
R^{AB}(t) = \frac{i}{\hbar}\, \theta(t) 
\int_{-\infty}^{\infty}\!\! \frac{\rmd\omega}{\pi} \rme^{-i\omega t} \tanh\left(\frac{\beta\hbar\omega}{2}\right) 
\, \tilde C^{AB}_+(\omega),
\label{eq:FDT-t}
\eeq
where we reinstated $\hbar$ to make 
the classical limit $\hbar\to0$ of the quantum FDT  transparent.
Indeed, in this case,  
one finds
\beq\label{FDT_classical-8}
R^{AB}(t) = - \beta\, \theta(t) \, \frac{\rmd}{\rmd t}C^{AB}(t),
\eeq
which is the classical FDT. As expected, this limit is recovered 
for $T = \beta^{-1} \gg \hbar \omega_{\rm typ}$ where $\omega_{\rm typ}$ is some typical energy scale of the quantum problem. 
The quantum FDT can be cast in a compact form in the frequency domain by Fourier transforming Eq.~\reff{eq:FDT-t}:
\beq
\hbar~\mbox{Im} \tilde R^{AB}(\omega) =
\tanh\left(\frac{\beta\hbar\omega}{2}\right)  \,
\tilde C^{AB}_+(\omega).
\label{eq:FDT-omega}
\eeq
Remarkably, knowing $R^{AB}$ and $C_+^{AB}$ for a pair of observables $A$ and $B$ allows 
the determination of the inverse temperature $\beta$ of the system in equilibrium via Eqs.~\reff{eq:FDT-t} 
and \reff{eq:FDT-omega}, whatever the observables $A$ and $B$ are. In a sense, 
the FDT provides a viable method for ``measuring" the temperature of a system, based on (local) 
measurements of correlations and response functions. 

Out of thermal equilibrium and in particular right after a quench in a isolated system the FDT  
is not expected to hold.
It is tempting, however,  to test whether FDRs 
such as Eqs.~\reff{eq:FDT-t} and \reff{eq:FDT-omega} can be used to define a single "macroscopic" temperature  (or maybe a 
few), at least long after the quench and in the stationary regime. 
This approach 
turned out to be particularly fruitful for understanding the physics of the thermalization of classical dissipative systems with slow 
dynamics~\cite{Teff-reviews,Cugliandolo-review}.
In particular, clarifying the relation between this temperature and the one defined from one-time observables 
\cite{Rossini10_short,Rossini10_long,Carleo-Schiro} 
via Eq.~\reff{Teff_quench} is definitely an important issue. In case some sort of thermalization occurs long after the quench, 
all these temperatures should become not only equal but also independent of the quantities used to define them.  

Depending on the specific quantum isolated system or model under consideration 
a stationary state may or may not be attained at long times.
In several studies presented in the literature it was shown that a number of quantum isolated systems with short-range interactions 
reach a stationary state~\cite{Rigol,Cazalilla_Iucci09,Fioretto,Calabrese11,Rossini10_short},  
while some fully-connected models~\cite{sciolla,schiro,EKW-09} and some mean-field approximations to models with short-range
interactions~\cite{AA-02,BCS-osc,CalGam}
keep a non-stationary behavior.  
For the isolated quantum Ising chain a stationary state is indeed reached after the
quench and we will therefore focus on this relatively simple case.
Because of time-translational invariance of the stationary state we can equivalently consider 
two-time quantities in the time or in the frequency domain.
According to the strategy outlined above, we define an effective inverse temperature 
$\beta^{AB}_{\rm eff}(\omega)$  by enforcing the quantum FDR 
relation~\reff{eq:FDT-omega}, \ie,  
\beq
\displaystyle \hbar~\mbox{Im} \tilde R^{AB}(\omega) =
\tanh\Big(\frac{\beta^{AB}_{\rm eff}(\omega) \, \hbar\, \omega}{2}\Big)  \,
\tilde C^{AB}_+(\omega),
\label{eq:FDT-omega-Teff}
\eeq
where we consider $C^{AB}_+(t)$ and $R^{AB}(t)$ within the stationary regime.
In complete generality $\beta^{AB}_{\rm eff}(\omega)$   defined from 
Eq.~\reff{eq:FDT-omega-Teff} depends both on the particular choice of the quantities $A$ and $B$ which the correlation and the 
response function refer to and on the frequency $\omega$. Indeed, 
as the functional dependence of $R^{AB}(\omega)$ and $C^{AB}_+(\omega)$ on $\omega$ are, 
in principle, unrelated out of equilibrium it is necessary to allow for such a frequency dependence of $\beta^{AB}_{\rm eff}$ in 
Eq.~\reff{eq:FDT-omega-Teff}. 
The study of this dependence on $\omega$ 
provides an important piece of information on the dynamical 
scales of the system with respect to a given pair of observables $A$ and $B$: heuristically, thermalization within a certain 
time scale would be indeed signaled by a $\beta^{AB}_{\rm eff}(\omega)$ which becomes almost constant
within the corresponding range of frequencies~\cite{Teff-reviews,Cugliandolo-review}.
This kind of analysis encompasses and generalizes in several respects 
previous studies in this direction. 
For example, in Ref.~\cite{Mitra11}
an effective temperature was extracted 
for a system of one-dimensional bosons, after a quench of their interaction, by looking at 
the zero-frequency and zero-momentum limit of the FDR associated with the two-point  density-density correlation function. Interestingly enough, 
such a temperature turned out to characterize the
long-time and large-distance properties of the system after the adiabatic 
application of a periodic potential, for which a thermal-like behavior was found. 
In full generality, this analysis in the frequency domain can be extended and complemented by the analogous one in the time domain. Indeed,
the time dependence of the response function $R^{AB}$ --- which, in equilibrium, is connected to  $\tilde C_+^{AB}(\omega)$ via 
Eq.~\reff{eq:FDT-t} --- can be obtained from the Fourier transform of Eq.~\reff{eq:FDT-omega-Teff}.
Due to the integration over $\omega$, the result of the possible variation of $\beta_{\rm eff}^{AB}$ with $\omega$ is "weighted" by the 
frequency dependence of $\tilde C_+^{AB}$ and therefore different "modes" contribute differently to the resulting time dependence 
of the response function. In order to highlight the possible emergence of "dominant" modes, one can define still another effective 
temperature  $\beta_{\rm eff}^{\ast AB}$, based on the FDT in the time domain~\reff{eq:FDT-t}, \ie, 
\beq
\displaystyle R^{AB}(t>0) = \frac{i}{\hbar} 
\int_{-\infty}^{\infty} \frac{\rmd\omega}{\pi} \rme^{- i \omega t} 
\tanh\Big(\frac{{\beta_{\rm eff}^{\ast AB}} \, \hbar \, \omega}{2}\Big) 
\, \tilde C^{AB}_+(\omega)
\label{eq:FDT-t-Teff}
\eeq
where, in contrast to Eq.~\reff{eq:FDT-omega-Teff}, the inverse effective temperature 
$\beta_{\rm eff}^{\ast AB}$  on the r.h.s is assumed to be independent of $\omega$.
Note that Eqs.~\reff{eq:FDT-omega-Teff}  and \reff{eq:FDT-t-Teff} are not equivalent, unless 
$\beta^{\ast AB}_{\rm eff}$ is allowed to depend on the frequency. 
In addition, while for given $R^{AB}(t)$ and $\tilde C_+^{AB}(\omega)$ it is always possible to define the frequency-dependent 
inverse temperature $\beta_{\rm eff}^{AB}(\omega)$ from Eq.~\reff{eq:FDT-omega-Teff},
there might be no value of ${\beta_{\rm eff}^{\ast AB}}$  for which the integral in the r.h.s.~of Eq.~\reff{eq:FDT-t-Teff} reproduces properly 
the functional form of the time dependence of the given response function $R^{AB}(t)$. 
Note that the frequency-dependent effective temperature $\beta^{AB}_{\rm eff}(\omega)$ defined in Eq.~\reff{eq:FDT-omega-Teff} is 
analogous to the mode-dependent effective temperature $\beta_k$ introduced previously in the literature~\cite{Rigol},
especially in connection with the generalized Gibbs ensemble. However, whereas the latter refers to the asymptotic averages of one 
time-quantities, 
the former accounts for the dynamical properties of the system in the stationary state. In the case of integrable models --- 
as we will see below for the specific case of the isolated quantum Ising chain --- these mode-dependent temperatures can be 
recovered as the frequency-dependent one
$\beta^{AB}_{\rm eff}(\omega)$ obtained by studying the correlation and response functions of specific observables.

Within the stationary regime, it is rather natural to consider the behavior of the response and correlation functions at well-separated 
times. In fact, in classical coarsening and glassy systems this is the regime of structural relaxation in which partial equilibration of the 
slow (and non-equilibrium) degrees of freedom occurs~\cite{Teff-reviews,Cugliandolo-review}. In 
consequence, we will focus on the effective temperature that emerges when one 
tries to relate the long-time stationary response and correlation functions after the quench via the long-time limit of the fluctuation-dissipation theorem in Eq.~\reff{eq:FDT-t}. In particular, for large $t$ the integral in Eq.~\reff{eq:FDT-t} is expected to be dominated 
by small values of $\omega$ and therefore one can expand the hyperbolic tangent in a power series, which returns a sum
over the odd time derivatives of $C_+^{AB}(t)$:
\beq
\label{Eq-beff-timedomain}
R^{AB}(t) = \frac{2 i}{\hbar} \sum_{{\rm odd\;}n=1}^{\infty} c_n 
\left( \frac{ i \beta \, \hbar}{2}\right)^n  \frac{\rmd^n C_+^{AB}(t)}{\rmd t^n} \quad\;\;\;\; \mbox{where} \quad\;\;
c_n \equiv \frac{1}{n!}\left. \frac{\rmd^n \tanh x}{\rmd x^n}\right|_{x=0} .
\eeq
By inserting in this equation the expressions of the stationary response and correlation functions, 
$R^{AB}$ and $C^{AB}$, after the quench one obtains an implicit definition of an effective temperature $\beta = \beta_{\rm eff}^{\ast AB}$ 
which, hopefully, does not depend on time at the leading order and therefore provides a good definition of the temperature in the long-time limit. 
In Sec.~\ref{sec:ET} we will present an explicit determination of this temperature. 
At this point it is worth mentioning that if the correlation function on the r.h.s.~of Eq.~\reff{Eq-beff-timedomain} decays as a power law at long 
times, then the leading order of the r.h.s.~is 
indeed provided by the term with $n=1$ and the possible temperature which one defines from this relation coincides with the one that one can 
infer from imposing (in the long-time limit) the validity of the  
fluctuation-dissipation theorem for a \emph{classical} system, as in Eq.~\reff{FDT_classical-8}. However, as we will see in 
Sec.~\ref{sec:results}, oscillatory terms do actually modulate the algebraic decay 
of $C^{AB}$; accordingly, the leading term on the r.h.s.~of Eq.~\reff{Eq-beff-timedomain} does not coincide with the first term of the expansion 
and the effective temperature $\beta = \beta_{\rm eff}^{\ast AB}$  inferred from Eq.~\reff{Eq-beff-timedomain} receives contributions from 
the derivatives of these oscillatory terms.  Heuristically, the long-time behavior of the stationary 
response and correlation functions is expected to be determined by the low-frequency limit of their Fourier transform. In view of this fact, in the 
following we will be interested in understanding whether it 
is possible to recover the same effective thermal description from  the FDRs in the frequency 
and the time domains, at least in the low-frequency and long-time regimes. This is clearly possible in equilibrium where $\beta$ is a constant. 
More precisely, in the non-equilibrium case we 
will compare the low-frequency limit of the effective temperature defined via Eq.~\reff{eq:FDT-omega-Teff}, \ie,
\beq
\lim_{\omega\to 0^+}
\beta^{AB}_{\rm eff}(\omega)
= \lim_{\omega\to 0^+}
\frac{2}{\hbar\omega} \mbox{arctanh} 
\left( \frac{\hbar\mbox{Im} \tilde{R}^{AB}(\omega)} {\tilde{C}_+^{AB}(\omega)} 
\right)
\label{eq:Teff-small-omega}
\eeq
with the value $\beta = \beta_{\rm eff}^{\ast AB}$ obtained on  the basis of Eq.~\reff{Eq-beff-timedomain} according to the procedure 
described thereafter. We mention here that in Sec.~\ref{Sec:Mz} we will consider 
global quantities which are obtained by summing over all lattice sites and which are characterized by the fact that their correlation 
function $C_+$ does not vanish in the stationary regime even for well 
separated times. In these cases $\lim_{\omega \to 0^+} \tilde C_+(\omega)\neq \tilde C_+(\omega=0)$, and care has to be taken in 
evaluating the denominator of Eq.~\reff{eq:Teff-small-omega}.
Finally, since one expects the quantum behavior to be relevant on the short-time
scale whereas decoherence takes over at longer time differences, 
we will also consider  the 
effective temperature extracted 
from the classical FDT in Eq.~\reff{FDT_classical-8}, \ie,
\beq\label{FDT_classical}
\displaystyle {
T^{AB}_{{\rm{cl, eff}}}}
= - \lim_{t \to \infty} \frac{1}{R^{AB}(t)}
\frac{\rmd C^{AB}_+(t)}{\rmd t },
\eeq
with $C_+^{AB}$ and $R^{AB}$ taken in the stationary regime long after the quench. 

In passing, we mention that an effective temperature can also be defined on the basis of 
the relation between the linear response $\chi$ (susceptibility) to a \emph{constant} external perturbation $h_A$ and the time-
independent fluctuations in the thermodynamic conjugate quantity $A$ \cite{log-10}. Indeed, in thermal equilibrium, the classical 
FDT in Eq.~\reff{FDT_classical-8} with $A=B$ can be integrated in time,
\beq
\chi \equiv \left. \frac{\delta \langle A(t) \rangle}{\delta h_A}\right|_{h_A=0} = 
 \int_{-\infty}^t \!\!\rmd t'\, R^{AA}(t-t') = \tilde R^{AA}(\omega=0) =  \beta C^{AA}(t=0) = \beta [\langle A^2 \rangle - \langle A \rangle^2 ]
\eeq
(where we assume $C^{AA}(t=-\infty) = 0$) which is indeed an alternative form of the classical equilibrium fluctuation-dissipation 
theorem. Note that, being the external perturbation $h_A$ constant in time, the susceptibility $\chi$ does not actually depend on the 
time $t$ at which $\langle A(t) \rangle$ is measured.
Out of equilibrium and in the stationary regime one can therefore introduce the additional effective temperature 
\beq
T_{{\rm cl, st}} \equiv \frac{C^{AA}_+(t=0)}{\tilde R^{AA}(\omega=0)},
\label{eq:Teff-cl-st}
\eeq
which has been used in Ref.~\cite{log-10} in order to test the thermalization of a specific isolated quantum system.
Differently from the other quantities discussed so far, however, this effective temperature does not allow the 
investigation of the dynamic behavior of the system within different time- or frequency-regimes, as it involves only quantities which 
have been integrated out either in time or in frequency.  

\section{The Ising model and its dynamics after a quench of the transverse field}
\label{sec:Ising}

In this Section we briefly present the model, we recall its equilibrium phase diagram,
and we discuss some of the properties of its dynamics after a  quantum quench.

\subsection{The model}
\label{sec:model}

We consider the quantum Ising chain in a transverse field $J \Gamma > 0$ 
described by the Hamiltonian
\beq\label{H_Ising}
\hat{H}(\Gamma) = - J~\sum_{i=1}^L \left[ \hat{\sigma}^x_{i} \hat{\sigma}^x_{i+1} 
+ \Gamma \, \hat{\sigma}^z_{i} \right],
\eeq
where $\hat{\sigma}^{x,y,z}_i$ are the standard Pauli matrices acting on the $i$-th site of the chain, 
which commute at different sites. We assume periodic boundary conditions $\hat{\sigma}^x_{L+1} = \hat{\sigma}^x_{1}$ 
and we take the length $L$ of the chain to be even.  In what follows  we set $J$, $\hbar$, and $k_B =1$, \ie, 
we measure time in units of $\hbar/J$ and temperature in units of $J/k_B$.
It is well-known that the Hamiltonian \reff{H_Ising}  can be diagonalized by performing three subsequent 
transformations. Firstly, we introduce Jordan-Wigner creation and annihilation fermionic operators 
$\hat{c}^{\dag}_j$, $\hat{c}_j$~\cite{mccoy,lieb}
which satisfy canonical anticommutation relations $\{\hat c^\dag_i, \hat c_j\} = \delta_{ij}$, $\{\hat c_i, \hat c_j\} = \{\hat c^\dag_i, \hat c^\dag_j\} = 0$ and in terms of which
\beq
\label{JW_fermions}
\displaystyle\hat{\sigma}^+_j 
= 
\frac{\hat{\sigma}^x_j + i \hat{\sigma}^y_j}{2} 
=
\prod_{l=1}^{j-1}\left[ 1 - 2 \hat{c}^{\dag}_l \hat{c}_l \right] ~\hat{c}_j .
\eeq
The first identity implies $\hat{\sigma}^{x}_j = \hat{\sigma}^+_j + (\hat{\sigma}^+_j)^{\dag}$ 
and 
\beq\label{sz-c}
\hat{\sigma}^{z}_j = 1 - 2 \hat{c}^{\dag}_j \hat{c}_j . 
\eeq
Having expressed all $\hat{\sigma}^x$ and $\hat{\sigma}^z$
in terms of fermionic operators, the Hamiltonian becomes
\beq\label{quadratic-H}
\hat{H}(\Gamma) = - \sum_{i=1}^{L-1} \left[ \hat{c}^{\dag}_{i} \hat{c}_{i+1} 
+ \hat{c}^{\dag}_{i} \hat{c}^{\dag}_{i+1} + h.c. \right] - \Gamma ~\sum_{i=1}^{L}~ \left[  \hat{c}_{i} \hat{c}^{\dag}_{i} - \hat{c}^{\dag}_{i} \hat{c}_{i} \right]
+ (-1)^{N_F} \left[ \hat{c}^{\dag}_{L} \hat{c}_{1} 
+ \hat{c}^{\dag}_{L} \hat{c}^{\dag}_{1} + h.c. \right],
\eeq
where $N_F = \sum_{i=1}^L  \hat{c}^{\dag}_i \hat{c}_i$ 
is  the number of fermions in the chain.
The last term in this equation can be accounted for by extending the sum in the first term up to $i=L$ after having defined 
$\hat c_{L+1} \equiv (-1)^{N_F+1}\hat c_1$, which amounts to assuming periodic boundary conditions  for the chain if $N_F$ is odd 
and anti-periodic ones if $N_F$ is even.  
The Hamiltonian~\reff{quadratic-H} conserves the parity of fermions 
and we restrict to the even sector which contains
the ground state. Note that restricting to one of the two
sectors is justified only when one considers expectation values of operators 
which are defined in terms of products of an even number of fermionic operators,
\ie,  such that they do not change the parity of the state that they act on.

The Hamiltonian $\hat{H}(\Gamma)$ in Eq.~\reff{quadratic-H}, being quadratic,  can be conveniently expressed after a Fourier 
transformation
\beq\label{Fourier_Transform}
\displaystyle \hat{c}_j = \frac{1}{\sqrt{L}} \sum_{k} \rme^{i k j} \hat{c}_k
\quad \mbox{with} \quad k =\pm\frac{\pi (2n+1)}{L} \quad\mbox{and}\quad n = 0,\dots,\frac{L}{2}-1,
\eeq
where the sum runs over all allowed values of $k$.  
Note that in this expression we indicate both the fermionic operator $\hat c$ on the l.h.s.~and its Fourier transform on the r.h.s.~with 
the same symbol, the difference being made clear by the spatial  ($i$, $j$) or momentum  ($k$, $l$) indices and by the context.  

Finally, the Hamiltonian is diagonalized by a Bogoliubov rotation
\beq\label{B_transform}
\left( \begin{array}{c}
\hat{\gamma}_k^{\Gamma}  \\
\hat{\gamma}_{-k}^{\Gamma \; \dag}  \end{array} \right) =
\left( \begin{array}{cc}
\cos \theta_k^{\Gamma} & - i \sin \theta_k^{\Gamma}  \\
- i \sin \theta_k^{\Gamma} & \cos \theta_k^{\Gamma}  \end{array} \right)
\left( \begin{array}{c}
\hat{c}_k  \\
\hat{c}_{-k}^{\dag}  \end{array} \right) =
{\cal R}( \theta_k^{\Gamma}) \left( \begin{array}{c}
\hat{c}_k  \\
\hat{c}_{-k}^{\dag}  \end{array} \right) ,
\eeq
where $\{ \hat{\gamma}_k^{\Gamma}\}$ represent fermionic
quasi-particles that satisfy the 
canonical anticommutation relations
$\{ \hat{\gamma}_k^{\Gamma}, \hat{\gamma}_{k'}^{\Gamma}\}=0$ and 
$\{ \hat{\gamma}_k^{\Gamma}, (\hat{\gamma}_{k'}^{\Gamma})^{\dag}\}= \delta_{k,k'}$, ${\cal R}$ is a unitary rotation matrix and
\beq
\tan (2 \theta_{k}^{\Gamma})
= \frac{\sin k}{\Gamma-\cos k}.
\label{eq:tanth}
\eeq
For $k>0$ this relation has to be inverted with $2\theta^\Gamma_k \in [0,\pi]$, whereas the values of $\theta^\Gamma_k$ for $k<0$ are obtained by using the property $\theta^\Gamma_{-k}  = - \theta^\Gamma_k$.
In terms of these quasi-particles the Hamiltonian $\hat{H}$ in Eq.~\reff{quadratic-H} reads
\beq\label{H_diagonal_Ising}
\displaystyle \hat{H}^{+}(\Gamma) = 
\sum_{k>0}\epsilon_k(\Gamma) \left(\hat{\gamma}^{\Gamma\,\dag}_{k} 
\hat{\gamma}^{\Gamma}_{k} +\hat{\gamma}^{\Gamma\,\dag}_{-k} \hat{\gamma}_{-k}^{\Gamma} - 1 \right),
\eeq
where 
\beq
\epsilon_k(\Gamma)=2\sqrt{\Gamma^2-2 \Gamma \cos k +1}
\label{eq:energy}
\eeq
is the dispersion law of the quasi-particles. In Eq.~\reff{H_diagonal_Ising} the superscript $+$ of $\hat{H}$ indicates  
that  $\hat{H}^+$ is the projection of the full Hamiltonian $\hat{H}$ in Eq.~\reff{quadratic-H} onto the sector with an even
number of fermions and that antiperiodic boundary conditions are enforced 
by choosing the wave-vectors $k$ as in Eq.~\reff{Fourier_Transform}. 
(Hereafter the superscript $+$ is understood.)
%
%
The ground state $\ket{0}_{\Gamma}$ of the chain is the vacuum of the quasi-particles, defined by 
$\hat{\gamma}_k^{\Gamma} \ket{0}_{\Gamma}=0$, 
$\forall k$,
which takes the form
\beq\label{GS_fermions}
\ket{0}_{\Gamma} = \prod_{k>0} \, 
(\cos \theta_k^{\Gamma} + i \sin \theta_k^{\Gamma}\; \hat{c}_{k}^{\dag}\hat{c}_{-k}^{\dag}) \ket{\tilde{0}} \propto \prod_{k>0} \rme^{i (\tan \theta^\Gamma_k )\, \hat{c}_{k}^{\dag}\hat{c}_{-k}^{\dag}}\ket{\tilde{0}}, 
\eeq
as a function of the fermions $\hat{c}_k$, 
where $ \ket{\tilde{0}}$ is the vacuum of the fermions $\hat{c}_{k} \ket{\tilde{0}}=0$, $\forall k$. %
Hence, $\ket{0}_{\Gamma}$ has the structure of  a superposition of pairs $\hat{c}_{k}^{\dag}\hat{c}_{-k}^{\dag}$, \ie, 
of pairs of fermions with opposite momenta. 
%
%
At zero temperature and in the thermodynamic limit, 
the system is characterized by a quantum
phase transition at $\Gamma=1$, where the gap of the dispersion relation $\epsilon_k(\Gamma)$ 
closes. The quantum phase transition separates a paramagnetic 
phase (PM, $\Gamma>1$) with vanishing order parameter $\la\hat{\sigma}^x_i\ra$ 
from a ferromagnetic phase (FM, $\Gamma < 1$) with spontaneous symmetry breaking
$\langle\hat{\sigma}_i^x\rangle \neq 0$ and long-range order along the $x$ direction. 
%
%
However, the long-range order disappears as soon as  the temperature $T$ takes non-vanishing values.
As far as the transverse magnetization $\hat{\sigma}^z$ is concerned, instead,  $\langle \hat{\sigma}_i^z\rangle \neq 0$ 
for all $\Gamma > 0$ and all $T >0$.

\subsection{Equilibrium and non-equilibrium dynamics}
\label{Sec:dyn_quench}

Thanks to the transformations \reff{JW_fermions}, \reff{Fourier_Transform}
and  \reff{B_transform} the Hamiltonian $\hat H$ defined in Eq.~\reff{H_Ising} 
takes the quadratic diagonal form of Eq.~\reff{H_diagonal_Ising},  which makes the model and its dynamics
exactly solvable: indeed, in terms of the quasi-particle operators $\hat{\gamma}_k^{\Gamma}$ 
associated with $\hat{H}(\Gamma)$ one has access to all (thermo)dynamical properties. 
Within the Heisenberg picture these quasi-particles have a simple evolution 
\beq
\label{eq:evolution}
\left( \begin{array}{c}
\hat{\gamma}_k^{\Gamma}(t)  \\
\hat{\gamma}_{-k}^{\Gamma \, \dag}(t)  \end{array} \right) =
\left( \begin{array}{cc}
\rme^{- i \epsilon_k(\Gamma) t} & 0  \\
0 &\rme^{ i \epsilon_k(\Gamma) t}  \end{array} \right)
\left( \begin{array}{c}
\hat{\gamma}_k^{\Gamma}  \\
\hat{\gamma}_{-k}^{\Gamma \, \dag}  \end{array} \right) \equiv 
{\cal U}( \epsilon_k^{\Gamma},t)
\left( \begin{array}{c}
\hat{\gamma}_k^{\Gamma}  \\
\hat{\gamma}_{-k}^{\Gamma \,  \dag}  \end{array} \right),
\eeq
[where $\epsilon_k^\G \equiv \epsilon_k(\G)$] independently of the initial state of the system. 
Accordingly, the number operator  $\hat{n}_k^{\Gamma} \equiv \hat{\gamma}_k^{\Gamma\,\dag} \hat{\gamma}_k^{\Gamma}$ of each 
kind of quasi-particle does not evolve in time $\hat{n}_k^{\Gamma}(t) = \hat{n}_k^{\Gamma}$ and 
its expectation value 
$\langle \hat{n}_k^{\Gamma}(t) \rangle$ on an arbitrary measure (\eg, on the initial state) is a constant of motion. 
%
%

\begin{figure}[h]
\centering
\vspace{0.5cm}
 \includegraphics[width=0.2\textwidth,angle=90]{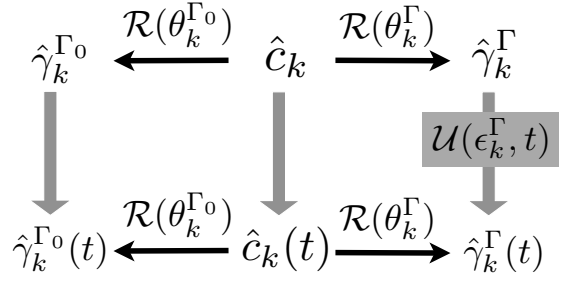} 
\caption{Schematic representation of the relations between the fermionic operators $\hat c_k$ and the quasi-particles $\hat\gamma_k^\Gamma$ and $\hat \gamma_k^{\Gamma_0}$ 
of the final and initial Hamiltonians, respectively. Black arrows indicate the linear mapping provided by the Bogoliubov rotation Eq.~\reff{B_transform}, whereas the grey 
vertical arrows indicate the time evolution, which takes the form~\reff{eq:evolution} in the basis of the quasi-particles $\hat\gamma_k^\Gamma$ of the final Hamiltonian $\hat H(\Gamma)$.}
\label{fig:mapping}
\end{figure}

In a quench, the system is prepared at $t=0$ in the ground state 
$\ket{0}_{\Gamma_0}$ of $\hat{H}(\Gamma_0)$, 
while it is subsequently allowed to evolve, isolated, according to the Hamiltonian $\hat{H}(\Gamma)$.  
The quench from $\Gamma_0$ to $\Gamma$ injects  into the system an extensive amount of 
energy 
%
%
which is henceforth conserved. (The statistics of this energy has been recently studied, \eg, in Refs.~\cite{Silva,GambSilva}.)
The dynamic observables one is interested in can typically be expressed in terms of the operators $\{\hat \sigma^a_i\}_{i,a}$ and via Eqs.~\reff{JW_fermions} and \reff{sz-c} in terms  of the fermions  $\{ \hat c_k, \hat c_{-k}^\dagger\}_k$ in momentum space. 
A convenient way to calculate the associated dynamic correlations 
after the quench
consists in expressing the (time-dependent) operators $\{ \hat c_k(t), \hat c_{-k}^\dagger(t)\}_k$  in terms of the operators 
$\{\hat{\gamma}_k^{\Gamma_0}\}_k$ which diagonalize the original Hamiltonian $\hat{H}(\Gamma_0)$. 
The merit of this procedure is evident when calculating expectation 
values over $\ket{0}_{\Gamma_0}$,
because $\{ \hat{\gamma}_k^{\Gamma_0}\}_k$ act trivially on their vacuum $\ket{0}_{\Gamma_0}$. %
On the other hand, the dynamics after the quench takes a particular 
simple form [see Eq.~\reff{eq:evolution}] if the operators one is interested in 
are  expressed in terms of the quasi-particles of the final Hamiltonian $\hat H(\Gamma)$. 
Figure~\ref{fig:mapping} summarizes schematically the relations between the various operators: 
black arrows indicate the 
transformations ${\cal R}$ which connect them, given explicitly in Eq.~\reff{B_transform}.  The mapping in the direction opposite to 
the one indicated by an arrow is realized by the inverse transformation ${\cal R}^{-1} = {\cal R}^\dagger$. The grey vertical arrows 
indicate the time evolution, which takes the form of Eq.~\reff{eq:evolution} in the basis of the quasi-particles 
$\{\hat \gamma_k^{\Gamma}, \hat \gamma_{-k}^{\Gamma\dagger}\}$.
In order to solve the dynamics of the model, 
one first expresses these quasi-particles 
$\{\hat \gamma_k^{\Gamma}, \hat \gamma_{-k}^{\Gamma\dagger}\}$ in terms of 
$\{\hat \gamma_k^{\Gamma_0}, \hat \gamma_{-k}^{\Gamma_0\dagger}\}$, which requires a total rotation 
${\cal R}( \theta_k^{\Gamma}) {\cal R}^{\dag}( \theta_k^{\Gamma_0}) = {\cal R}( \delta_k (\Gamma,\Gamma_0))$ of the suitable angle 
$\delta_k(\Gamma,\Gamma_0) \equiv \theta_k^{\Gamma}-\theta_k^{\Gamma_0}$, as indicated in Fig.~\ref{fig:mapping}.
Then the quasi-particles $\{\hat \gamma_k^{\Gamma}, \hat \gamma_{-k}^{\Gamma\dagger}\}$ 
are evolved according to Eq.~\reff{eq:evolution} in order to obtain  $\{\hat \gamma_k^{\Gamma}(t), \hat \gamma_{-k}^{\Gamma\dagger}(t)\}$. In terms of the latter, 
the time-dependent operators $\{ \hat c_k(t), \hat c_{-k}^\dagger(t)\}$ are eventually 
expressed according to Eq.~\reff{B_transform} via a rotation ${\cal R}^\dagger(\theta_k^\Gamma)$.
%

Combining these various transformations, the time-dependent operators $\{\hat c_k(t), \hat c_{-k}^\dagger(t)\}$ are given in terms of 
$\{\hat \gamma_k^{\Gamma_0}, \hat \gamma_{-k}^{\Gamma_0\dagger}\}$ by
\beq
\displaystyle \left( \begin{array}{c} 
 \hat{c}_k (t)  \\[2mm]
\hat{c}_{-k}^{\dag}(t)  \end{array} \right) = 
{\cal R}^{\dag}( \theta_k^{\Gamma}){\cal U}( \epsilon_k^{\Gamma},t)
{\cal R}( \delta_k (\Gamma,\Gamma_0)) 
\left( \begin{array}{c}
\hat{\gamma}_k^{\Gamma_0}  \\[2mm]
\hat{\gamma}_{-k}^{\Gamma_0\;\dag}  \end{array} \right)
 \equiv
\left( \begin{array}{cc}
u_k^{\Gamma,\Gamma_0}(t) & - [v_k^{\Gamma,\Gamma_0}(t)]^*  \\[2mm]
 v_k^{\Gamma,\Gamma_0}(t) & [u_k^{\Gamma,\Gamma_0}(t)]^*
 \end{array} \right)
\left( \begin{array}{c}
\hat{\gamma}_k^{\Gamma_0}  \\[2mm]
\hat{\gamma}_{-k}^{\Gamma_0 \; \dag}  \end{array} \right) 
\ ,
\label{Eq:Dynamics_quench}
\eeq
where
\beq
\begin{cases}
\displaystyle u_k^{\Gamma,\Gamma_0}(t) = \rme^{-i \epsilon^\G_k t}   \cos \theta_k^{\Gamma} \cos( \theta_k^{\Gamma}- \theta_k^{\Gamma_0})
+ \rme^{i  \epsilon^\G_k t}  \sin \theta_k^{\Gamma} \sin( \theta_k^{\Gamma}- \theta_k^{\Gamma_0}) ,
\\
\displaystyle v_k^{\Gamma,\Gamma_0}(t) = i \rme^{- i \epsilon^\G_k t}\sin \theta_k^{\Gamma} \cos( \theta_k^{\Gamma}- \theta_k^{\Gamma_0})
-  i \rme^{i  \epsilon^\G_k t} \cos \theta_k^{\Gamma} \sin( \theta_k^{\Gamma}- \theta_k^{\Gamma_0}) ,
\end{cases}
\label{eq:ukvk}
\eeq
and $\epsilon^\G_k \equiv \epsilon_k(\Gamma)$. This mapping allows one to express the average 
\beq
\label{def:aver}
\langle \bullet \rangle = {}_{\Gamma_0}\langle 0 | \bullet | 0 \rangle_{\Gamma_0}
\eeq 
over the initial condition $\ket{0}_{\Gamma_0}$
in terms of $u_k^{\Gamma,\Gamma_0}(t)$ and 
$v_k^{\Gamma,\Gamma_0}(t)$ defined above.  

After the quantum quench, all the observables but the
integrals of motion (and possible functions of them) show a non-stationary behavior.
After a transient (studied in Refs.~\cite{Igloi00,Igloi11} for the chain with free boundaries)
the system reaches an asymptotic stationary regime. 
The typical time scale of this transient depends on the observable under study and on the 
initial and final values of the parameters, \ie, on $\Gamma_0$ and $\Gamma$, respectively.
For some observables the approach to the stationary value occurs 
via an algebraic decay in time. For other
observables, instead, this decay is exponential and becomes faster upon increasing the energy injected in the system,
which actually increases upon increasing $|\Gamma-\Gamma_0|$.

\subsection{Effective temperatures for the Ising model}
\label{Sec:Teff_Ising}

In  this Section we 
specialize the various definitions of effective temperatures proposed in the literature for quantum quenches and 
discussed in 
Section~\ref{sec:Teff_static}, in the case of the isolated quantum Ising chain. 
Because of the  unitary dynamics, the energy of the system is conserved. 
Accordingly, if a thermal behavior emerges long after the quench, 
the (statistical) expectation value of the energy
calculated on the corresponding ensemble has to match the (quantum-mechanical) expectation value of the energy of the 
system right after the quench ~\cite{Jaynes}. 
This suggests 
comparing the energy ${}_{\Gamma_0}\!\langle 0 |\hat{H}(\Gamma) | 0 \rangle_{\Gamma_0}$ right after the quench 
with an equilibrium thermal average $\langle\hat{H}(\Gamma)\rangle_{T=T_{\rm eff}^E}$ and  
defining an effective temperature $T_{\rm eff}^{E}(\Gamma,\Gamma_0)$ in such a way that these two averages coincide.
By expressing $\hat H(\Gamma)$ in Eq.~\reff{H_diagonal_Ising} in terms of $\hat \gamma_k^{\Gamma_0}$ (see 
Fig.~\ref{fig:mapping}), one readily finds~\cite{Rossini10_short,Rossini10_long,Calabrese11}
\beq
{}_{\Gamma_0}\!\langle 0 |\hat{H}(\Gamma) | 0 \rangle_{\Gamma_0} = - \int_0^{\pi}\! \frac{\rmd k}{2\pi} \ \epsilon_k(\Gamma) \!  
\cos \Delta_k(\Gamma,\Gamma_0) ,
\label{eq:EnQ}
\eeq
where $\Delta_k(\Gamma,\Gamma_0) \equiv 2 \delta_k(\Gamma,\Gamma_0)$ satisfies [see Eq.~\reff{eq:tanth}]
\beq
\cos \Delta_k(\Gamma ,\Gamma_0) = 
\frac{4 ~[\Gamma \Gamma_0 - (\Gamma+\Gamma_0) 
\cos k + 1]}{\epsilon_k(\Gamma) \epsilon_k(\Gamma_0)}.
\label{eq:Delta_k}
\eeq
In Eq.~\reff{eq:EnQ} we took the thermodynamic limit $L \to \infty$, which is assumed henceforth, 
and which allows one to replace $\sum_{k>0} \to L\int_0^\pi \rmd k/(2\pi)$. 
The angle $\Delta_k(\Gamma,\Gamma_0)$ in the previous equation is a crucial quantity, as it encodes the 
dependence on the initial state and fixes the (non-thermal) statistics of the excitations created at $t=0$.
Indeed, $\cos \Delta_k $ determines the expectation value $\langle \hat n_k^\Gamma \rangle$ over the initial state 
$|0\rangle_{\Gamma_0}$  of the population $\hat n_k^\Gamma \equiv \hat \gamma_k^{\Gamma\dagger}\hat\gamma_k^\Gamma$ of the quasi-particles of $\hat H(\Gamma)$ in the $k$-th mode
\beq
\langle \hat n_k^\Gamma  \rangle = \langle \hat n_{-k}^\Gamma  \rangle  = \frac{1 - \cos \Delta_k}{2},
\label{eq:cos-initial}
\eeq
which follows by direct calculation from Eq.~\reff{B_transform}.
It is convenient to mention here that if the chain with Hamiltonian $\hat H(\Gamma)$ is in equilibrium within  
a Gibbs ensemble $\hat{\rho}(T)=\exp[-\hat{H}(\Gamma)/T]/Z(T)$ at temperature $T$,  
the average occupation number 
$\langle \hat{n}_k^{\Gamma} \rangle = \langle \hat{n}_{- k}^{\Gamma} \rangle =
1/\big[1+\rme^{\epsilon_k(\Gamma)/T}\big]$ can be formally obtained from the expression \reff{eq:cos-initial} (valid for the quench), with the substitution 
\beq
\cos \Delta_k \mapsto \tanh(\epsilon_k(\Gamma)/(2T)). 
\label{eq-neq-conn}
\eeq
We will see below that this formal mapping is actually effective for a variety of observables, 
as it can be verified from direct calculations. 
In particular, the average energy within such an ensemble can be expressed as 
\beq
\langle\hat{H}(\Gamma)\rangle_T = \mbox{Tr}\, [\hat{H}(\Gamma) \hat{\rho}(T)] = 
- \int_0^{\pi}\! \frac{\rmd k}{2\pi} \ \epsilon_k(\Gamma) \! 
\tanh\frac{\epsilon_k(\Gamma)}{2 T}.
\label{eq:thexp}
\eeq
As in the case of the average occupation number, this expression 
can also be obtained from the corresponding one
after a quench in Eq.~\reff{eq:EnQ} via 
the formal substitution in Eq.~\reff{eq-neq-conn}.
According to the strategy discussed in Sec.~\ref{sec:Teff_static}, one can implicitly define a
global effective temperature  $T_{\rm eff}^{E}(\Gamma,\Gamma_0)$ from the equality of 
energy averages 
\beq
{}_{\Gamma_0}\!\langle 0 |\hat{H}(\Gamma) | 0 \rangle_{\Gamma_0} = 
\langle\hat{H}(\Gamma)\rangle_{T = T_{\rm eff}^{E}(\Gamma,\Gamma_0)} ,
\label{Teff_energy}
\eeq
where the l.h.s.~is given by Eq.~\reff{eq:EnQ} and the r.h.s by Eq.~\reff{eq:thexp}.

In addition, it is also possible to define a mode-dependent effective  temperature 
$T_{\rm eff}^{k}$~\cite{Rigol,Rossini10_short,Rossini10_long,Calabrese11} 
by requiring 
the integrands in Eqs.~\reff{eq:thexp} and \reff{eq:EnQ} to be equal,
\ie, by defining $T_{\rm eff}^{k}$ such that 
\beq
\cos \Delta_k(\Gamma,\Gamma_0) = \tanh
\frac{\epsilon_k(\Gamma)}{2 T_{\rm eff}^k(\Gamma,\Gamma_0)}.
\label{Teff_GGE}
\eeq
This is nothing but the temperature that controls the population $\hat n_k^\Gamma$ 
of the $k$-th mode and indeed $T^k_{\rm eff}$ could be equivalently derived  by imposing that each mode $k$
is populated according to a Fermi distribution with temperature  $T_{\rm eff}^{k}(\Gamma,\Gamma_0)$ such that
$\langle \hat{n}_k^{\Gamma} \rangle = \langle \hat{n}_{- k}^{\Gamma} \rangle =
1/\big[1+\rme^{\epsilon_k(\Gamma)/T_{\rm eff}^{k}}\big]$.
Note that in the thermodynamic limit $T_{\rm eff}^{k}(\Gamma,\Gamma_0)$ 
becomes a continuous function of $k$. This means that
the diagonal and quadratic structure of $\hat{H}(\Gamma)$ (\ie, the integrability
of the model) naturally introduces an infinity (an extensive number) of  ``microscopic" temperatures,
each one associated with a particular integral of motion $\langle \hat{n}_k^{\Gamma} \rangle$. 
The relevance of these temperatures is transparent by
recalling that all expectation values are eventually determined by 
functions of $\langle \hat{n}_k^{\Gamma} \rangle$. However, for generic observables,
these functions are typically combinations of multidimensional integrals, 
determinants, oscillatory factors in times, etc. and  one cannot rule out {\it a priori} an emergent  
effective thermal behavior. 

\subsection{Why a critical quench?}
\label{subsec:why}

Although all the considerations up to now are completely general, in what follows we will
focus on the specific case of critical quenches,  \ie, $\Gamma=1$. 
This choice is motivated by the following heuristic arguments:
\begin{enumerate}
\item There is some evidence that  the 
expectation values of certain observables 
 in the long-time limit after a quench to $\Gamma=1$ do indeed coincide with the ones 
 of a thermal state at
temperature $T_{\rm eff}^E(\Gamma=1,\Gamma_0)$.  
This  does not hold for quenches towards a phase with a gap (\ie, with $\Gamma\neq 1$). 
We will recall some of these results in the next Section. 
It is then natural to investigate up to which extent the apparent ``thermalization"
found for such one-time quantities at $\Gamma=1$ carries over to two-time
observables in the same stationary regime. The generalized
FDRs precisely provide the tool to accomplish this goal.
\item One can heuristically argue 
that the correlations between the entire system
and a subpart of it might effectively act as a 
thermal bath for the subsystem. In this case, a gapless spectrum 
is expected to favor this effect, 
because energy exchanges at all scales 
are facilitated by the absence of a gap.  
\item Since our  
analysis will be primarily done as a function of the frequency $\omega$ and eventually focus on the $\omega\to0^+$ limit, it is 
natural to start the investigation in the absence of an energy gap in the spectrum. 
The gap might in fact determine a low-frequency threshold below which the
spectral representation of some observables becomes trivial.
Moreover,  depending on the specific observable considered, the presence of the
gap may 
introduce different frequency scales and non-analyticities in 
the Fourier transform of correlation and response functions.  
This issue definitely merits attention but it requires a dedicated study which is the natural continuation of the investigation
presented here for $\Gamma=1$.
\item The population $\langle \hat n_k^\Gamma\rangle$ of the $k$ modes as a function of $k$ 
[see Eqs.~\reff{eq:Delta_k} and \reff{eq:cos-initial}] 
for quenches to (and from) the critical point is qualitatively different 
from the one involving
the phases $\Gamma,\Gamma_0\neq 1$ with a gap in the spectrum, as shown in Fig.~\ref{fig:nk}.
\begin{figure}[h]
\centering
 \includegraphics[width=0.325\textwidth]{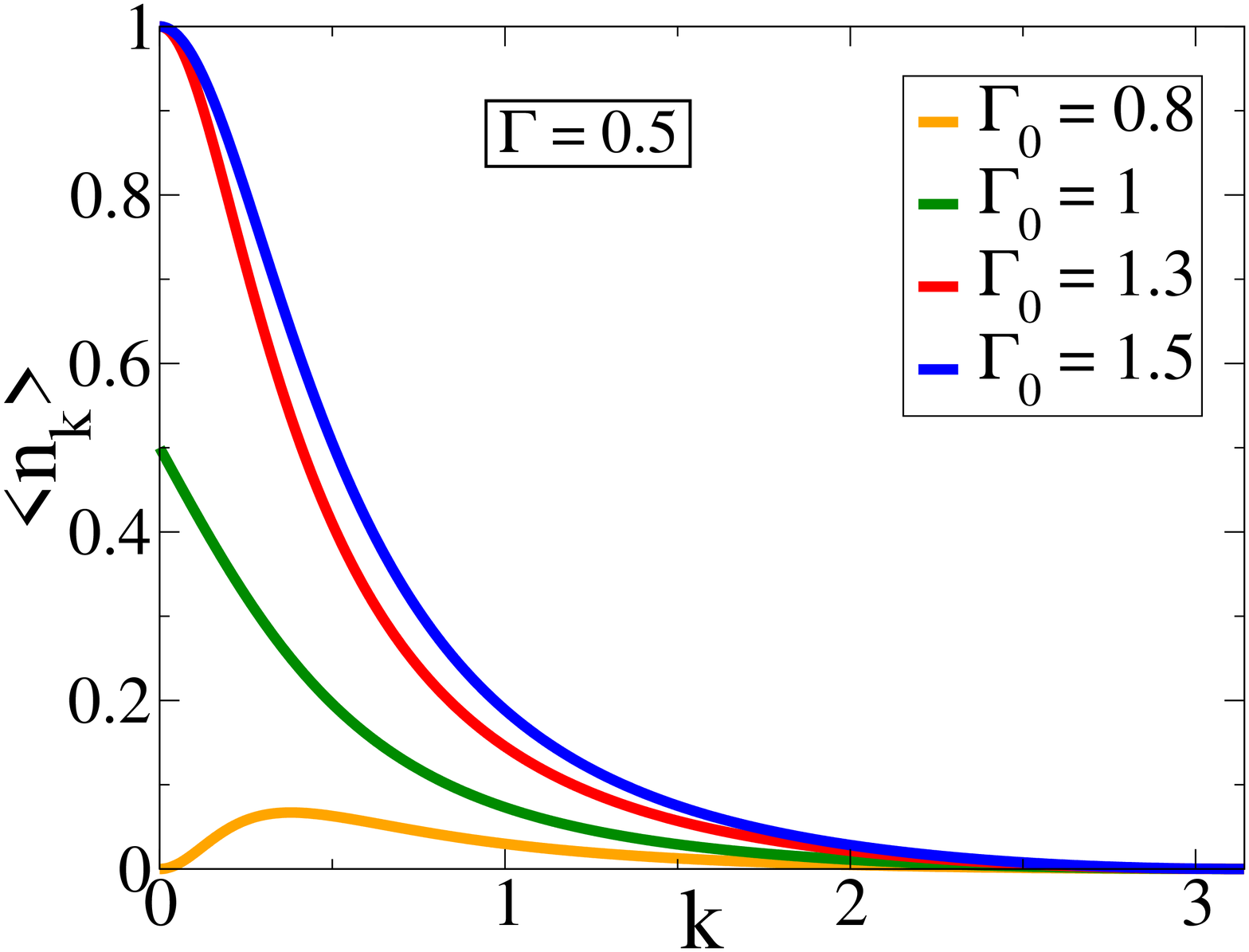} 
 \includegraphics[width=0.325\textwidth]{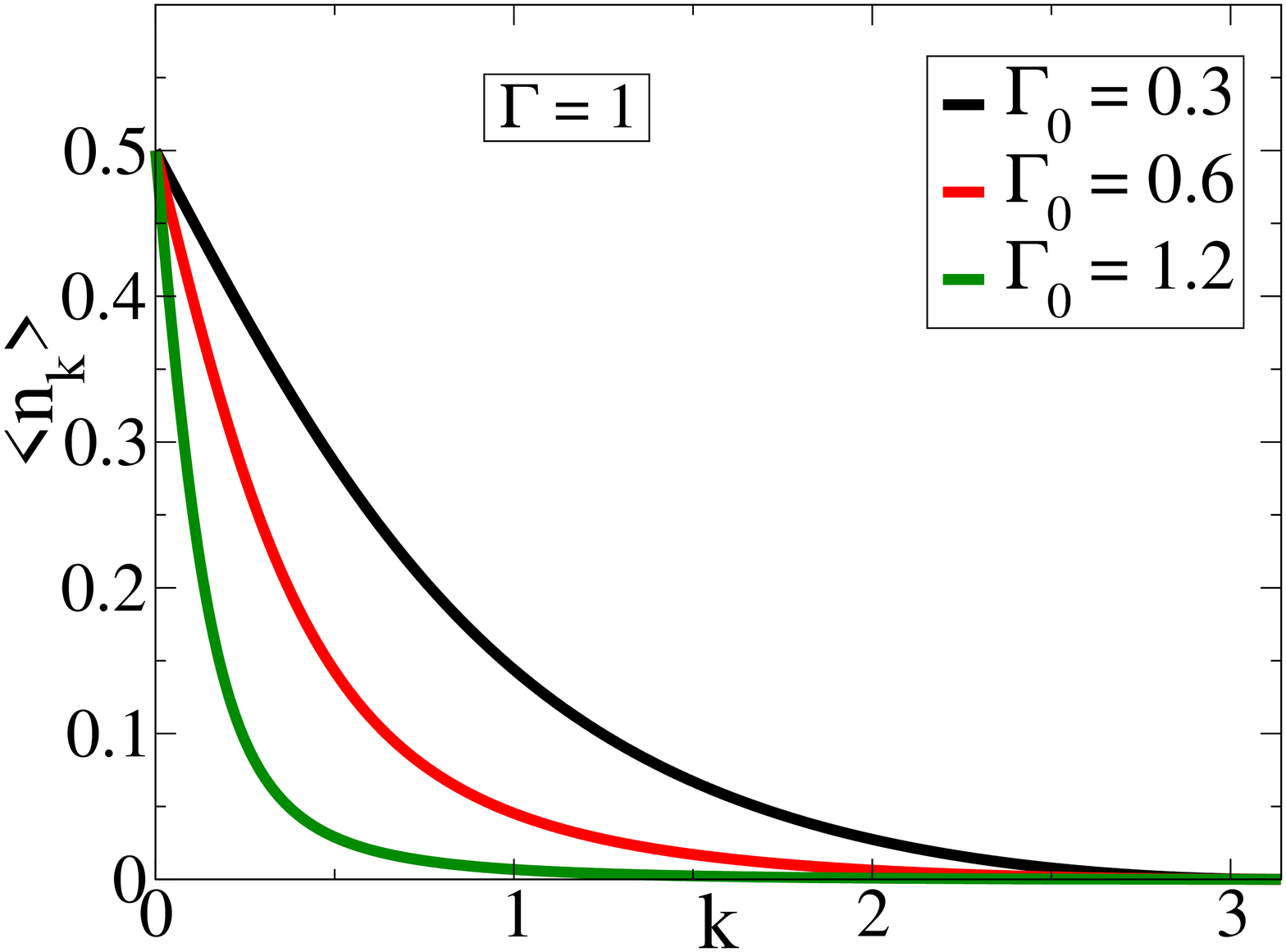} 
  \includegraphics[width=0.325\textwidth]{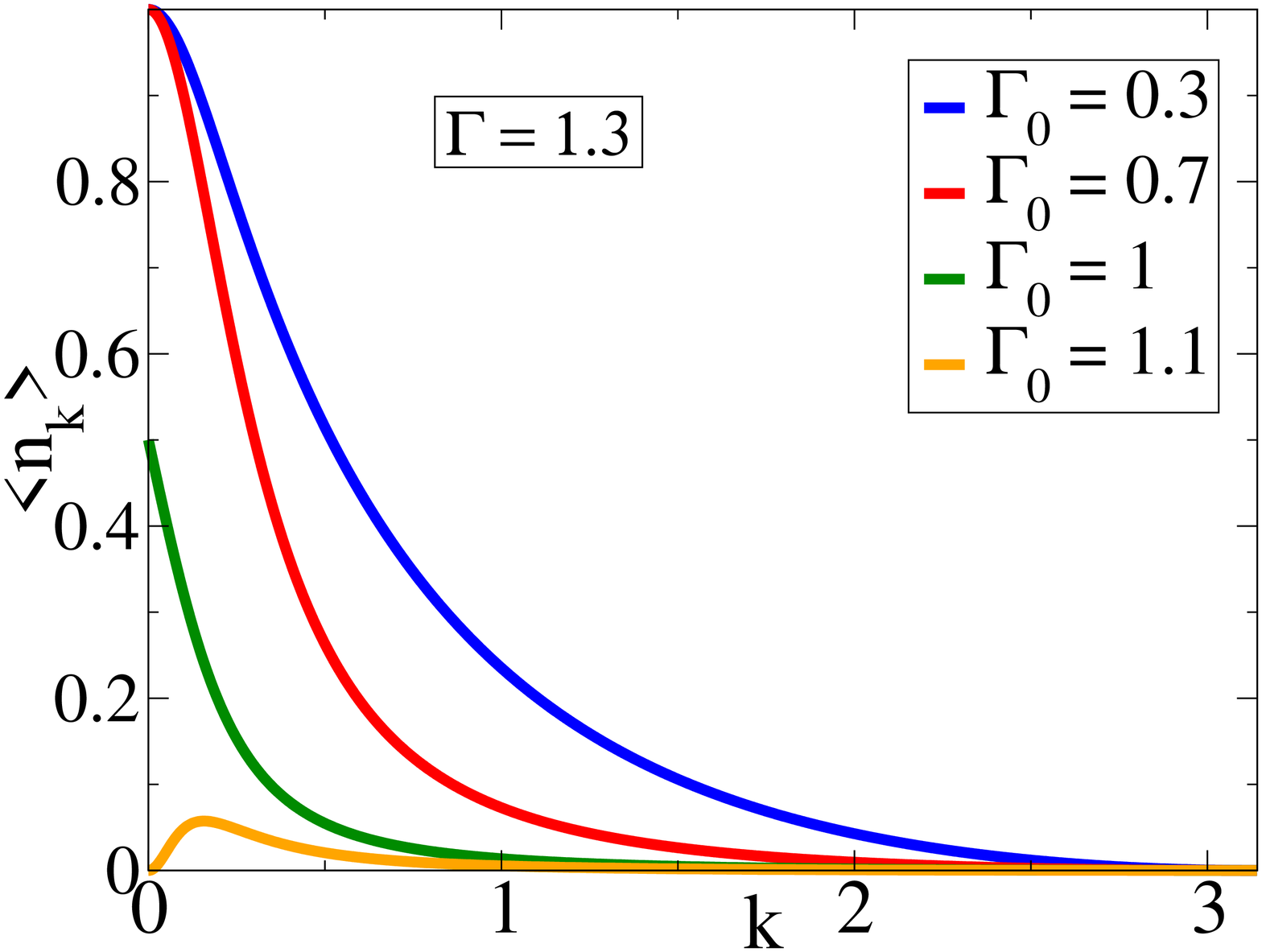} 
\caption{Quasi-particle occupation number $\langle\hat{n}_k\rangle$ as a function of
$k$ after the quench, for quenches towards the ferromagnetic phase $\Gamma=0.5$ (left panel),
the critical point $\Gamma=1$ (central panel), and the paramagnetic phase $\Gamma=1.3$ (right panel),
and various initial conditions. We note that quenches within the same (gapped) phase are characterized
by a low density of excitations $\hat{n}_k$ at low energy, while quenches across
the critical point are characterized by low energy modes with $\langle\hat{n}_k\rangle > 1/2$
which correspond to negative effective temperatures $T^k_{\rm eff}$.
}
\label{fig:nk}
\end{figure}
%
%
Indeed, the distributions of $\langle\hat{n}_k\rangle$ for quenches at the critical point (or from it)
have two properties in common with the equilibrium Fermi-Dirac distribution:
\begin{itemize}
\item It is a monotonically decreasing function of the energy (\ie, of $k$ for $k>0$).
\item It varies within the range $0\leq \langle\hat{n}_k\rangle\leq\frac12$ and therefore does not require the 
introduction of negative $T^k_{\rm eff}$.
\end{itemize}
When these conditions are not satisfied (for quenches to the gapped phases), states exist at higher energy
that have a larger overlap with the initial conditions than others at lower energy, 
and thus turn out to be 
more probable. This is in contrast with the behavior at the critical point and with any model
at Gibbs equilibrium, where the probability of a given state is monotonically decreasing with its energy.
\item The long-time, large-distance dynamical properties of a $d$-dimensional isolated quantum system after a quench from a ground state can be studied in terms of a suitable $d+1$-dimensional problem in a slab \cite{Calabrese06}. For a one-dimensional system quenched at a critical point with linear dispersion relation (\ie, dynamic exponent $z=1$), this mapping has far-reaching consequences because the corresponding $1+1$-dimensional problem is described by a boundary Conformal Field Theory (CFT) on the continuum. Interestingly enough, it turns out that the dynamic properties predicted within this approach are the same as those of the same CFT in equilibrium at a certain, finite temperature $T$, which should therefore naturally emerge after the quench of a generic one-dimensional system at its critical point. (Note, however, that the possible emergence of this thermal behavior depends on some rather general properties of the initial state~\cite{Cardy-talk}.) 

\end{enumerate}

\subsection{Dynamic observables}

In the following we will focus primarily on the transverse magnetization $\hat{\sigma}_i^z$ and on the order parameter
$\hat{\sigma}_i^x$, and  
we will denote by $C_\pm^z(t)$ and $C_\pm^x(t)$ the
corresponding autocorrelation functions. In addition, we will also 
consider the global magnetization 
$\hat{M}(t)=1/L\sum_{i=1}^{L} \hat{\sigma}_i^z$ and 
the corresponding correlation function $C_\pm^M(t)$.
These observables are distinguished by an important 
property~\cite{Rossini10_long,polkovnikov2010}:
$\hat{\sigma}_i^x$  is non-local with respect to the
quasi-particles in the sense that it has non-vanishing matrix elements
with most of the states of the Hilbert space.  On the contrary, $\hat{\sigma}_i^z$ 
is local in the same variables, in the sense that
it couples only few states. This distinction is 
more transparent if one recalls the expressions of these operators in terms of the Jordan-Wigner fermions 
of Eq.~\reff{JW_fermions}: $\hat{\sigma}_i^z$ is a quadratic
function of the $\hat{c}_k$ and therefore also
of the excitations $\hat{\gamma}_k^{\Gamma}$, while $\hat{\sigma}_i^x$
is the product of a string of fermions and therefore it is non-local in the
operators which diagonalize the Hamiltonian. 

Before presenting our results we
briefly summarize what is known about the  dynamics of $C_\pm^{z,x}$ at  
the critical point $\Gamma=1$ of the Ising chain in a transverse field.
At equilibrium ($\Gamma_0=\Gamma$)  the time decay of 
$\langle \hat{\sigma}_i^z(t+t_0)\hat{\sigma}_i^z(t_0)\rangle$
as a function of $t$ with fixed $t_0$ is algebraic $\sim |t|^{-3/2}$ at $T=0$, whereas
$\sim |t|^{-1}$ at finite temperature~\cite{Rossini10_short}.   
In the isolated system after the quench
($\Gamma_0\neq\Gamma$), instead, the
stationary decay of $C_+^z = \frac{1}{2} \langle \{ \hat{\sigma}_i^z(t+t_0),\hat{\sigma}_i^z(t_0) \} \rangle
-  \langle \hat{\sigma}_i^z(t+t_0) \rangle \langle  \hat{\sigma}_i^z(t_0)  \rangle$ 
is still algebraic, but 
with a different exponent $\sim |t|^{-2}$~\cite{Igloi00}.  
When its initial condition is chosen to be fully polarized
along the $z$-direction (corresponding to $\Gamma_0= \infty$),   
$C_-^z = \frac{1}{2} \langle [ \hat{\sigma}_i^z(t+t_0),\hat{\sigma}_i^z(t_0) ] \rangle$ 
follows the same power-law decay $\sim |t|^{-2}$, 
as one can infer from the results of Ref.~\cite{Karevski}.

The expectation value $\langle \hat{\sigma}^x_i(t) \rangle $ of the order parameter  $\hat{\sigma}^x_i(t)$ 
decays to zero in the long-time limit for all $\Gamma\neq\Gamma_0$. This is the same as  
 in thermal equilibrium at $T>0$, which is characterized by the absence of long-range order.
The equal-time two-point correlation function $\langle\hat{\sigma}_i^x(t_0)\hat{\sigma}_j^x(t_0)\rangle$, instead, 
displays  as a function of $t_0$ an exponential relaxation towards its stationary value.
This was first argued in Ref.~\cite{Calabrese06} on the basis of semi-classical methods 
and Conformal Field Theory (CFT) and
later shown to hold exactly via a suitable analysis of the model on the lattice~\cite{Calabrese11}.  
Moreover, 
$|\langle\hat{\sigma}_i^x(t+t_0)\hat{\sigma}_i^x(t_0)\rangle|$ was found~\cite{Rossini10_short,Rossini10_long} to decay exponentially $\propto \exp\{-t/\tau_Q\}$ as a function of $t$ (for fixed $t_0$), 
where the scale $\tau_Q$ turns out to coincide 
numerically --- at least for small quenches --- with the one $\tau_E$ which characterizes the time decay in equilibrium at a temperature $T\simeq T_{\rm eff}^E$ (with $T_{\rm eff}^E$ defined according to 
Eq.~\reff{Teff_energy}).
The observed exponential relaxation is
in contrast with the  equilibrium correlations of the 
order parameter at $T=0$, which decay algebraically both in space and time.   
To the best of our knowledge, instead, $C^x_-$ and therefore the response function $R^x$ 
have  not  been analyzed so far. 

\section{Results}
\label{sec:results}

In this Section we present and  discuss our results about the behavior of a variety of correlation and 
linear response functions. This study allows us to assess the possible relevance of an effective thermal 
description of the dynamics following a critical quantum quench of the Ising model described in 
Sec.~\ref{sec:Ising}. 

\subsection{Transverse magnetization}
\label{Sec:sigmaz}

We start our analysis by considering the expectation 
value of $\hat{\sigma}^z(t)$ in the long-time stationary state after the quench
(in view of translational invariance, we can drop the site index $i$ from the notation; 
the case of a time-dependent $\Gamma(t)$ was studied in full generality in Ref.~\cite{BMD70}): 
\beq
 \langle \hat{\sigma}^z \rangle_{\mathcal Q} \equiv 
 \displaystyle \lim_{t\to\infty} \langle \hat{\sigma}^z (t) \rangle 
 = 
 \int_0^{\pi} \frac{\rmd k}{\pi}~ 
\cos(2\theta_k^{\Gamma})  \cos \Delta_k(\Gamma,\Gamma_0),
\label{eq:Qavsz}
\eeq
where here and in what follows the thermodynamic 
limit $1/L \sum_{k>0} \to  \int_0^\pi \rmd k/(2\pi)$ 
is taken. 
(This expression for $\langle \hat{\sigma}^z \rangle_{\mathcal Q}$ can be obtained straightforwardly by considering the time-independent contributions which emerge upon expressing the fermions in Eq.~\reff{sz-c} in terms of $\hat \gamma^{\G_0}_k$ via Eq.~\reff{Eq:Dynamics_quench}, see also Refs.~\cite{CEF12,Rossini10_long,Igloi00}.)
As it is shown in Ref.~\cite{Rossini10_long} this asymptotic value differs in general from the one 
$\langle \hat \sigma^z\rangle_T$
that this observable would have in a Gibbs 
thermal ensemble at the temperature $T=T_{\rm eff}^E$ set by the energy of the initial state according to 
Eq.~\reff{Teff_energy}. 
However, at the critical point, it turns out that~\cite{Rossini10_long}
\beq
 \langle \hat{\sigma}^z \rangle_{\mathcal Q} =  \langle \hat{\sigma}^z \rangle_{T=T_{\rm eff}^E},
 \label{eq:eqsz}
\eeq
which, by itself, would suggest an effective thermalization within a Gibbs ensemble at temperature $T_{\rm eff}^E$. 
This is due to the fact that both for critical quenches in the stationary regime and at thermal 
equilibrium with $\Gamma=1$, the expectation value of $\hat \sigma^z$ is related to the one of the 
Hamiltonian: $\langle \hat{\sigma}^z \rangle = -\langle \hat{H} \rangle/2$. This relationship can be established 
by comparing Eq.~\reff{eq:EnQ} and Eq.~\reff{eq:Qavsz}, with the help of Eqs.~\reff{eq:costhcrit} and \reff{eq:energy},  which give $\epsilon_k(\G) = 4 \cos(2\theta^\G_k)$ for $\Gamma=1$. 
Accordingly,  Eq.~\reff{eq:eqsz} follows immediately from Eq.~\reff{Teff_energy}. 
[In passing, we anticipate here that at the critical point, 
$\langle \hat{\sigma}^z \rangle_{\mathcal Q} = - E'(0)$,
where $E'(0)$ is given in Eq.~\reff{eq:Ep0}.]

%
\begin{figure}[h]
\centering
 \includegraphics[width=0.435\textwidth]{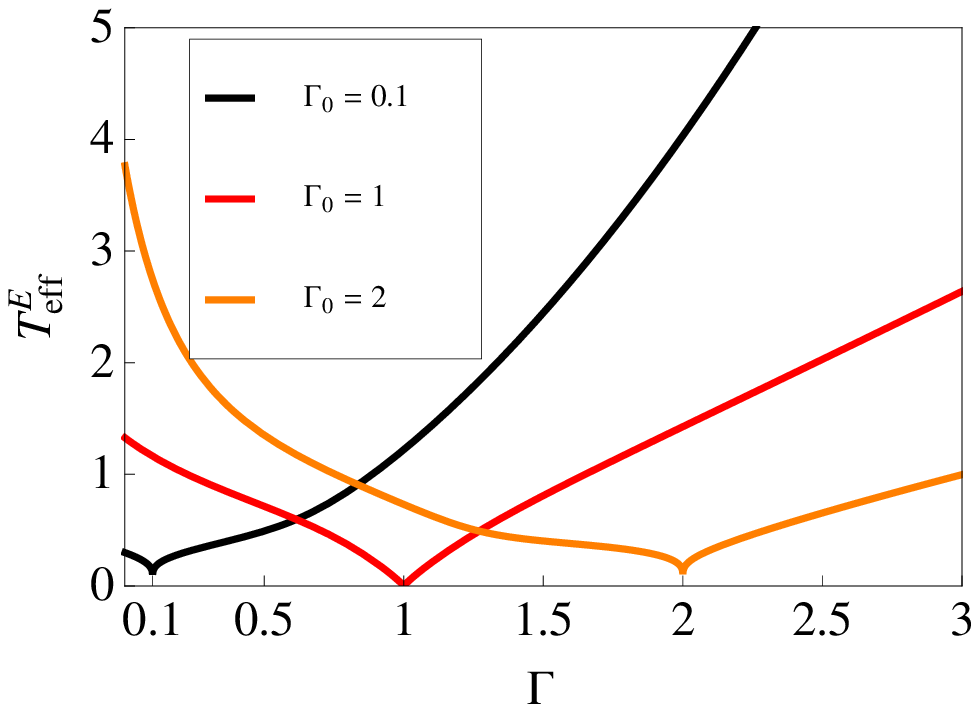} 
 \includegraphics[width=0.4\textwidth]{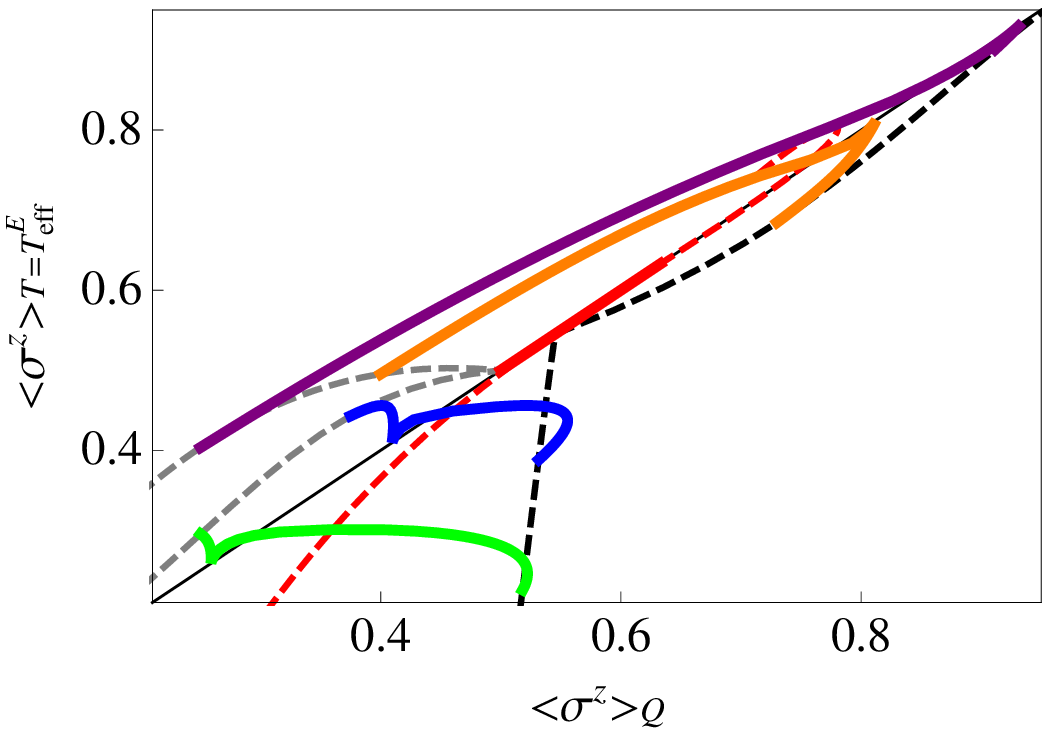} 
\caption{%
Left panel: Effective temperature $T^E_\eff(\G,\G_0)$ [see Eq.~\reff{Teff_energy}] as a function of $\G \in [0,3]$, for fixed 
values of $\G_0=0.1$, 1, and 2. As expected, the equilibrium value $T^E_\eff=0$ is recovered for $\G \to \G_0$.  
Right panel: Comparison between the stationary expectation value $\langle \hat{\sigma}^z 
\rangle_{\mathcal Q}$  [see Eq.~\reff{eq:Qavsz}]
after the quench and the thermal expectation value $\langle \hat{\sigma}^z  \rangle_T$ in equilibrium at temperature 
$T = T_{\rm eff}^E(\G,\G_0)$.  The solid lines are obtained by varying $\G_0\in [0,5]$, with fixed values of $\G = 0.5$ (green), 0.75 (blue), 1 
(red), 1.25 (orange), 2 (purple), from bottom to top along the diagonal. The dashed lines, instead, correspond to fixed values of $\G_0 = 0$ 
(grey leftmost curve), 1 (red central), 5 (black rightmost) and are obtained by varying $\G$.
For critical quenches $\Gamma=1$, $\langle \hat{\sigma}^z  \rangle_{T=T_{\rm eff}^E}$ and  
$\langle \hat{\sigma}^z \rangle_{\mathcal Q}$ take the same value upon varying $\G_0$ and indeed the corresponding curve lies along the diagonal.
}
\label{fig:sigmaz}
\end{figure}
%
The comparison between the expectation value obtained after the quench and
the thermal one can be highlighted as follows (see also Fig.~9 of Ref.~\cite{Rossini10_long} for an equivalent analysis).  For a fixed 
initial value $\G_0$,  Eq.~\reff{Teff_energy} provides an implicit 
equation for $T^E_{\rm eff}(\G,\G_0)$ which can be solved as a function of $\G$. The plot of this effective temperature is presented in 
the left panel of Fig.~\ref{fig:sigmaz} as a function of $\G$ varying 
between 0 and 3,  for fixed $\G_0=0.1$, 1, and 2. As expected, the equilibrium value $T^E_\eff = 0$ is recovered when
$\G$ approaches the value of $\G_0$ corresponding to that particular curve. In 
addition, the curves clearly show that the effective temperature $T^E_\eff$ increases as the "distance" $|\G-\G_0|$ from the 
equilibrium condition increases.
The knowledge of this effective temperature $T^E_\eff(\G,\G_0)$ for a given pair of values $(\G,\G_0)$ allows one to calculate the 
expectation value that  a generic observable 
would have at equilibrium in 
a Gibbs ensemble with temperature $T=T^E_\eff(\G,\G_0)$. 
On the other hand, one can also determine the long-time 
expectation value (if any)  of the same observable after a quench from $\G_0$ to 
$\G$. In case thermalization occurs, these two values have to coincide upon varying $\G$ and $\G_0$. On the right panel of 
Fig.~\ref{fig:sigmaz} we present this test for $\hat \sigma^z$, by plotting the 
thermal expectation value $\langle \hat{\sigma}^z \rangle_{T=T_{\rm eff}^E}$ (vertical axis) versus the one after the quench  
$\langle \hat{\sigma}^z \rangle_{\mathcal Q}$ (horizontal axis). The various 
solid curves are obtained by varying $\G_0$ between 0 and  5, with fixed values of $\G = 0.5$ (green), 0.75 (blue), 1 (red), 1.25 
(orange), 2 (purple), from bottom to top along the diagonal. The dashed 
lines, instead, are obtained upon varying $\G$ between 0 and  5, with fixed values of $\G_0 = 0$ (grey leftmost curve), 1 (red central), 
5 (black rightmost). For generic values of $\G$ and $\G_0$ the 
various curves clearly show that the average after the quench and the thermal average at temperature $T^E_\eff$ do not coincide, 
with the sole exception of quenches at the critical point $\G=1$, for 
which the corresponding solid curve in the right panel of Fig.~\ref{fig:sigmaz} lies completely along the diagonal of the plot (thin 
black line). 
As we anticipated in the previous Section, this result,
apparently confirmed by an analogous observation in CFT~\cite{Calabrese06}, was taken as an evidence for a sort of thermalization  in the model, at least for this particular 
choice of parameters.
However, it has been recently pointed out~\cite{CEF12,Cardy-talk} that the emergence of a well-defined, single effective temperature in CFT is actually a consequence of the particular choice of the initial condition in Ref.~\cite{Calabrese06}.

According to our general discussion in Sec.~\ref{sec:eff-temp} it is natural to investigate the issue of this apparent thermalization also 
in the light of the FDRs and of the various ``effective temperature" parameters that can be extracted from 
them and which, in principle, depend on $t$ or $\omega$. These parameters are 
expected to provide a more sensitive tool to probe the possible asymptotic thermal behavior. 

The two-time symmetric connected correlation and linear response functions for generic
$\Gamma$ and $\Gamma_0$ are given by (see App.~\ref{app:sz} for the details of the calculation):
\begin{align}
\label{CZ_general}
 C_{+}^{z}(t+t_0,t_0) &= \displaystyle  4  
 \int_0^{\pi}  \frac{\rmd k}{\pi}  \int_0^{\pi}  \frac{\rmd l}{\pi}  ~\RE \left[ v_k(t+t_0) \, 
 v_k^*(t_0) \, u_l(t+t_0) \, u_l^*(t_0) \right] , 
\\
\label{RZ_general}
R^{z}(t+t_0,t_0) &= \displaystyle - 8\, \theta(t)
 \int_0^{\pi}  \frac{\rmd k}{\pi}  \int_0^{\pi}  \frac{\rmd l}{\pi} ~ \IM \left[ v_k(t+t_0) \, v_k^*(t_0) \, u_l(t+t_0) \, u_l^*(t_0) \right],
\end{align}
with $v_k(t)\equiv v_k^{\Gamma,\Gamma_0}(t)$ and $ u_k(t) \equiv u_k^{\Gamma,\Gamma_0}(t)$ given by Eq.~\reff{eq:ukvk}.
For critical quenches ($\Gamma=1$, see App.~\ref{Appendix_sz_critical}) 
and in the stationary regime $t_0\to\infty$, one finds
\begin{align}
C_+^z(t) \equiv \lim_{t_0\to\infty} C_+^z(t+t_0,t_0) &= J_0^2(4t) - E^2(4t) + J_1^2(4t)  - [E'(4 t)]^2 ,\label{CZ} \\[2mm]
R^z(t) \equiv \lim_{t_0\to\infty} R^z(t+t_0,t_0) &=4\,\theta(t) [J_0(4t) E(4t) -J_1(4t) E'(4 t) ], \label{RZ}
\end{align}
where here and in the following $J_{\alpha}(\tau)$ indicates the
Bessel function of the first kind and order $\alpha$ (see,\eg, chapter 10 in Ref.~\cite{HMF}), whereas 
\beq
E(\tau) \equiv \int_0^\pi  \frac{\rmd k}{\pi} \sin( \epsilon_k \tau/4)   \cos \Dt_k (1,\Gamma_0) 
\label{eq:defE-mt},
\eeq
[see Eq.~\reff{eq:app-defE} and the plot of $E(\tau)$ in Fig.~\ref{fig:E-F-app}] with $\epsilon_k \equiv \epsilon_k(\Gamma=1)= 4 \sin (k/2)$ [see Eq.~\reff{eq:energy}]. 
The initial condition enters these expressions only via 
$\cos \Delta_k$ [see Eq.~\reff{eq:Delta_k}]. 
Remarkably, at the critical point $\Gamma=1$, this quantity turns out to depend on $\Gamma_0$ only through the ratio [see Eq.~\reff{eq:cosDcrit}]
\beq
\Y=\left(\frac{1+\Gamma_0}{1-\Gamma_0}\right)^2 > 1.
\label{eq:defY}
\eeq
Note that $\Y$ and consequently the correlation and response functions $C^z_+$ and $R^z$ are invariant under the transformation $\Gamma_0 \mapsto \Gamma_0^{-1}$ 
which maps a paramagnetic initial condition into a ferromagnetic 
one and vice versa. 
However, this is true only for the \emph{stationary} part of $C^z_+$ and $R^z$. Indeed, from the definition of $\theta_k$ 
in Eq.~\reff{eq:tanth} it follows that $\tan (2\theta_k^{1/\Gamma}  + 2\theta_k^{\Gamma}) = - \tan k$ and therefore, taking into 
account that Eq.~\reff{eq:tanth} implies $2\theta_k^{\Gamma=1} = (\pi - k)/2$,  one finds
$2\theta_k^{1/\Gamma}  + 2\theta_k^{\Gamma} = \pi - k$, which yields
$\Delta_k(\Gamma=1 ,\Gamma_0) = - \Delta_k(\Gamma=1 ,\Gamma_0^{-1})$ and thus a change of
the sign of $\sin \Delta_k$. 
With the help of the results in App.~\ref{app:sz}, especially Eq.~\reff{eq:Cp-app1}, it is possible to see that the non-stationary 
terms in $R^z$ and $C_+^z$ do actually depend on $\sin \Delta_k$ and, as a result, 
they are not invariant under the mapping 
$\Gamma_0 \mapsto \Gamma_0^{-1}$. As we focus below only on the stationary regime, we can 
restrict our analysis to initial conditions in the ferromagnetic phase $\Gamma_0 < 1$.

\begin{figure}[h]
\centering
 \includegraphics[width=0.45\textwidth]{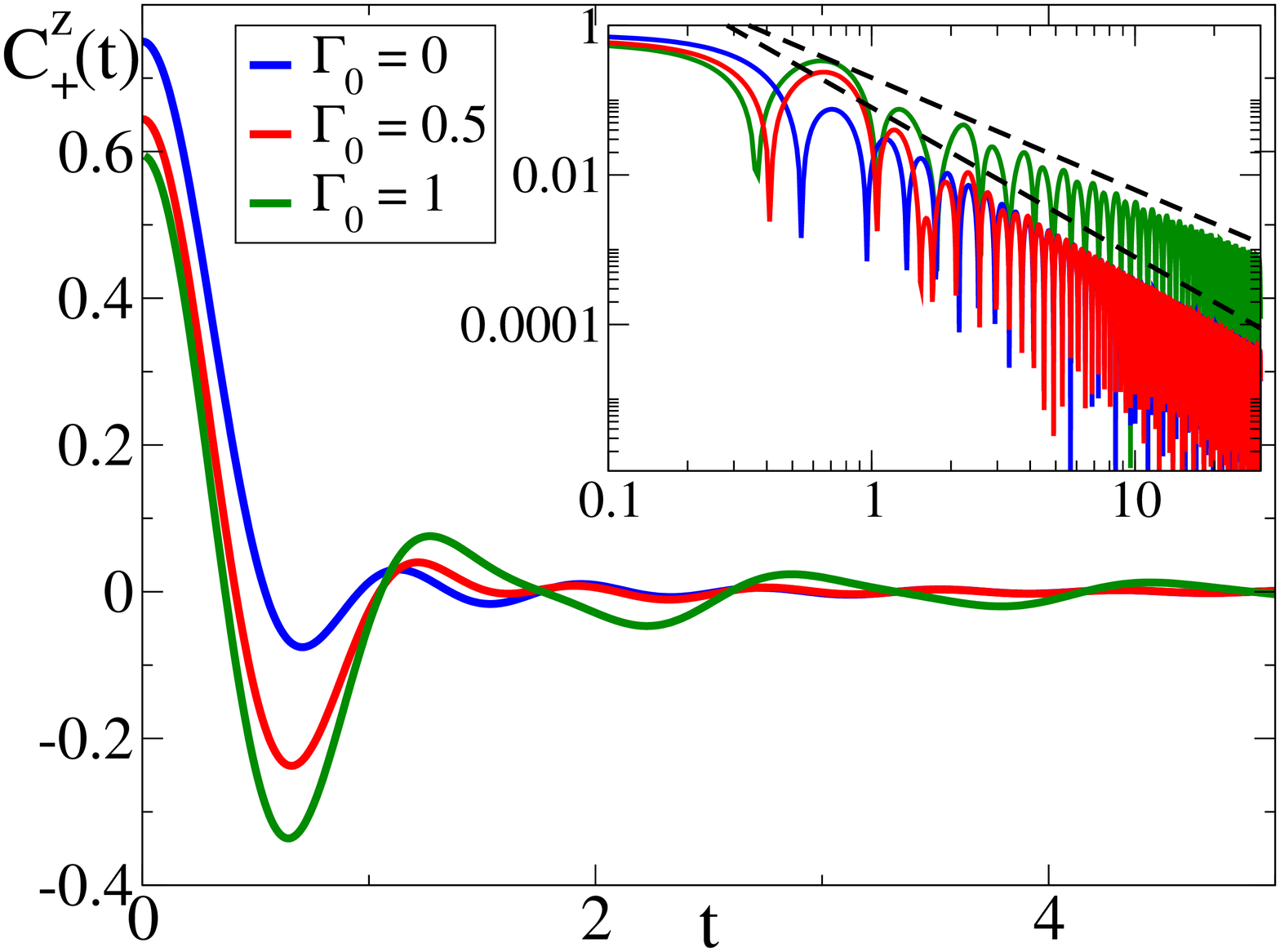} 
 \includegraphics[width=0.45\textwidth]{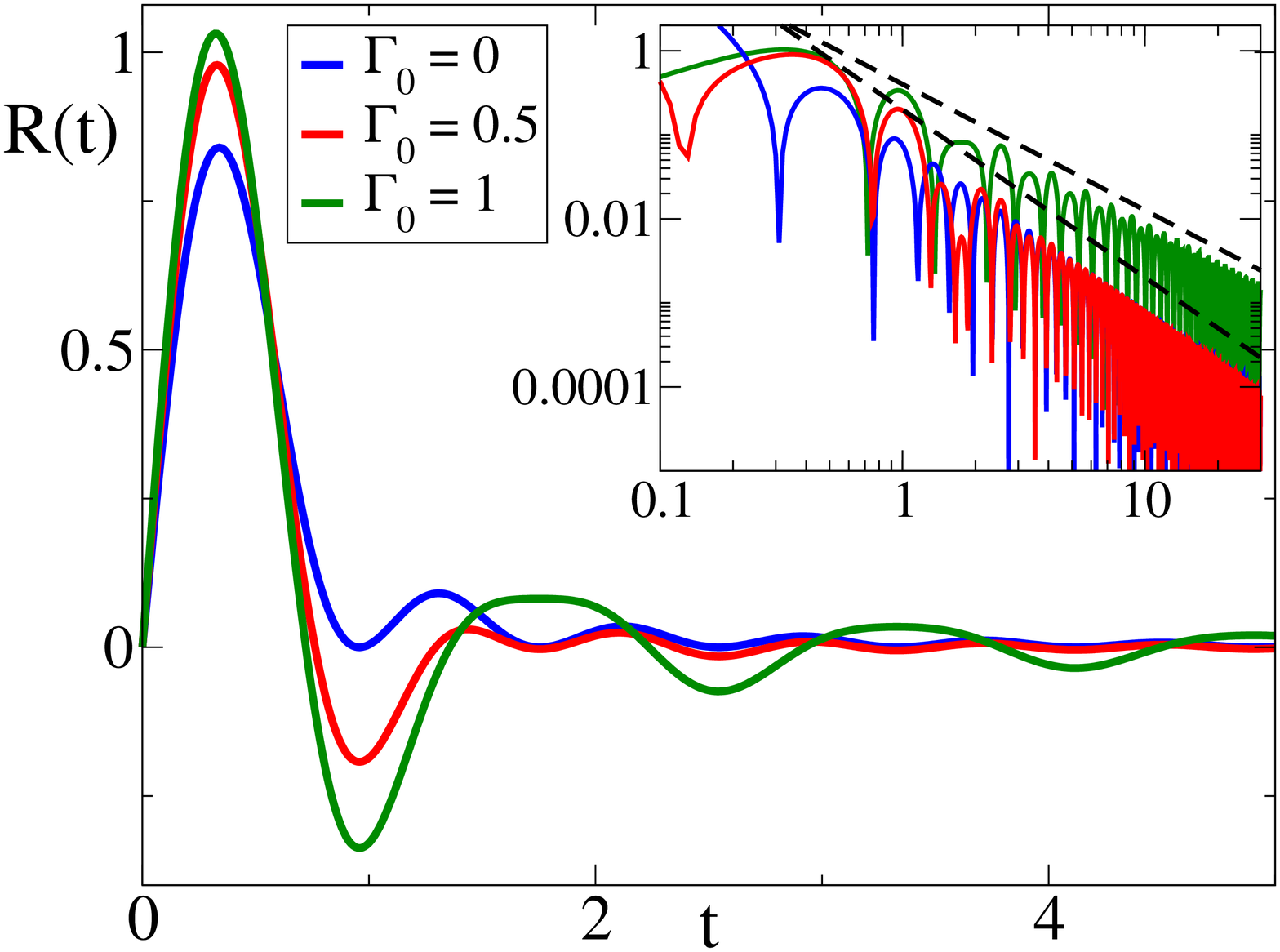} 
\caption{Time dependence of the correlation $C^z_{+}(t)$ (left panel) 
and linear response functions $R^z(t)$ (right panel)
 of the local transverse magnetization $\hat{\sigma}^z_i$ [see Eqs.~\reff{CZ} and \reff{RZ}] 
 in the stationary regime after a quench to the critical point $\Gamma=1$.
The blue, red and green solid lines correspond to $\Gamma_0=0$, 0.5 and 1 (equilibrium at $T=0$).  
The insets in the left and right panel highlight  the algebraic decay of  
$|C^z_{+}(t) |$ and $|R^z(t) - 1/(4  \pi\Y t^2)|$, 
respectively, which are compatible with the asymptotic forms~\reff{CZ_long_time} and \reff{RZ_long_time} for $\G_0\neq \G=1$ (blue and red solid lines), 
indicated by the lowermost thin dashed lines $\sim t^{-2}$ in these double logarithmic plots. 
In the case of equilibrium at zero temperature $\G_0=\G=1$ (green solid line), both $|C^z_{+}(t) |$  and 
$|R^z(t)|$ decay still algebraically, but with the different law $\sim t^{-3/2}$ indicated by the upper thin dashed lines in the insets. 
See App.~\ref{app:asympt} and the main text for a detailed
discussion of these functions and their asymptotic behavior.
}
\label{fig:CRZTime}
\end{figure}

In Fig.~\ref{fig:CRZTime} we plot the stationary correlation [Eq.~\reff{CZ}] and linear response [Eq.~\reff{RZ}] functions (left and right panel, 
respectively) as a function of time, in the case of the quench to the critical point $\G=1$, for $\G_0 = 0$ 
(fully polarized case), 0.5 and 1 (equilibrium at $T=0$). 
In both panels, the inset highlights in a double logarithmic scale the long-time algebraic decay of these functions 
(modulated by oscillatory terms), indicated by the thin dashed lines. 
If the system is initially prepared either deeply in the ferromagnetic phase 
$\G_0=0$ or, equivalently, in the highly paramagnetic phase $\Gamma_0=\infty$ (both corresponding to $\Y=1$), 
$E(\tau) = J_1(\tau)$ (see App.~\ref{app:crit-polarized}) and Eqs.~\reff{CZ} and \reff{RZ} become
\begin{align}
C_+^z(t) &= J^2_0(4 t) - \frac{1}{4} [ J_0(4t) - J_2(4 t) ]^2, \label{eq:Cpz-fullpol-mt} \\[2mm]
R^z(t) &= 2 \, \theta(t) J_1(4t) [J_0(4 t) + J_2(4 t)].
\end{align}
These results are consistent with the expressions reported in Ref.~\cite{Igloi00}
for the symmetric correlation function $C_+^z$ and in Ref.~\cite{Karevski} for the response
function $R^z$ after quenches starting from the fully polarized state $\Gamma_0=\infty$, which are both generalized by our 
Eqs.~\reff{CZ} and \reff{RZ}. For a generic value of $\Gamma_0$ (\ie, $\Y\neq 1$) 
$E(\tau)$ cannot be expressed in terms of known special functions, though 
its asymptotic behavior in the long-time limit  $t \gg 1$
can be determined analytically and yields
[see App.~\ref{app:asympt}, in particular Eqs.~\reff{eq:app-asy-Cz} and \reff{eq:app-asy-Rz}]:
\begin{align}
\label{CZ_long_time}
C^z_+(t) &= - \frac{1}{8\pi t^2} \cos (8t) + {\cal O}(t^{-3}), \\
R^z(t) &= \frac{1}{4\pi t^2} \left[  \Y^{-1} - \sin (8t)  \right] +{\cal O}(t^{-3}),
\label{RZ_long_time}
\end{align}
with $\Y$ given in Eq.~\reff{eq:defY}. These expressions assume $\Y^{-1} \neq 0$, \ie, $\G_0\neq\G=1$: indeed, in equilibrium $\G_0=\G=1$ at zero temperature the relaxation is qualitatively different and the leading algebraic decay of both $C^z_+$ and $R^z$ [which, in this case, can be expressed in terms of Bessel and Struve special functions, see Eqs.~\reff{eq:Cp-app-crit-eq} and \reff{eq:Rz-app-crit-eq}] turns out to be $\sim t^{-3/2}$, as discussed in App.~\ref{app:asympt}  [see, in particular, Eqs.~\reff{eq:app-asy-Cz-eq} and \reff{eq:app-asy-Rz-eq}] and highlighted in the inset of Fig.~\ref{fig:CRZTime}. In equilibrium at finite temperature, instead, such a decay is $\sim t^{-1}$~\cite{Rossini10_long}.
%
%

In order to explore the possible definitions of effective temperatures based on the behavior of $C_+^z$ and $R^z$ in the frequency domain, we consider the Fourier transform of Eqs.~\reff{CZ} and \reff{RZ}, according to the definitions in Eq.~\reff{eq:FT}.  
%
\begin{figure}[h]
\centering
 \includegraphics[width=0.45\textwidth]{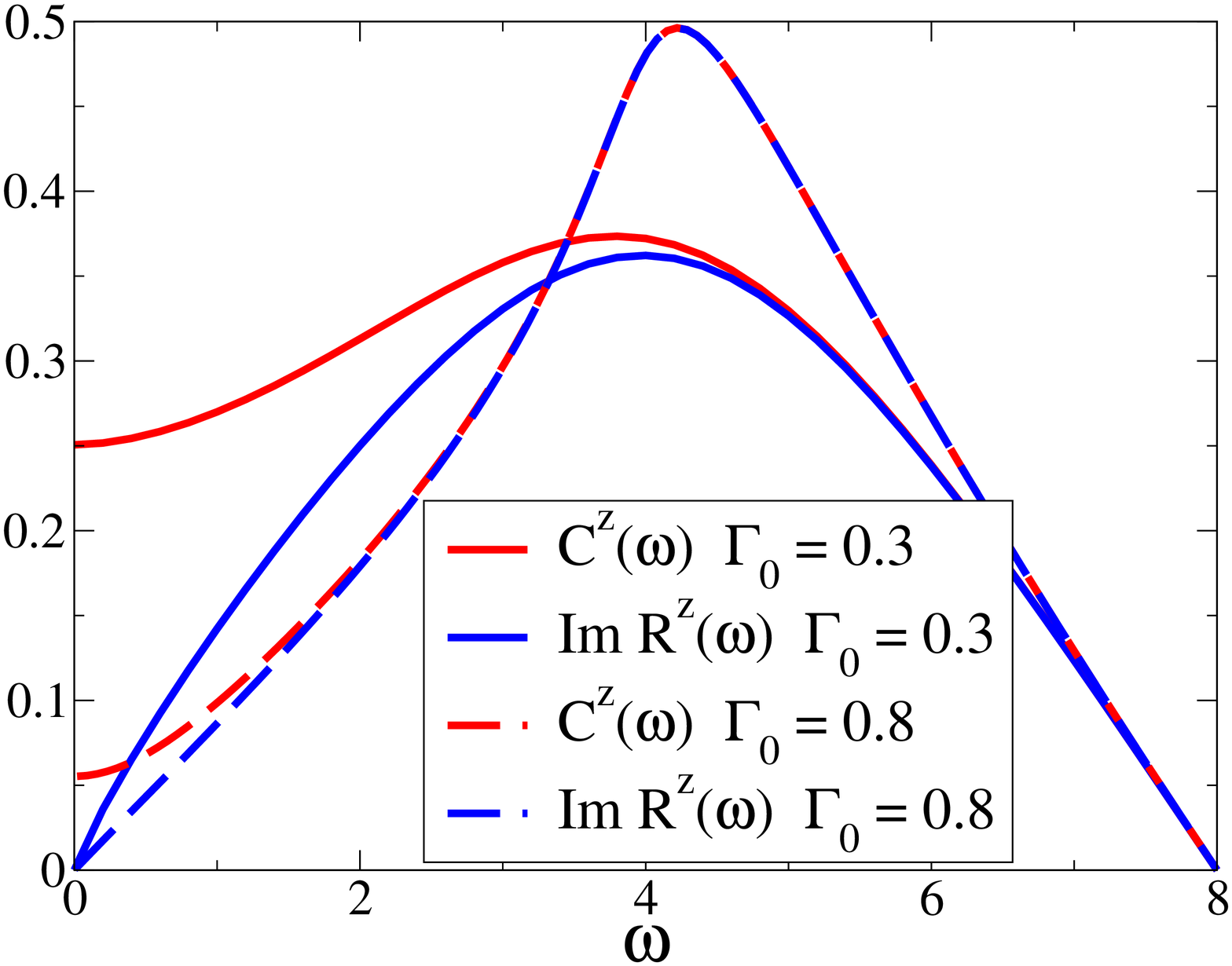} 
 \includegraphics[width=0.45\textwidth]{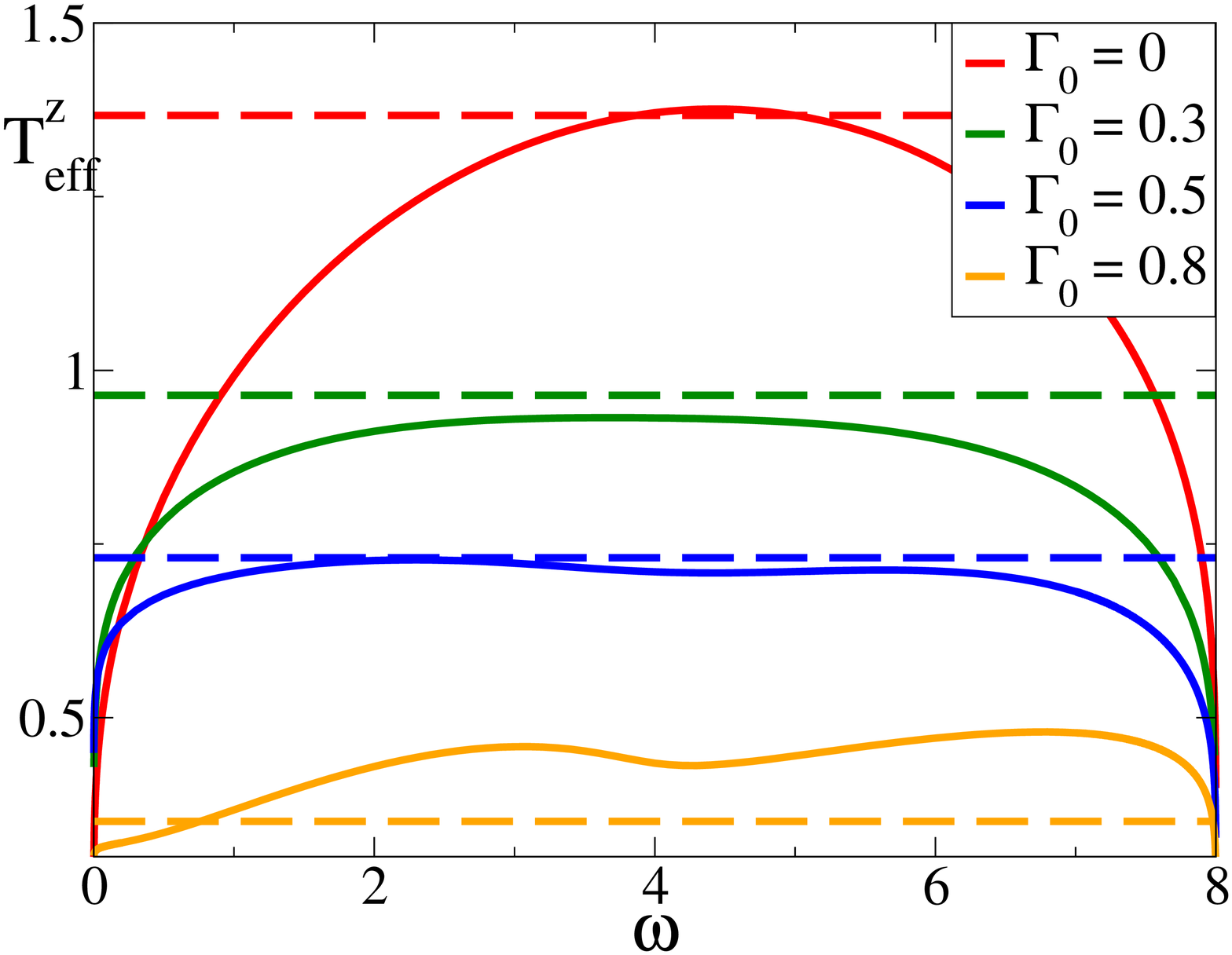} 
\caption{Left panel: Dependence of the correlation function ($\tilde{C}^z_{+}$) and the imaginary part of the linear 
response function ($\IM\,\tilde{R}^z$) of $\hat{\sigma}^z$ 
on the frequency $\omega$, for $\Gamma_0=0.3$ (solid lines) and $0.8$ (dashed lines) 
at the critical point $\Gamma=1$.  
$\tilde{C}^z_{+}(\omega)$ and $\IM\,\tilde{R}^z(\omega)$  are shown  by a red (uppermost solid and dashed) and a 
blue (lowermost solid and dashed) line, respectively,  and they vanish identically for $|\omega| > \omega_{\rm max} =8$. [Note that the values of 
these functions for $\omega<0$ can be inferred from Eq.~\reff{eq:sym}.] 
Right panel: 
Effective temperatures $T_{\rm eff}^z(\omega) = 1/\beta_{\rm eff}(\omega)$ defined on the basis of 
Eq.~\reff{eq:FDT-omega-Teff}, for $\Gamma=1$
and various values of $\Gamma_0$. 
From top to bottom the red, green, blue and yellow solid curves 
correspond, respectively, to 
$\Gamma_0 = 0$, 0.3, 0.5, and 0.8. The corresponding dashed horizontal lines indicate the values of the effective temperature 
$T_{\rm eff}^E(\Gamma=1,\Gamma_0)$ determined  on the basis of the expectation value of the energy from Eq.~\reff{Teff_energy}. 
The comparison shows that there is no special relationship between these two possible effective temperatures, even though a 
thermal behavior was apparently observed when studying one-time quantities, {\it i.e.} with respect to $T_{\rm eff}^E$ 
(see Fig.~\ref{fig:sigmaz}). Note that for 
$0.35 \lesssim \Gamma_0 \lesssim 2.9$, $T_{\rm eff}^z$ exhibits a mild concavity  as a function of $\omega$.
}
\label{fig:beffZ-omega}
\end{figure}
%
%
Due to the quadratic structure of these expressions in $\exp(\pm i \epsilon_{k,l})$ --- in terms of which  the trigonometric 
functions involved in the definitions of $J_\alpha$ and $E$, see Eqs.~\reff{eq:Cp-app-crit}--\reff{eq:app-defEp}, are written --- 
the corresponding Fourier transforms
receive contributions only from real values of the frequency $\omega$ which coincide either with the sum $\epsilon_k+\epsilon_l$ or 
with the difference $\epsilon_k-\epsilon_l$ of the energies $\epsilon_{k,l}$ of two quasi-particles, depending on the range of
$\omega$.
In turn, this structure is due to the fact that the observable $\hat{\sigma}^z$ under study is a quadratic form of the fermionic  excitations 
$\{\hat \gamma_k^{\Gamma_0},\hat \gamma_k^{\Gamma_0\dagger}\}$. 
Note that for $|\omega| > \omega_{\rm max}\equiv 2\epsilon_{k=\pi}$ the Fourier transforms $\tilde{C}_+^z(\omega)$ and  
$\tilde{R}^z(\omega)$ of $C_+^z(t)$ and $R^z(t)$, respectively, do not receive any contributions from the integrals, because 
$|\epsilon_k \pm \epsilon_l| < 2\epsilon_{k=\pi}$ due to the existence of the upper bound at $k = \pi$ of the dispersion relation 
$\epsilon_k(\Gamma)$. 
Accordingly, this results into a finite cut-off frequency $\omega_{\rm max}$ in the spectral representation of $C^z_+$ and $R^z$, 
and the corresponding Fourier transforms $\tilde C^z_+(\omega)$ and $\tilde R^z(\omega)$ vanish identically for 
$|\omega| > \omega_{\rm max}$.
In the stationary state of the isolated system one expects the dynamics to be invariant under time reversal, which implies 
$C^z_\pm(t) = C^z_\pm(-t)$ for the specific observable we are focussing on here; in turn, this implies the 
following symmetry properties
\beq
\tilde{C}_+^{AB}(\omega)=\tilde{C}_+^{AB}(-\omega) \quad
\mbox{and} \quad  \mbox{Im } \tilde{R}^{AB}(\omega)= - \mbox{Im } \tilde{R}^{AB}(-\omega)
\label{eq:sym}
\eeq
for the Fourier transforms of the symmetric correlation and response functions. 
 In view of them, below we will restrict our analysis to the case 
$\omega>0$. In Fig.~\ref{fig:beffZ-omega} (left panel) we present the result 
for $\tilde{C}_+^z(\omega)$ (red) and $\tilde{R}^z(\omega)$ (blue)
for $\Gamma_0=0.3$ (solid lines) and $0.8$ (dashed lines),
which we obtained by numerical integration of Eqs.~\reff{CZ} and \reff{RZ}. 
These functions can be used to extract 
the frequency-dependent effective temperature 
$T_{\eff}^z(\omega)$, or equivalently $\beta_{\eff}^z(\omega)$, 
from the FDR given in Eq.~\reff{eq:FDT-omega-Teff}.
The dependence of the resulting $T_{\eff}^z(\omega)$ on the frequency $\omega$ 
is shown as solid lines in
Fig.~\ref{fig:beffZ-omega} (right panel) 
for various values of $\Gamma_0$, \ie, from top to bottom, 
$\Gamma_0 = 0$ (red), 0.3 (green), 0.5 (blue), and 0.8 (yellow). 
The dashed horizontal lines indicate, in the same order from top to bottom, the 
corresponding (frequency-independent) values of the 
effective temperature $T^E_{\rm eff} = T^E_{\rm eff}(\Gamma=1,\Gamma_0)$
 determined on the basis of  Eq.~\reff{Teff_energy}, as in Ref.~\cite{Rossini10_long}, and 
in terms of which a thermal-like behavior was found for the asymptotic expectation value 
of $\hat \sigma^z_i$ (see Fig.~\ref{fig:sigmaz}). We point out that, 
beyond the fact that they seemingly tend to approach each other 
for $\omega\simeq 4$, as discussed further below, there is no obvious relationship between the effective temperature 
$T^E_{\rm eff}$ and the frequency-dependent 
$T_{\eff}^z(\omega)$, which even develops a mild concavity 
as a function of $\omega$ for $0.35 \lesssim \Gamma_0  \lesssim 2.9$. 
Note that $T_{\eff}^z(\omega)$ vanishes both for $\omega \to \omega_{\rm max}$ [with $\omega_{\rm max} = 2 \epsilon_{k=\pi}
(\Gamma=1) = 8$ at the critical point] and for  $\omega \to 0$, where it does so as
\beq
T_{\eff}^z(\omega \to 0) \simeq  2 \frac{1 + (\Y-2) \arctan(\sqrt{\Y-1})/\sqrt{\Y-1}}{(1-1/\Y) } \left(\ln \frac{1}{\omega}\right)^{-1}. 
\label{eq:TeffZwto0}
\eeq
The comparison between this analytic expression and the actual behavior of $T_{\eff}^z(\omega \to 0)$ calculated numerically is 
shown in the inset of Fig.~1 of Ref.~\cite{Foini11}.
[Apart from this asymptotic behavior, $\tilde{C}_+^z(\omega)$, $\tilde{R}^z(\omega)$ and $T_{\eff}^z(\omega)$ have been 
calculated numerically on the basis of Eqs.~\reff{CZ} and \reff{RZ}. Note, however, that these Fourier transform can be expressed as 
convolutions of those of $J_\alpha(\tau)$ and $E(\tau)$, the latter being discussed in App.~\ref{app:sub-M-FT}, see 
Eqs.~\reff{app:FT-E} and \reff{eq:ReEwPi}.]
The analysis of FDRs for the global transverse magnetization $\hat M(t)$, presented in the next Section, 
suggests that the vanishing of the temperature $T_{\eff}^z(\omega \to 0) $ can be traced back 
to the fact that the correlation and response functions for $\omega \to 0$ are not only determined by the low-$k$ modes, but they 
actually receive a contribution from the energy difference $\epsilon_k-\epsilon_l$ between high-energy modes with $k,l\simeq \pi$, 
which are indeed characterized by $T^{k\simeq\pi}_{\rm eff} \simeq 0$ (see the definition in Sec.~\ref{Sec:Teff_Ising}).  
Accordingly, one can heuristically think of the behavior of $\tilde C^z(\omega)$ and $\tilde R^z(\omega)$ for $\omega\simeq 0$ and $\omega \simeq \pi$ as being essentially determined by the presence of the lattice, which introduces a non-linear dispersion relation $\epsilon_k(\Gamma=1)$ of the quasi-particles together with an upper bound $\pi$ to the possible values of $|k|$. In terms of this picture, 
the observation made above that $T_{\eff}^z(\omega)$ in Fig.~\ref{fig:beffZ-omega} (right panel) 
approaches $T^E_{\rm eff}$ for $\omega \simeq 4$ can be explained as a consequence of the fact that far from $\omega=0$ and $\omega=8$ 
more states contribute to the correlation and response function of $\hat\sigma^z$
which are less affected by the presence of the lattice, being in a sense closer to the continuum limit with linear dispersion relation, for which thermalization is expected (see point 5 in Sec.~\ref{subsec:why}).

For increasingly narrower quenches with $\Gamma_0\to 1$,
$T_{\eff}^z$ vanishes uniformly over all frequencies, as expected from the fact that the 
equilibrium value $T=0$ has to be recovered for $\Gamma_0\to\Gamma=1$.

We conclude that, although $\langle \hat{\sigma}^z_i\rangle_{\mathcal Q}$ 
takes a thermal value~\cite{Rossini10_long}, the dynamics of
$\hat{\sigma}^z_i$ is not compatible with an equilibrium thermal behavior
that would require all these temperatures to be equal within a Gibbs description. 
This case clearly demonstrates that assessing the emergence of a "thermal behavior" solely on the basis 
of one-time quantities might be misleading. 

\subsection{Global transverse magnetization}
\label{Sec:Mz}

We consider here the global transverse magnetization 
\beq
\label{eq:defMtext}
\hat{M} = \frac{1}{L} \sum_{i=1}^L \hat{\sigma}^z_i.
\eeq
The generic two-time connected correlation and response functions of
$\hat{M}(t)$ can be expressed as (see App.~\ref{app:gtM}):
\begin{align}
 C_{+}^{M}(t+t_0,t_0) &=
 8 \int_0^{\pi}  \frac{\rmd k}{\pi}~ \RE\Big[ v_k^{\ast}(t_0)v_k(t+t_0)u_k^{\ast}(t_0)u_k(t+t_0) \Big], \\
R^{M}(t+t_0,t_0) &=
 -  16\,\theta(t)  \int_0^{\pi}  \frac{\rmd k}{\pi}~ \IM\Big[ v_k^{\ast}(t_0)v_k(t+t_0)u_k^{\ast}(t_0)u_k(t+t_0) \Big],
\end{align}
where $v_k(t)\equiv v_k^{\Gamma,\Gamma_0}(t)$ and $ u_k(t) \equiv u_k^{\Gamma,\Gamma_0}(t)$ are 
given by Eq.~\reff{eq:ukvk}. We point 
out that, in order to have a non-vanishing value when taking the thermodynamic limit of the corresponding expressions on the lattice,  we have multiplied the fluctuations of 
$\hat{M}$ by a factor $L$, as it is required for 
global observables whose fluctuations are otherwise suppressed as $L$ increases, see Eq.~\reff{CS_CA_M}.
For critical quenches ($\Gamma=1$) the correlation and response functions in the stationary regime 
$t_0\to\infty$ read (see App.~\ref{app:gtM})
\begin{align}
C^M_+(t)  \equiv \lim_{t_0\to\infty} C_+^M(t+t_0,t_0) &= C +  \frac{J_0(8t)+J_2(8t)}{2}  + F(8t)  + F''(8t), \label{CM}\\[2mm]
 R^M(t)   \equiv \lim_{t_0\to\infty} R^M(t+t_0,t_0)  &= 4 \,\theta(t)[E(8t) + E''(8t)] , \label{RM}
\end{align}
where, in addition to the function $E$ defined in Eq.~\reff{eq:defE-mt}, we introduced the function $F$:
\beq
F(\tau) \equiv \int_0^\pi\frac{\rmd k}{\pi} \cos(\epsilon_k \tau/4) \cos^2\Delta_k(1,\Gamma_0) , \label{eq:defF-mt}
\eeq
[see Eq.~\reff{eq:app-defF} and the plot of $F(\tau)$ in Fig.~\ref{fig:E-F-app}] 
with $\epsilon_k = \epsilon^{\Gamma=1}_k = 4 \sin(k/2)$ and $C=(1+\sqrt{\Y})^{-2}$ [see Eqs.~\reff{eq:defC-app} and \reff{eq:valC}]. 
%
\begin{figure}[h]
\centering
 \includegraphics[width=0.45\textwidth]{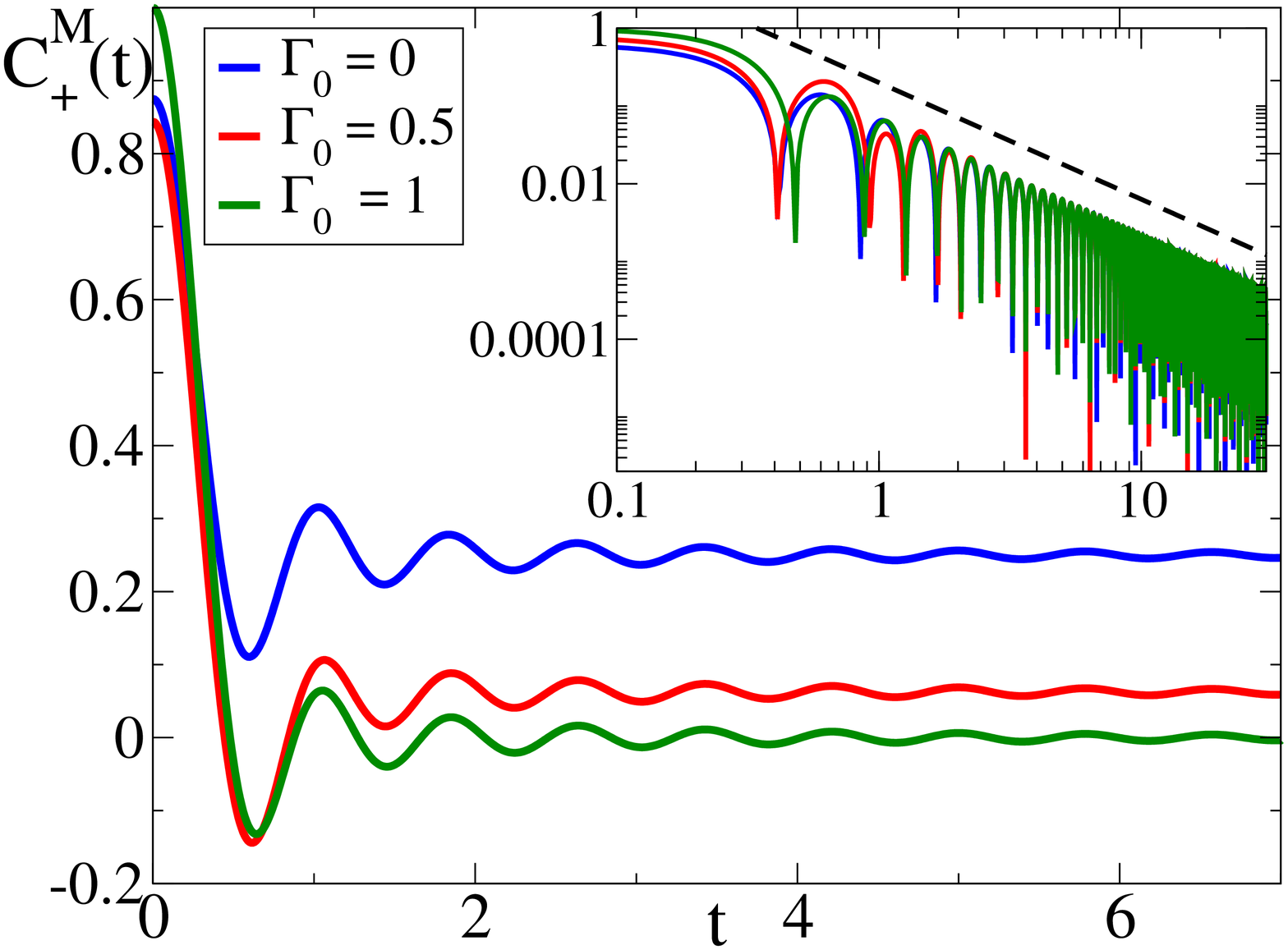} 
 \includegraphics[width=0.45\textwidth]{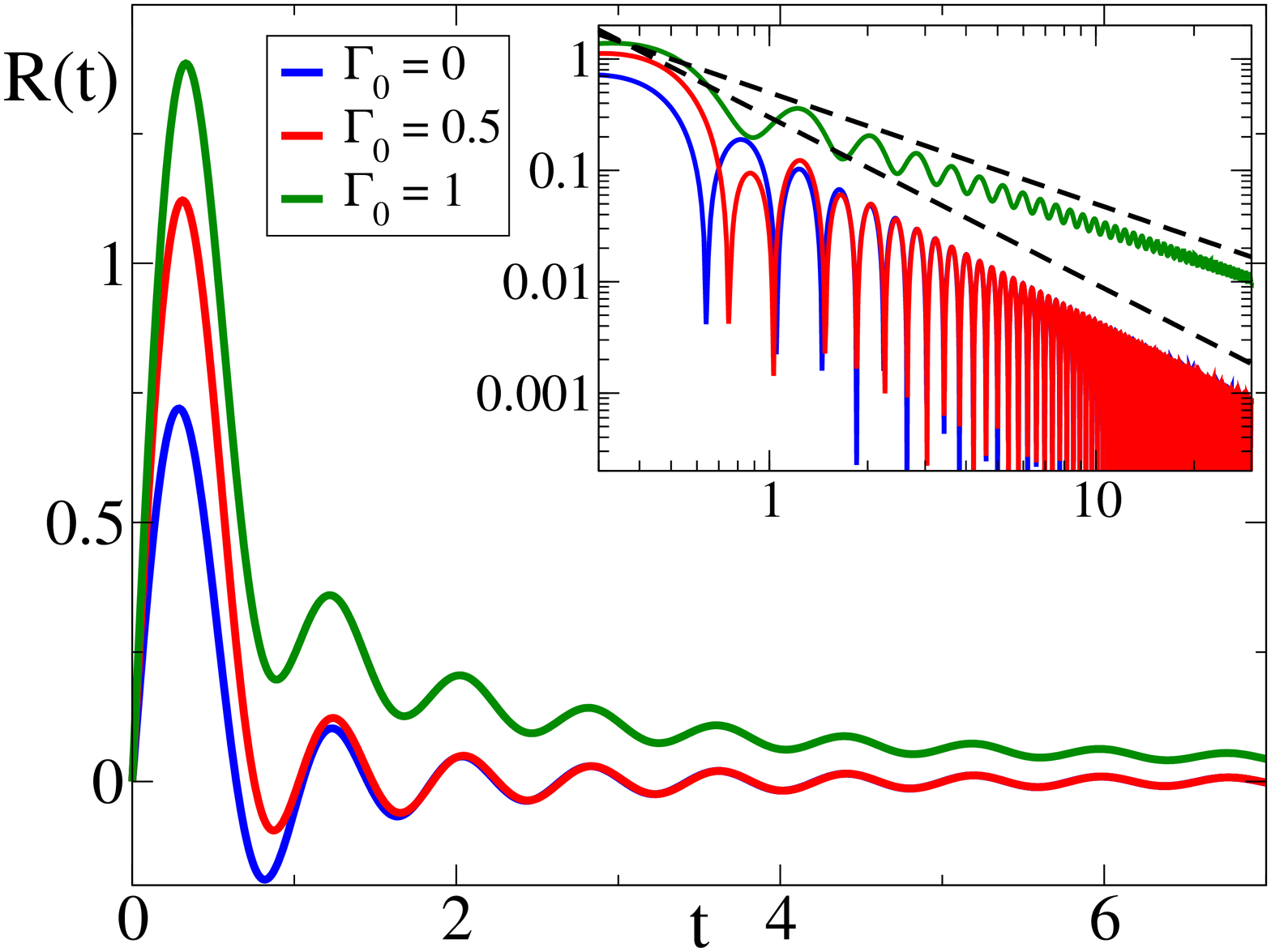} 
\caption{%
Time dependence of the correlation $C^M_{+}(t)$ (left panel) 
and linear response functions $R^M(t)$ (right panel)
 of the local transverse magnetization $\hat{M}$ [see Eqs.~\reff{CM} and \reff{RM}] 
in the stationary regime after the quench to the critical point $\Gamma=1$.
The blue, red and green solid lines (top to bottom on the left panel, bottom to top on the right one) 
correspond to $\Gamma_0=0$, 0.5 and 1 (equilibrium at $T=0$).  
The insets in the left and right panel highlight, in a double logarithmic scale,  the algebraic decay of 
$|C^M_{+}(t) - C|$ and $|R^M(t)|$, respectively,
which are compatible with the asymptotic forms~\reff{CMlong_time} and \reff{RMlong_time} for $\G_0\neq \G=1$ (blue and red solid lines)
indicated by the thin dashed line in the left panel and by the lowermost thin dashed line in the right panel, both $\sim t^{-3/2}$. 
In the case of equilibrium at zero temperature $\G_0=\G=1$ (green solid lines), $|C^M_{+}(t) |$ still decays 
as $ t^{-3/2}$ whereas $|R^M(t)|$ decays more slowly, as indicated by the uppermost thin dashed line $\sim t^{-1}$ 
in the inset of the right panel. 
See App.~\ref{app:M-long-time} and the main text for a detailed
discussion of these functions and their asymptotic behavior.
}
\label{fig:CRMTime}
\end{figure}
%
%
Interestingly enough, implementing the formal substitution of Eq.~\reff{eq-neq-conn}, \ie,
$\cos\Delta_k \to\tanh(\beta\epsilon_k/2)$ in the definition \reff{eq:defC-app} of the constant $C$ (see Eq.~\reff{CM}) does not render its equilibrium value 
obtained with the equilibrium density matrix~\cite{Niemeijer}. 
This indicates that the dynamics of the model cannot be described solely in terms of the occupation numbers 
$\langle \hat n_k^\Gamma\rangle$ (see Sec.~\ref{Sec:Teff_Ising}) --- as the GGE does --- because possible correlations 
$\langle \hat n_k^\Gamma\hat n_{-k}^\Gamma \rangle$ between  $k$ and $-k$ in the initial state can play an important role, at least for 
certain quantities~\cite{Cazalilla_Iucci09, polkovnikov2010}.
The constant $C$ is in fact the result of time-independent correlations of the form $c(k_1,k_2) = \langle \hat{n}_{k_1} \hat{n}_{k_2}\rangle - 
\langle \hat{n}_{k_1}\rangle\langle \hat{n}_{k_2}\rangle$ \cite{Mazur-69}. 
While these terms do not vanish only if $k_1=k_2$ 
within a statistical ensemble --- such as Gibbs or GGE --- which treats $\{\hat{n}_k\}_k$ as statistically 
independent variables, after the quench the summation on $k_{1,2}$ of $c(k_1,k_2)$ which yield $C$
receives contributions also from the term with $k_2=-k_1$, \ie, from $c(k_1,-k_1) = c(k_1,k_1)$ due to the 
particular structure of the initial state.
Generically, this latter contribution is subleading compared to the former as the system size increases and the thermodynamic limit is approached. However, this is not the case for $\hat M$, due to its global nature.

In Fig.~\ref{fig:CRMTime} we plot the time dependence of the stationary (connected) correlation [Eq.~\reff{CM}] and response [Eq.~\reff{RM}] 
functions of the global magnetization $\hat M$ (left and right panel, respectively) for a quench to the critical point $\G=1$, with $\G_0 = 0$ (fully 
polarized case), 0.5 and 1 (equilibrium at zero temperature). In both panels, the inset highlights in a double logarithmic scale the long-time 
algebraic decay (modulated by oscillatory terms) of these functions, indicated by the thin dashed lines. 

As in the case of $\hat \sigma^z_i$ discussed in Sec.~\ref{Sec:sigmaz}, the initial condition enters the expressions of $C^M_+$ and $R^M$ in 
Eqs.~\reff{CM} and \reff{RM} only via the value of $\Y$ defined in Eq.~\reff{eq:defY}.  
In the relevant cases of an initial condition deep in the ferromagnetic or paramagnetic phase, $\Gamma_0=0$ or $\Gamma_0=\infty$, one has 
$\Y =1$, $F(\tau) = [J_0(\tau)-J_2(\tau)]/2$, and Eqs.~\reff{CM} and \reff{RM} can be expressed completely in terms of Bessel functions [see 
Eqs.~\reff{eq:CMp-pol} and \reff{eq:CMm-pol} and right before Eq.~\reff{eq:Cpz-fullpol-mt}]
\begin{align}
C^{M}_+(t)  
&= \frac{1}{4} + \frac{5}{8}J_0(8 t)  + \frac{1}{2} J_2(8 t) - \frac{1}{8} J_4(8 t), \\[2mm] 
R^{M}(t)  &=  \theta(t) [ J_1(8 t)  +  J_3(8 t)].
\end{align}
For generic $\Gamma_0$ ($\neq \Gamma = 1$, \ie, $\Y^{-1}\neq 0$),
the long-time decay of the stationary correlation and response of the global transverse magnetization 
for $t\gg 1$ are even slower than the ones of the corresponding quantities for the local transverse magnetization and is given by 
\begin{align}
C_{+}^M(t) &= C + \frac{1}{8\sqrt{\pi}t^{3/2}}  \sin(8t - \pi/4) + {\cal O}(t^{-5/2}), \label{CMlong_time} \\
R^M(t) &= - \frac{1}{4\sqrt{\pi}t^{3/2}} \cos(8t - \pi/4) + {\cal O}(t^{-5/2})
\label{RMlong_time}
\end{align}
[see Eqs.~\reff{eq:CMp-asy}, \reff{eq:CMm-asy}, and \reff{eq:valC}].
Note that the constant term $C$ in Eqs.~\reff{CMlong_time} and \reff{CM} is not relevant 
for the purpose of studying the FDRs but it 
shows that  the cluster property of the correlation 
function --- which would require the connected correlation 
functions to vanish for well-separated times --- does not hold because of the global nature of 
the quantity under study.
The long-time behavior in Eqs.~\reff{CMlong_time} and \reff{RMlong_time} 
is similar to the one observed at equilibrium at finite temperature~\cite{Niemeijer}.
The case $\Gamma_0=\Gamma=1$, \ie, $\Y^{-1} = 0$ corresponds to the equilibrium 
dynamics at zero temperature $T=0$ for which the correlation and response function can 
be expressed as in Eqs.~\reff{eq:CMp-eq} and  \reff{eq:CMm-eq} in terms of known 
Struve and Bessel special functions, as discussed  in App.~\ref{app:gtM}. In addition, as shown 
in App.~\ref{app:M-long-time}, the leading algebraic decay of $R^M$  
changes into $\sim t^{-1}$, whereas the one of $C_+^M$ is the same as in Eq.~\reff{CMlong_time} with $C=0$.
We also note that --- differently from the case of $\hat \sigma^z_i$ discussed in Sec.~\ref{Sec:sigmaz} --- the leading-order 
decay $\sim t^{-3/2}$ of both $C^M_+$ and $R^M$ for $\G_0\neq\G=1$ 
is actually the same as in equilibrium at $T\neq 0$, which is discussed in Ref.~\cite{Niemeijer}.
Motivated by this observation, one could be tempted to extract  the effective temperature of these dynamics by matching the 
features of the correlations and response functions after the quench with those of the same quantities in equilibrium at finite 
temperature, somehow extending to the present case the approach which was used in, \eg, the early studies of 
Refs.~\cite{Rossini10_short,Rossini10_long}.
However, while the prefactor of the leading-order decay of both $C^M_+$ and $R^M$ in equilibrium
depends upon $T$ \cite{Niemeijer}, in the case of the quench 
the dependence on $\Gamma_0$ appears only at the next-to-leading order, given that
the long-$t$ limit of the quantities in Eqs.~\reff{CMlong_time} and 
\reff{RMlong_time} does not retain memory of the
initial condition beyond the value of the constant $C$. 

\begin{figure}[h]
\centering
 \includegraphics[width=0.45\textwidth]{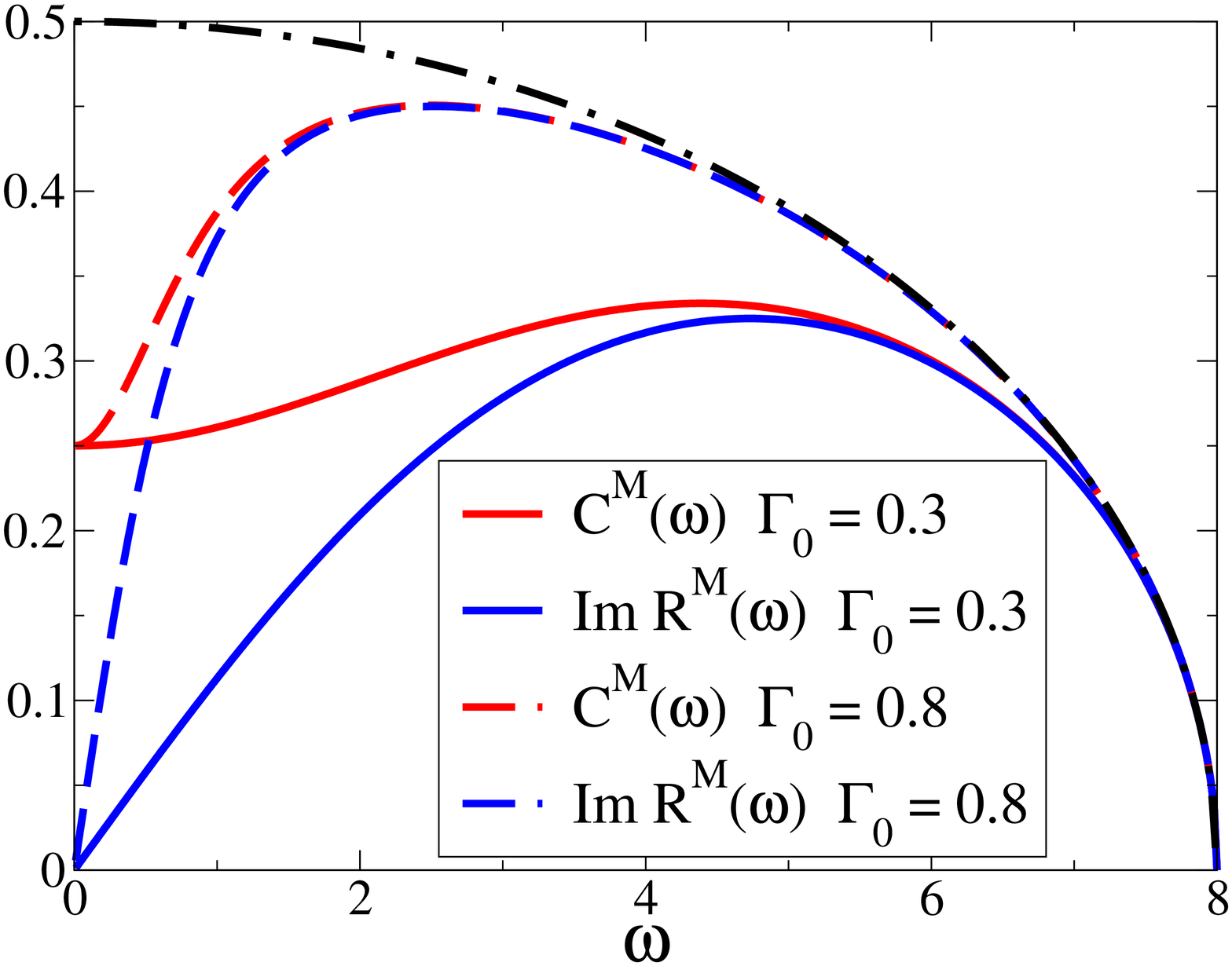} 
 \includegraphics[width=0.45\textwidth]{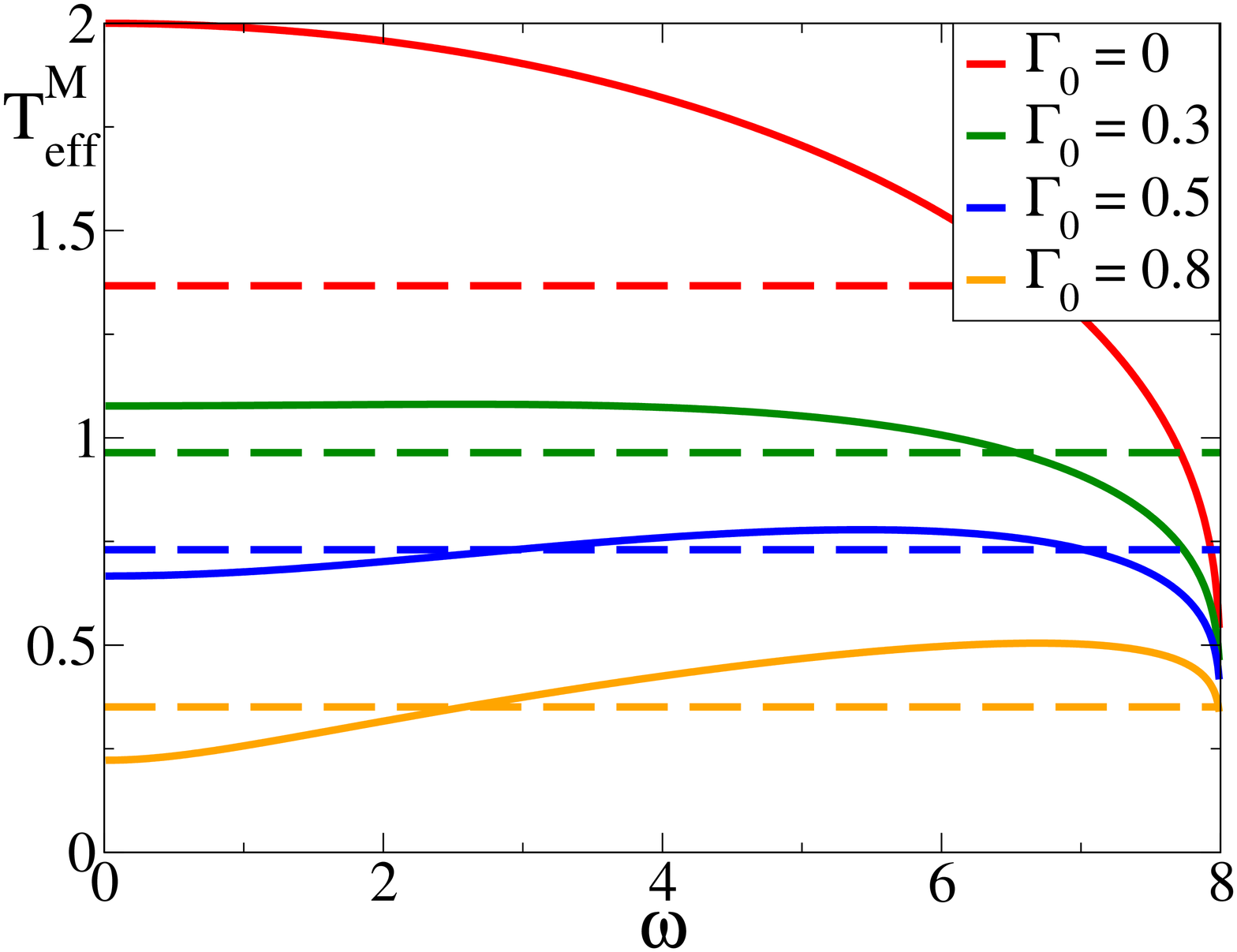} 
\caption{Left panel: Correlation function ($\tilde{C}^M_{+}$) and imaginary part of the linear response 
function ($\IM\,\tilde{R}^M$) of the global magnetization $\hat M$ [see Eq.~\reff{eq:defMtext}]
as functions of the frequency $\omega$, for $\Gamma_0=0.3$ (solid lines) and $0.8$ (dashed lines) 
at the critical point $\Gamma=1$.  
$\tilde{C}^M_{+}(\omega)$ and $\IM\,\tilde{R}^M(\omega)$  are shown respectively by a red 
(uppermost solid and dashed) and a 
blue (lowermost solid and dashed)  line.  
The dot-dashed black line shows the limiting value $1/2 \sqrt{1-(\omega/8)^2}$
of $\tilde C_+^M$ and $\mbox{Im} \tilde R^M$ for $\Gamma_0\to1$.
Due to the property stated in \reff{eq:sym} we consider here only 
$\omega>0$.
Right panel: 
Effective temperatures $T_{\rm eff}^M(\omega) = 1/\beta^M_{\rm eff}(\omega)$ defined on the basis of 
Eq.~\reff{eq:FDT-omega-Teff}, for 
$\Gamma=1$ and various initial conditions $\Gamma_0$. 
From top to bottom the red, green, blue and yellow solid curves 
correspond, respectively, to 
$\Gamma_0 = 0$, 0.3, 0.5, and 0.8. The corresponding dashed horizontal lines indicate the values of the effective temperature 
$T_{\rm eff}^E(\Gamma=1,\Gamma_0)$ determined  on the basis of the expectation value of the energy from Eq.~\reff{Teff_energy}. 
As in 
the case of Fig.~\ref{fig:beffZ-omega}, the comparison shows that there is no special relationship between these two possible 
effective 
temperatures, even though a thermal behavior was apparently observed in the analysis of one-time quantities  
(see Fig.~\ref{fig:sigmaz}). Note 
that for $0.27 \lesssim \Gamma_0 \lesssim 3.7$, $T_{\rm eff}^M$ exhibits a mild non-monotonic behavior as a function of $\omega$ 
and that $T_{\rm eff}^M$ has a different qualitative behavior compared to  $T_{\rm eff}^z$ reported in Fig.~\ref{fig:beffZ-omega}.
}
\label{fig:TeffM-omega}
\end{figure}

In order to define a frequency-dependent effective temperature associated with $\hat M$, we focus below on 
$C_+^M$ and $R^M$ in the frequency domain.
Differently from $\tilde{C}_{+}^z(\omega)$ and $\tilde{R}^z(\omega)$,
 the Fourier transforms of  Eqs.~\reff{CM} and \reff{RM} 
receive a contribution only from frequencies $\omega$ that equal 
$\pm 2\epsilon_k$ [in view of Eq.~\reff{eq:sym} we will restrict below to $\omega>0$].
This means that 
each frequency $\omega$ ``selects" a 
mode $k$ such that $\omega=2\epsilon_k$ (see App.~\ref{app:sub-M-FT} for additional details),
and the high-energy modes with $k\simeq\pi$ do not contribute to the low-frequency behavior. 
These Fourier transforms reduce to integrals over $k$ of terms of the form $\delta(\omega - 2 \epsilon_k)$ which can be easily 
worked out:
\begin{align}
\tilde{C}_{+}^M(\omega) &= \frac{1}{(\sqrt{\Y}+1)^2}\delta(\omega) +  
\frac{1}{4} \sqrt{1- (\omega/8)^2} \;
\frac{1+(2 \Y -1) (\omega/8)^2 }{1+(\Y -1) (\omega/8)^2}\;\theta(8-|\omega|), 
\label{eq:CMomega}\\
\IM ~\tilde{R}^M(\omega) &= \frac{\omega}{16} \sqrt{1-(\omega/8)^2} 
 \; \frac{\sqrt{\Y}}{\sqrt{1+(\Y -1) (\omega/8)^2}}\;\theta(8-|\omega|).
 \label{eq:RMomega}
\end{align}
Note that, as in the case discussed in Sec.~\ref{Sec:sigmaz}, these Fourier transforms vanish for 
$|\omega| > \omega_{\rm max} = 8$. 
[For the purpose of defining the effective temperature according to Eq.~\reff{eq:FDT-omega-Teff} we focus here --- as we did in 
Sec.~\ref{Sec:sigmaz} --- only on the imaginary part of $\tilde R^M(\omega)$; its real part can be obtained from the Kramers-Kronig 
relation \cite{Appel}, as discussed in App.~\ref{app:sub-M-FT}.]
These expressions allow us to determine the effective temperature $T_{\eff}^{M}(\omega)$ on the basis of the FDR in 
Eq.~\reff{eq:FDT-omega-Teff}. Remarkably, this temperature turns out to coincide 
with the mode-dependent temperature $T_{\eff}^{k}$ --- which characterizes the GGE in Eq.~\reff{Teff_GGE} for the 
Ising model \cite{Rossini10_long,Calabrese11,rieger2011,CEF12} --- calculated for the value $k_\omega$ of $k$ which is selected by 
$\omega$, \ie, such that $\omega = 2 \epsilon_{k_\omega}$. 
Indeed, if in the definitions of $E(\tau)$ and $F(\tau)$ [see Eqs.~\reff{eq:defE-mt} and \reff{eq:defF-mt}]
which appear in Eqs.~\reff{CM} and \reff{RM} one expresses 
$\cos \Delta_k$  in terms of $T_{\eff}^{k}$, the Fourier transform yields
$[\IM\, \tilde{R}^M(\omega)]/\tilde{C}_{+}^M(\omega) = \tanh(\omega/(2 T_{\rm eff}^{k_\omega}))$.
Accordingly, comparing with Eq.~\reff{eq:FDT-omega-Teff}, one concludes that $T_\eff^M(\omega) = T_{\rm eff}^{k_\omega}$ or, equivalently,
\beq
T_{\eff}^{k}  = T_{\eff}^{M}(\omega=2\epsilon_k).
\label{eq:TGGE}
\eeq

Figure~\ref{fig:TeffM-omega} (left panel) shows the frequency dependence 
of $\tilde{C}_+^M(\omega)$ (red) and $\IM\,\tilde{R}^M(\omega)$ 
given, respectively, by Eqs.~\reff{eq:CMomega} and \reff{eq:RMomega},
for $\Gamma_0=0.3$ (solid lines) and $0.8$ (dashed lines). Note that
$\lim_{\omega\to 0^+} \tilde{C}_+^M(\omega) = 1/4$ whereas $\IM\,\tilde{R}^M(\omega\to 0)=(\sqrt{\Y}/16) \omega + {\cal O}(\omega^2)$ and 
both functions vanish for $\omega \to \omega_{\rm max}$ as $(\omega_{\rm max}-\omega)^{1/2}/4 + 
{\cal O}((\omega_{\rm max}-\omega)^{3/2})$.
This analysis  in the frequency domain allows us to 
define the frequency-dependent effective temperature 
$T_{\eff}^M(\omega)$ via the FDR in Eq.~\reff{eq:FDT-omega-Teff}.
The function $T_{\eff}^M(\omega)$ is shown as a solid line in
Fig.~\ref{fig:beffZ-omega} (right panel) 
for various values of $\Gamma_0$, \ie, from top to bottom, 
$\Gamma_0 = 0$ (red), 0.3 (green), 0.5 (blue), and 0.8 (yellow). 
The dashed horizontal lines indicate, in the same order from top to bottom, the 
corresponding (frequency-independent) values of the 
effective temperature $T^E_{\rm eff} = T^E_{\rm eff}(\Gamma=1,\Gamma_0)$
 determined as in Ref.~\cite{Rossini10_long} on the basis of  Eq.~\reff{Teff_energy}. 
 We point out that, as for $T_{\eff}^z(\omega)$ reported in the right panel of Fig.~\ref{fig:beffZ-omega}, there is no obvious 
 relationship between the effective temperature $T^E_{\rm eff}$ and the frequency-dependent $T_{\eff}^M(\omega)$. 
 $T_{\eff}^M(\omega)$
 displays a non-monotonic behavior as a function of $\omega$ for $0.27 \lesssim \Gamma_0  \lesssim 3.7$. Note that 
 $T_{\eff}^M(\omega)$ vanishes for $\omega=\omega_{\rm max}$, as $T_{\eff}^z(\omega)$ does, whereas, at variance with it, 
 $T_{\eff}^M(\omega)$ approaches a finite value
\beq
T_{\eff}^M \equiv \lim_{\omega\to 0^+} T_{\eff}^M(\omega) = \frac{2}{\sqrt{\Y}} = 2\frac{|1-\G_0|}{1+\G_0}
\label{eq:Teff0M}
\eeq
at low frequencies. 
[Note that taking the limit $\omega\to 0^+$ is necessary in order to avoid the contribution 
$\propto \delta(\omega)$ in $\tilde C^M_+(\omega)$, see Eq.~\reff{eq:CMomega}.]
This behavior is qualitatively different from the one of the frequency-dependent temperature $T^z_{\rm eff}(\omega)$ in 
Fig.~\ref{fig:beffZ-omega}, which vanishes for $\omega \to 0$: 
the low-frequency value of $ T_{\eff}^M(\omega)$ is solely determined by the 
low-energy modes which are characterized by a 
finite effective temperature $T_{\rm eff}^k$. This property is unique of quenches
to (and from) the critical point.

Quite naturally, one may expect to recover this value $T_{\eff}^M$ by 
considering the FDR in the time domain for large times, as we discussed in Sec.~\ref{Sec:Teff_Ising}:
by replacing $\beta$  by a constant effective value 
$\beta^*_{\rm  eff}$ in the r.h.s.~of Eq.~\reff{eq:FDT-t-Teff}, 
the integral can be written as series of odd time 
derivatives of $C_+^M(t)$, as in Eq.~\reff{Eq-beff-timedomain}.
Inserting the asymptotic long-time behavior of $R^M(t)$ and $C_+^M(t)$ [see Eqs.~\reff{CMlong_time} 
and~\reff{RMlong_time}] in the r.h.s.~and l.h.s.~of this 
expression, respectively, yields the relation $1 = \tanh(4 \beta^*_{\rm eff})$  at the leading order for 
$t\to\infty$ and therefore $T^*_{\rm eff} \equiv 1/\beta^*_{\rm eff} = 0$.
The fact that $T_{\eff}^M\neq T^*_{\rm eff}$ indicates that 
$\beta^M_{\rm eff}(\omega) = 1/T_{\eff}^M(\omega)$ cannot be approximated by an average 
constant in the integral on the r.h.s.~of Eq.~\reff{eq:FDT-t-Teff}. 
Indeed, since only the derivatives of the oscillating factor in Eq.~\reff{CMlong_time} 
contribute to the leading order of Eq.~\reff{eq:FDT-t-Teff}, $T^*_{\rm eff}$ 
can be interpreted
as a temperature associated with the oscillatory frequency $\omega=8$, corresponding 
to the threshold value $\omega_{\rm max}$.  Consistently with this interpretation,  
 the effective temperature $T_{\rm eff}^M(\omega)$ defined on the basis of Eq.~\reff{eq:FDT-omega-Teff} 
 (see the right panel of Fig.~\ref{fig:TeffM-omega})
 vanishes upon approaching the threshold $\omega_{\rm max}$, 
 as $T_{\rm eff}^M(\omega\to \omega_{\rm max}) \simeq - 
4/\ln(\omega_{\rm max} - \omega)$.  
The fact that this behavior turns out to be independent of $\Y$ (and therefore of $\Gamma_0$) is also consistent with 
the fact that the leading-order term of the long-time asymptotic behavior 
of $C_{+}^M(t)$ and $R^M(t)$ [see Eqs.~\reff{CMlong_time} and \reff{RMlong_time}] is --- up to the constant $C$ --- independent of $\Gamma_0$ and it is sensitive only to the largest 
frequencies (that, in turn, are associated with the largest energies).
Such threshold $\omega_{\rm max}$  results from the maximum of 
the dispersion relation  and the quadratic dependence 
of $\hat{M}$ on the fermionic excitations, as noted for $\hat{\sigma}^z$. 
From this analysis it is therefore unclear if one can recover the value $T_{\eff}^M=T_{\eff}^M(\omega\to0)$ of Eq.~\reff{eq:Teff0M} from an 
analysis of the response and correlation functions in the time domain.

As we anticipated at the end of Sec.~\ref{sec:Teff_dynamic}, an effective temperature can also be defined from the relation that
connects in a \emph{classical} equilibrium system the static susceptibility $\chi$ to the fluctuations of the quantity which $\chi$ refers to, 
according to Eq.~\reff{eq:Teff-cl-st} \cite{log-10}; 
note, however, that this definition probes essentially the static behavior of the system, as the time dependence of the response function is 
"averaged" by the integration over time, which is known to miss the important separation of time scales in glassy systems. 
By using the results of 
App.~\ref{app:gtM} and in particular Eqs.~\reff{eq:appCM-0-inf} and 
\reff{app:FT-R0}, the effective temperature $T_{\rm cl,st}$ for the global magnetization $\hat M$ can be expressed as
\beq
\begin{split}
T^M_{\rm cl,st} &= \frac{C_+^M(t=0) - C_+^M(t=-\infty)}{\tilde R^M(\omega=0)} \\
&= \pi \left[ 1 + \frac{1}{2\sqrt{\Y}(1+\sqrt{\Y})}\right] \frac{\sqrt{\Y}-1}{\Y K_e(1-\Y) - E_e(1-\Y)},
\end{split}
\label{eq:Teffchi}
\eeq
where $K_e$ and $E_e$ are the complete elliptic integrals of the first and second kind, respectively 
[see, \eg, chapter 19 in Ref.~\cite{HMF} and the definitions in Eqs.~\reff{eq:EI-i} and \reff{eq:EI-ii}].
%
%
%
%
\begin{figure}[h]
\centering
 \includegraphics[width=0.4\textwidth]{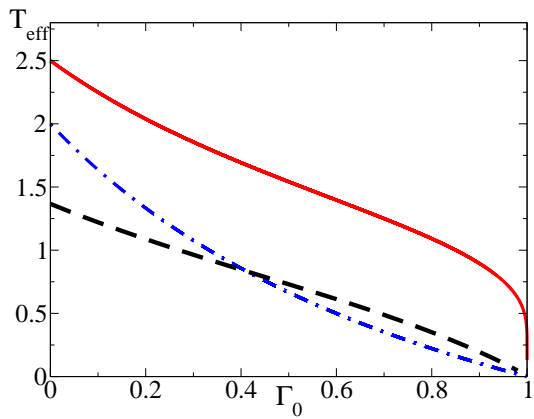} 
\caption{%
Comparison between the effective temperatures $T^M_{\rm eff}$ [see Eq.~\reff{eq:Teff0M}, dot-dashed blue line] and 
$T^M_{\rm cl,st}$ [see Eq.~\reff{eq:Teffchi}, red solid line] as functions of $\Gamma_0$ or, equivalently, $\G_0^{-1}$. For comparison, we report also the effective temperature $T^E_\eff$ (dashed black line) 
defined from the energy of the system. While all these temperatures vanish upon approaching the equilibrium case 
$\G_0 \to \G =1$, they are all generically different and in particular $T^M_{\rm eff}$ is always smaller than $T^M_{\rm cl,st}$. %
}
\label{fig:Teff-M}
\end{figure}
%
%
In Fig.~\ref{fig:Teff-M} we compare this effective temperature (solid line) with $T^M_{\rm eff}$ 
[see Eq.~\reff{eq:Teff0M}, dot-dashed line]
defined from the low-frequency limit  of the frequency-dependent temperature $T^M_{\rm eff}(\omega)$ (see 
Fig.~\ref{fig:TeffM-omega}, right panel), as functions of $\G_0$ (or, equivalently, $\G_0^{-1}$).
While both of them vanish upon approaching the equilibrium case $\Gamma_0 \to \Gamma=1$ (\ie, $\Y \to \infty$), they differ 
significantly for generic values of $\Gamma_0$, with $T^M_{\rm cl,st}$ approaching $5/2$ 
for $\Gamma_0\to 0$ (equivalently, $\G_0\to\infty$, \ie, the fully polarized case $\Y=1$),  whereas $T^M_{\rm eff} \to 2$ in the same limit. 
Accordingly, the temperature $T^M_{\rm cl,st}$ which relates the \emph{stationary}  fluctuations in the (stationary) 
state with the response to a constant external perturbation does not render the value $T_{\eff}^M$ which is associated with the low-frequency limit of the \emph{dynamical} properties of the system, and in particular it does not seem to have 
any clear relation with the various frequency-dependent temperatures $T_{\eff}^M(\omega)$ defined above for $\hat M$.
For the sake of comparison, in Fig.~\ref{fig:Teff-M} we report also the temperature $T^E_\eff$ indicated by the dashed line. 

\subsection{Order parameter}
\label{Sec:sigmax}

In this Section we focus on the correlation function 
\beq
\label{def:Cxx}
C^x (t,t_0) = \langle \hat{\sigma}^x_{i} (t_0+t) \, \hat{\sigma}^x_i (t_0)\rangle
\eeq
[following the notation introduced in Eq.~\reff{def:aver}] of the (local) order parameter $\hat \sigma_i^x$.
The stationary correlation and response functions of the local and global transverse 
magnetization $\hat \sigma^z$ discussed in Secs.~\ref{Sec:sigmaz} and \ref{Sec:Mz} turn out
to be invariant --- for quenches to the critical point $\Gamma=1$ --- under the mapping 
$\Gamma_0\mapsto\Gamma_0^{-1}$. This is due to the fact that  
the dependence of the dynamics on $\Gamma_0$  is brought about only by
$\cos \Delta_k$ [see, \eg, Eqs.~\reff{CZ}, \reff{RZ}, \reff{eq:defE-mt}, \reff{CM}, \reff{RM}, and \reff{eq:defF-mt}] and
$\cos\Delta_k(1,\Gamma_0)=\cos \Delta_k(1,\Gamma^{-1}_0)$ 
[see the discussion after Eq.~\reff{eq:defY}].
In the stationary state attained long after the quench
and for $\Gamma=1$ we find numerically that this
invariance also  holds for $C_{\pm}^x$. As a consequence, we will focus below on quenches originating 
from the ferromagnetic phase $0<\Gamma_0<1$.

\subsubsection{Computation of the correlation functions}

As it was already pointed out in Refs.~\cite{Rossini10_short,Calabrese06,Calabrese11,CEF12,ES12},  
the expectation value $\langle\hat{\sigma}^x(t)\rangle$ of  $\hat\sigma^x$ decays 
to zero at long times for any $\Gamma\neq\Gamma_0$.
This is comparable to the equilibrium thermal behavior at finite temperature $T>0$,
which is characterized by the absence of long-range order of the magnetization along the $x$ component. 
In order to verify the possible emergence of well-defined effective temperature(s) in the stationary state 
we focus on the  two-time correlations $C_{\pm}^x$, which we computed 
on the basis of the method proposed in Refs.~\cite{Rossini10_short,Rossini10_long} 
as an extension of a well-known approach in equilibrium~\cite{mccoy} to the case of the dynamics after the quench.
For periodic boundary conditions
the computation of  $C_{\pm}^x$ is non-trivial because
the operator $\hat{\sigma}^x_{i} (t_0+t) \, \hat{\sigma}^x_i (t_0)$ 
has non-zero matrix elements between states with different
$\hat{c}$-fermionic parity. Accordingly, for this observable, the assumption 
mentioned in Sec.~\ref{sec:model} about the restriction  to the even sector
is not justified. 
However, following Refs.~\cite{mccoy,Rossini10_long,lieb}, the correlation functions of 
this operator can be determined by computing a four-spin 
correlation function $C^{xx}$, which can be done within 
the (antiperiodic) even sector, because $C^{xx}$ conserves the parity.
On a chain of length $L$ this correlation function is defined as 
follows:
\beq
C^{xx} (t,t_0;L) = 
\langle \hat{\sigma}^x_{L+1} (t_0+t) \, \hat{\sigma}^x_{1} (t_0) \,
 \hat{\sigma}^x_{\frac{L}{2}+1} (t_0+t) \, 
 \hat{\sigma}^x_{\frac{L}{2}+1} (t_0) \rangle \;.
\label{eq:fourpoint}
\eeq
The spin $\hat{\sigma}^x_{L+1}$ is identified with the spin $\hat{\sigma}^x_{1}$
after that the full string of Jordan-Wigner fermions, from $1$ to $L$, has been inserted.
By using the cluster property and by taking the thermodynamic limit, one can recover $C^x_{\pm}(t+t_0,t_0)$ from this quantity:
\beq
\big[C^{x} (t+t_0,t_0) \big]^2 = \lim_{L \to \infty} C^{xx} (t,t_0;L) \;,
\eeq
where $C^{x} (t_1,t_2)  = C^{x}_{+} (t_1,t_2) + i \,C^{x}_{-} (t_1,t_2)$.
Following Ref.~\cite{Rossini10_long,mccoy} we introduce the operators 
 $\hat{A}_j (t) \equiv \hat{c}_j^\dagger (t) + \hat{c}_j (t)$ and
 $\hat{B}_j (t) \equiv \hat{c}_j^\dagger (t) - \hat{c}_j (t)$ in terms of the Jordan-Wigner fermions 
 [see Eq.~\reff{JW_fermions}]. Note that $(1- 2 \hat{c}^{\dag}_j\hat{c}_j) = \hat{A}_j  \hat{B}_j = - \hat{B}_j  \hat{A}_j$ 
and $\{  \hat{A}_j ,  \hat{B}_l \} =0$, $\forall j,l$.
 Then, recalling the transformation in Eq.~(\ref{JW_fermions}) we get
\beq
\begin{split}
C^{xx} (t,t_0;L) = &
\left.
 \langle 
[ \hat{B}_{\frac{L}{2}+1} (t_0+t)  \cdots  \hat{B}_{L} (t_0+t) ] 
[ \hat{A}_{\frac{L}{2}+2} (t_0+t)  \cdots  \hat{A}_{L+1} (t_0+t)  ]  \right. 
\\
& \quad \quad \quad \times [ \hat{B}_1 (t_0) \cdots  \hat{B}_{\frac{L}{2}} (t_0) ] 
[ \hat{A}_2 (t_0)  \cdots  \hat{A}_{\frac{L}{2}+1} (t_0)    ]   
\rangle
\end{split} 
\label{eq:ABcorr}
\eeq
where the expectation value, here and in the following, is over the initial state,
$[ \hat{B}_{i} (t)  \cdots  \hat{B}_{j} (t) ] = \prod_{l=i}^j \hat{B}_{l} (t) $
and the same holds for the product of operators $\hat{A}_i(t)$.
The next and final step  is to write the Pfaffian in Eq.~\eqref{eq:ABcorr} 
 in terms of the determinant 
 of a $2L \times 2L$ matrix~\cite{mccoy,Rossini10_long,lieb}:
\beq
\big[ C^{xx} (t,t_0;L) \big]^2 = \vspace*{2mm} \\ 
 \left| \begin{array}{cccc}
 \vspace{0.2cm}
\langle B B \rangle_{j_1,l_1}^{t_0+t,t_0+t}  & \langle BA \rangle_ {j_1,l_2}^{t_0+t,t_0+t}  &
\langle BB \rangle_{j_1,l_3} ^{t_0+t,t_0} & \langle BA \rangle_{j_1,l_4}^{t_0+t,t_0}  \\
\vspace{0.2cm}
- \langle BA \rangle_{l_1,j_2}^{t_0+t,t_0+t} & \langle AA \rangle_{j_2,l_2}^{t_0+t,t_0+t} &
\langle AB \rangle_{j_2,l_3}^{t_0+t,t_0} &  \langle AA \rangle_{j_2,l_4}^{t_0+t,t_0} \\
\vspace{0.2cm}
- \langle BB \rangle_{l_1,j_3}^{t_0+t,t_0} & - \langle AB \rangle_{l_2,j_3}^{t_0+t,t_0} &
\langle BB \rangle_{j_3,l_3}^{t_0,t_0} &  \langle BA \rangle_{j_3,l_4}^{t_0,t_0}  \\
\vspace{0.2cm}
- \langle BA \rangle_{l_1,j_4}^{t_0+t,t_0}  & - \langle AA \rangle_{l_2,j_4}^{t_0+t,t_0} &
- \langle BA \rangle_{l_3,j_4}^{t_0,t_0} &  \langle AA \rangle_{j_4,l_4}^{t_0,t_0}
\end{array} \right| \,
\label{eq:toeplitz}
\eeq
with $\langle X Y \rangle_{i,j}^{t_1,t_2} = \langle  \hat{X}_{[i]} (t_1) \hat{Y}_{[j]} (t_2) \rangle$ 
and
 $X,Y \in \{A,B\}$. The entries in Eq.~(\ref{eq:toeplitz})
 are block matrices, whose indices $j,l$ have subscripts which
indicate their range, with
$j_1, l_1 \in \{ \frac{L}{2}+1, \frac{L}{2}+2\dots, L\}$,
$j_2, l_2 \in \{\frac{L}{2}+2, \frac{L}{2}+3, \dots, L+1\}$,
$j_3, l_3 \in \{ 1, 2,\ldots, \frac{L}{2}\}$,
and $j_4, l_4 \in \{2, 3, \ldots, \frac{L}{2}+1\}$.
All the diagonal entries of the matrix in Eq.~\reff{eq:toeplitz} are zero,
since they do not enter the contractions.
The matrix elements take the form: 
\beq
\quad \left\{ \begin{array}{l}
\langle \hat{A}_j (t_1) \hat{A}_l (t_2) \rangle = \displaystyle \frac{1}{L} \sum_k \rme^{i k (j-l)}
\big[ u_k (t_1) + v_k (t_1) \big] \big[ u_k^* (t_2) + v_k^* (t_2) \big] , \\
\langle \hat{A}_j (t_1) \hat{B}_l (t_2) \rangle = \displaystyle \frac{1}{L} \sum_k \rme^{i k (j-l)}
\big[ u_k (t_1) + v_k (t_1) \big] \big[ u_k^* (t_2) - v_k^* (t_2) \big] , \\
\langle \hat{B}_j (t_1) \hat{A}_l (t_2) \rangle = \displaystyle \frac{1}{L} \sum_k \rme^{i k (j-l)}
\big[ v_k (t_1) - u_k (t_1) \big] \big[ u_k^* (t_2) + v_k^* (t_2) \big] , \\
\langle \hat{B}_j (t_1) \hat{B}_l (t_2) \rangle = \displaystyle \frac{1}{L} \sum_k \rme^{i k (j-l)}
\big[ u_k (t_1) - v_k (t_1) \big] \big[ v_k^* (t_2) - u_k^* (t_2) \big] ,
\end{array} \right. 
\label{eq:aacomput}
\eeq
and the sums over $k$ are consistent with the antiperiodic boundary conditions, \ie, they run
over the values of $k$ given in Eq.~\reff{Fourier_Transform}.
Some care is required in extracting numerically
the real and the imaginary part of $C^{x}(t)$
from the value of $\big[ C^{xx} (t,t_0;L) \big]^2$. 
In our case, starting form the initial value $C^x(0)=1$,  
we followed numerically the analytical solution in time. 
In passing, we mention that a different approach based on the solution of
differential equations has been proposed in Ref.~\cite{Perk} for computing 
similar correlation functions.

\subsubsection{Numerical results for the dynamics}

We computed $C_{\pm}^x(t+t_0,t_0)$  for an isolated chain with anti-periodic boundary conditions, 
$L=10^3$ and  $t_0=10$. It turns out that these values of $t_0$ and $L$  are sufficiently large for 
accessing both the stationary and the thermodynamic limit
(at least for values of $\G_0$ not too close to 1, see the discussion after Eq.~\reff{eq:ratA} further below),  
as we verified by comparing with results obtained 
for different values of $t_0$ and $L$. Accordingly, we focus below only on the 
dependence on $t$ of $C_\pm^x(t+t_0,t_0)$.

We analyzed the effect of the boundary conditions by studying also the dynamics of
an open chain with free boundary conditions, a problem that has been
thoroughly investigated in Refs.~\cite{rieger2011,Igloi11}.
This case differs from the one considered here
because in the open chain there is no distinction 
between the even and the odd sectors and translational 
invariance is absent. 
In spite of these differences, we found that, for fixed parameters, $C_{\pm}^x$
calculated in the periodic chain and in the bulk of the open chain 
perfectly coincide, at least for times such that finite size effects are not relevant.
At larger times, instead, the correlation function 
manifests finite-size effects which 
are expected to depend on the boundary conditions; however, the study of these effects
is beyond the purposes of this work. 

%
\begin{figure}[h]
\centering
 \includegraphics[width=0.49\textwidth]{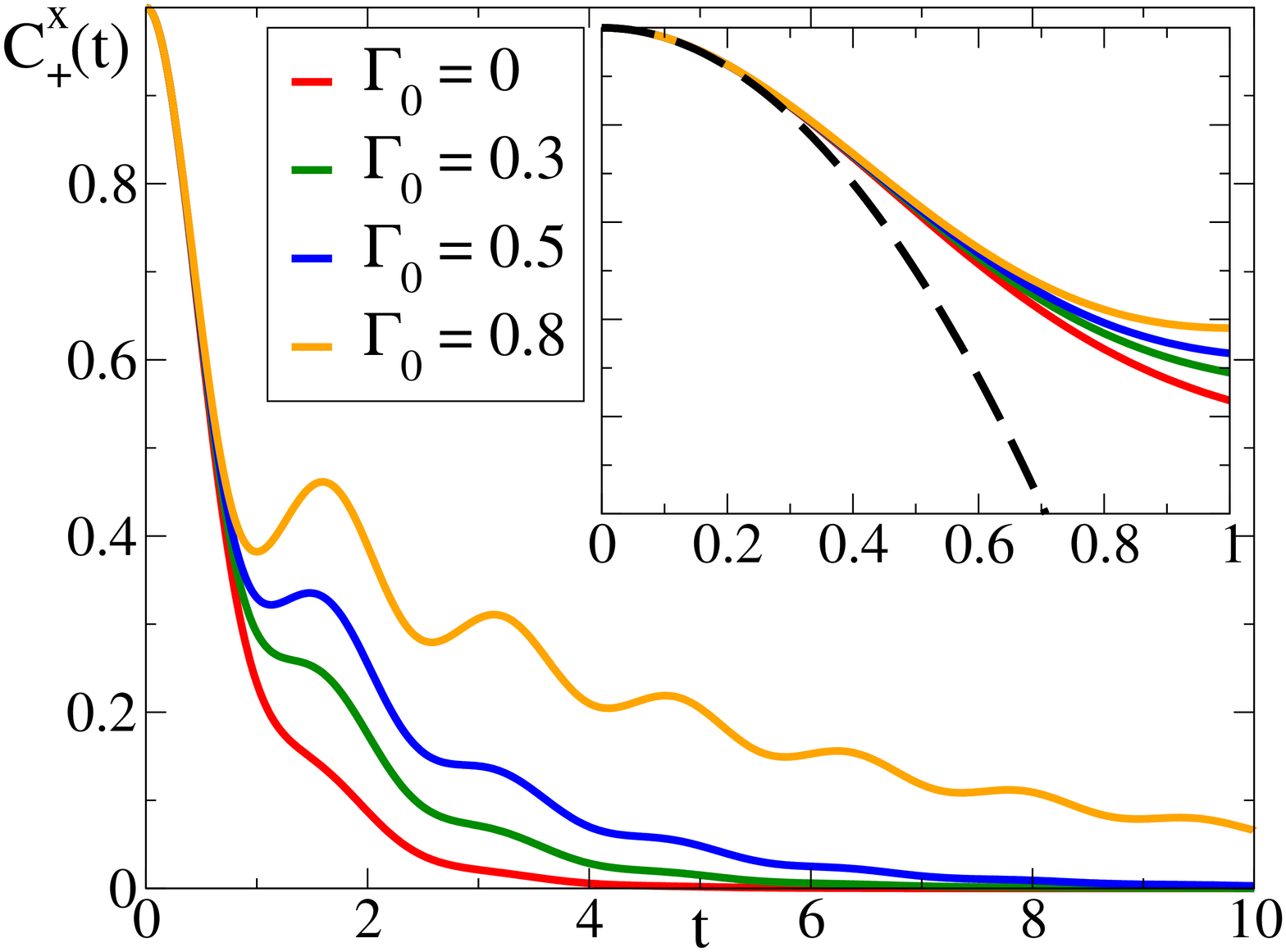}
 \includegraphics[width=0.49\textwidth]{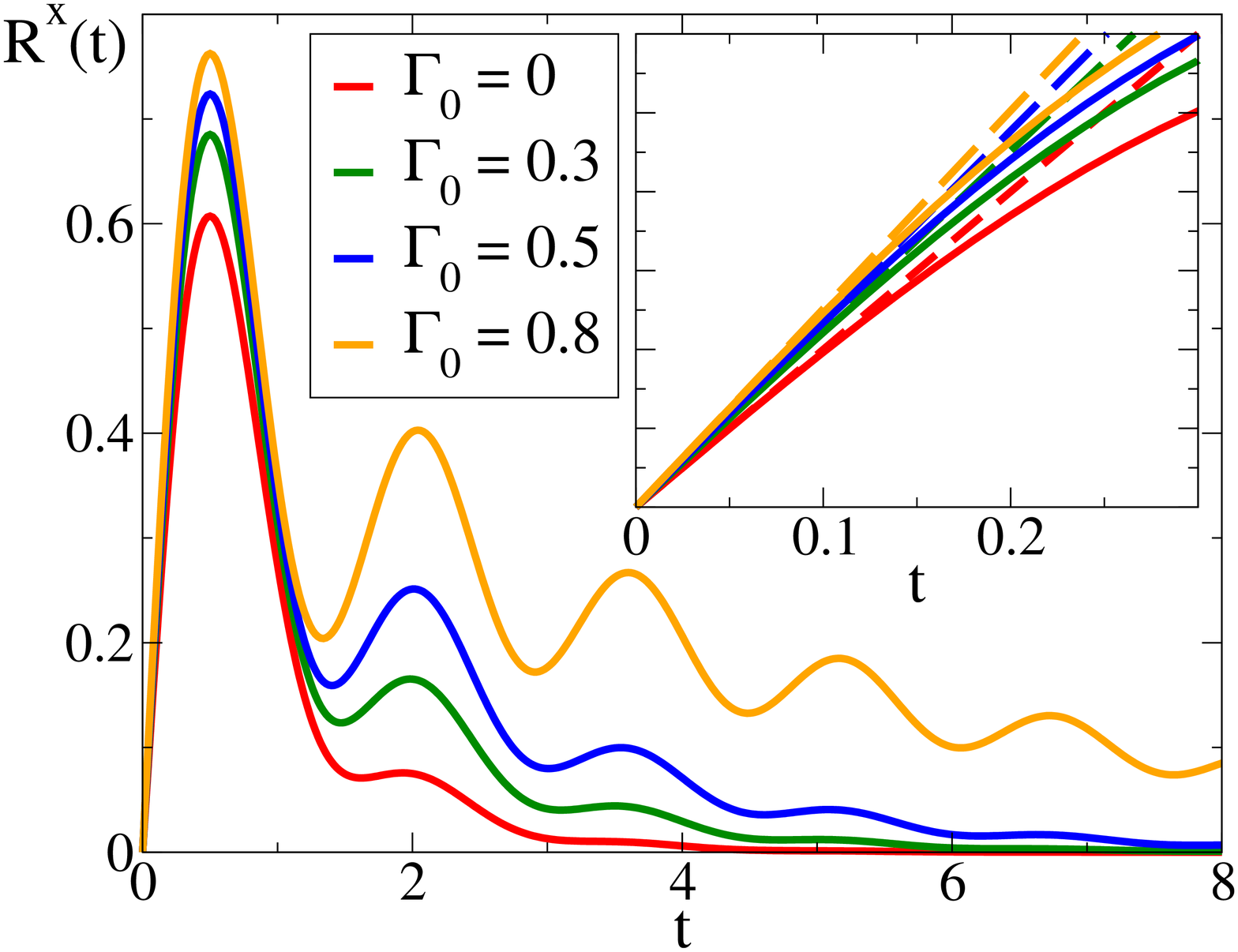}
\caption{Time dependence of the correlation 
$C_{+}^x(t)$ (left panel) and linear response $R^x(t)$ (right panel) functions of 
the order parameter, in the stationary state after a quench to the critical 
point $\G=1$ and for different values of $\G_0$. From bottom to top, red, green, blue and yellow lines correspond
to $\Gamma_0=0$, 0.3, 0.5, and 0.8,  respectively. 
The insets highlight the short-time
behavior of these functions  and the dashed curves correspond to the analytic 
expressions in Eqs.~\reff{eq:Cx-st} and  \reff{eq:Rx-st} with 
$\G = 1$. 
}
\label{fig:CX_RX_time}
\end{figure}
%

In Fig.~\ref{fig:CX_RX_time} we show the time evolution of 
$C_{+}^x(t)$ (left panel) and $R^x(t)=2 i \theta(t)C^x_-(t)$ 
[right panel, see Eq.~\reff{Kubo}] for various values of the initial
transverse field $\G_0$. (As anticipated, here and in the following we focus only on quenches
originating from the ferromagnetic phase $\Gamma_0<1$ because the numerical data 
provide strong evidence that the stationary dynamics with $\Gamma=1$ is invariant under the mapping 
$\Gamma_0\mapsto\Gamma_0^{-1}$.) The insets in both panels provide a closer view of the initial-time 
regime and the corresponding dashed lines represent (with $\G=1$) the short-time behavior 
\begin{align}
C^x_+(t) &= 1 - 2 (\G t)^2 + {\cal O}(t^4), \label{eq:Cx-st}\\
R^x(t) &= 4  \langle \hat{\sigma}^z \rangle_{{\cal Q}} \G t + {\cal O}(t^3), \label{eq:Rx-st}
\end{align}
determined analytically in App.~\ref{Appendix_order_parameter}, with $\langle \hat{\sigma}^z \rangle_{{\cal Q}} $
given in Eq.~\reff{eq:Qavsz}.

Qualitatively, the curves in Fig.~\ref{fig:CX_RX_time} are intriguingly similar to those found 
for dissipative disorder spin models~\cite{Culo,Lozza-etal}, in the sense that they consist of a superposition 
of a monotonicly decaying tail and oscillations with decreasing amplitude. However, in the 
disorder spin model treated in Refs.~\cite{Culo,Lozza-etal} there are the extra ingredients of aging 
phenomena and slow relaxation in the ordered phase, which are absent in the present problem.

For $t\gtrsim15$,  our numerical results are fitted very accurately by
\begin{align}
C^x_+(t) &\simeq \rme^{-t/\tau} A_C \left[ 1 + 
\frac{a_C}{ t^{1/2}} \ \sin (4t+\phi) \right], 
\label{eq:Cx-fit}\\
R^x(t) &\simeq \rme^{-t/\tau} 
A_R \left[1 - \frac{a_R}{ t^{1/2}} \ \cos (4t+\phi) \right],
\label{eq:Rx-fit}
\end{align}
where $A_{C,R}$, $a_{C,R}$, $\tau$, and $\phi$ are the parameters of the fit. 
These {\it Ansatze}  are
motivated by the observed rapid 
temporal decay of these functions,  modulated by damped oscillations.
%
\begin{center}
\begin{table}[h!b!p!]
    \begin{tabular}{ | c | c | c |}
    \hline
       $\Gamma_0$ & $\tau$ fitted  &	$\tau$ Eq.~(\ref{Tau}) \\ \hline
   0 &	0.785418 &	0.785398 \\ \hline
0.1 &	0.90101 &	0.901016 \\ \hline
0.2  & 1.04403 &	1.04403 \\ \hline
0.3 &	1.22641  &	1.22641 \\ \hline
0.4 &	1.46803 &	1.46803 \\ \hline
0.5 &	1.80465 &	1.80464 \\ \hline
0.6 &	2.30773 &	2.30771 \\ \hline
0.7 &	3.14395 &	3.14393 \\ \hline
0.8 &	4.81334 &	4.81332 \\ \hline
0.9 &	9.81476 &	9.8159 \\ \hline
    \end{tabular}
    \caption{Comparison between the characteristic time $\tau$ extracted by the best fit 
of Eq.~\reff{eq:Cx-fit} 
to the numerical data and the corresponding value
predicted by Eq.~\reff{Tau} for various values of $\Gamma_0$. 
The fit has been done within the time interval $t\in[20,60]$ for the correlation function $C_+^x(t)$.
We found good agreement between the analytical expression~(\ref{Tau}) and 
the fit of the numerical data also for the decay of the response function.
The major source of uncertainty in this fit is the systematic error associated with
the choice of the interval within which the data are fitted by Eqs.~\reff{eq:Cx-fit} and \reff{eq:Rx-fit}.  
In the worst case, this choice affects the estimate of $\tau$ in its third digit.
}
\label{tau_fitted}
\end{table}
\end{center}
%
%
Remarkably, the values of $\tau$ that provide the best fit to the numerical data 
turn out to coincide with great accuracy (see Table~\ref{tau_fitted}) with the time constant
\beq\label{Tau}
\tau^{-1} = - \int_0^\pi \frac{\rmd k}{\pi} \; 
\frac{\rmd\epsilon_k(\Gamma)}{\rmd k}\; \ln\cos\Delta_k(1,\Gamma_0) 
= \frac{4  \arctan(\sqrt{\Y-1})}{ \pi \sqrt{\Y-1}},
\eeq
calculated in Refs.~\cite{Calabrese11,CEF12}, which characterizes the exponential 
long-time decay of $ \langle\hat{\sigma}^x_i(t) \rangle \propto \exp(-t/\tau)$ and 
consequently of the equal-time correlation 
$\langle \hat{\sigma}^x_i(t)\hat{\sigma}^x_j(t)  \rangle \propto \exp(- 2 t/\tau)$ at large spatial separations $|i-j|\gg 4 t$. 
Although we are not aware of any analytical proof of this fact, we 
conjecture that the time constant $\tau$ of the long-time exponential decay of 
$\langle \hat{\sigma}^x_i(t+t_0)\hat{\sigma}^x_i(t_0)  \rangle \propto \exp(- t/\tau)$ in the stationary regime $t_0\to\infty$ is 
determined by Eq.~\reff{Tau}
beyond the numerical coincidence shown in Table~\ref{tau_fitted} \cite{footnote1}. 
Indeed, the analytic expression in Eq.~\reff{Tau} finds further support from the fact 
that it correctly reproduces the characteristic time $\tau_{\rm eq}$~\cite{Deift} 
which controls the long-time decay of the equilibrium correlation function $\langle \hat{\sigma}^x_i(t)\hat{\sigma}^x_i(0)  \rangle$ at 
temperature $T$ when, according to the formal mapping in Eq.~\reff{eq-neq-conn}, one substitutes $\cos \Delta_k$ in Eq.~\reff{Tau} 
with $\tanh(\epsilon_k/(2T))$. In addition, note that the non-equilibrium coherence time 
$\tau$ given by Eq.~\reff{Tau} decreases upon increasing  $|1-\Gamma_0|$, \ie, the energy injected into
the system and $\tau \sim |1-\Gamma_0|^{-1}$ for $\Gamma_0 \to 1  (=\Gamma)$. This fact is consistent with the equilibrium 
behavior $\tau_{\rm eq} \sim 1/T \to \infty$ for $T \to  0$;
finally, with this assumption, the relation between the long-time decay of 
$\langle \hat{\sigma}^x_i(t)\hat{\sigma}^x_i(s) \rangle$ for $t=s$ (studied in Refs.~\cite{Calabrese11,CEF12})  and for $t$, 
$s \to \infty$ with fixed $|t-s|$ (present study) is the same as the one generally 
expected for the dynamic correlations $\langle \Phi(r,t) \Phi(0,s) \rangle$ of the order parameter $\Phi(r,t)$ at point $r$ and time $t$ 
after a quench to a quantum \emph{critical} point of a one-dimensional 
model with a linear  dispersion relation $\epsilon_k \propto |k|$ (\ie, a 1+1-dimensional CFT)~\cite{Calabrese06}.  Analogously to 
Ref.~\cite{Calabrese06}, also in the present case we expect the non-linearity 
and the upper bound  of the dispersion relation to be responsible for the oscillating corrections in Eqs.~\reff{eq:Cx-fit} 
and \reff{eq:Rx-fit}.
%
%
\begin{figure}[h]
\centering
 \includegraphics[width=0.6\textwidth]{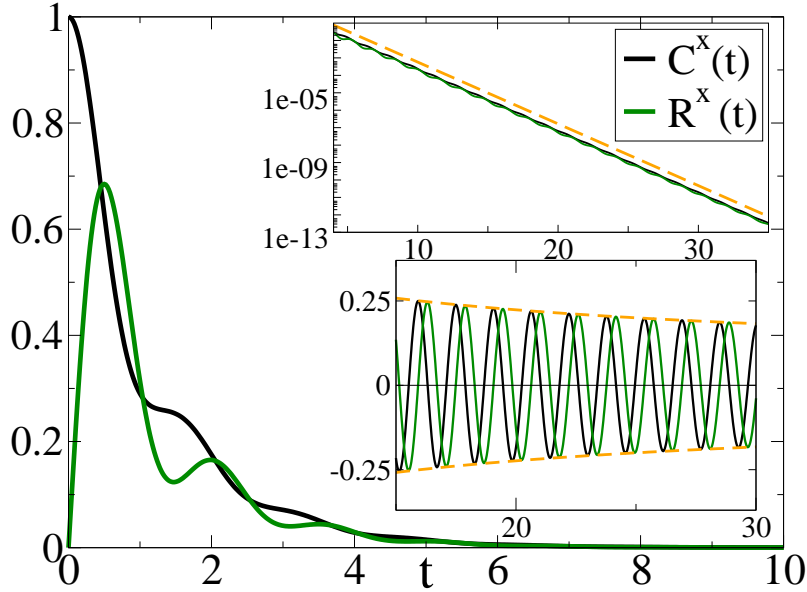}
\caption{Decay of the 
  correlation function $C_+^x$ (black line) and of the linear response 
  function $R^x$ (green line) of the order parameter, as a function of time $t$ for $\Gamma_0=0.3$. 
  Upper inset: zoom into the long-$t$ 
  decay that demonstrates the exponential relaxation with the characteristic 
  time $\tau$ defined in Eq.~\reff{Tau} (dashed yellow line). Lower inset:
  $(\rme^{t/\tau}C_+^x/A_C -1)/a_C$ (black) and $(\rme^{t/\tau}R^x/A_R-1)/a_R$  (green)
  {\it vs.}~$t$. The yellow dashed line represents the  $t^{-1/2}$
  envelope of the damped oscillations, in agreement with Eqs.~\reff{eq:Cx-fit} and~\reff{eq:Rx-fit}.
   }
\label{fig:RCz}
\end{figure}
%

\begin{figure}[h]
\centering
 \includegraphics[width=0.49\textwidth]{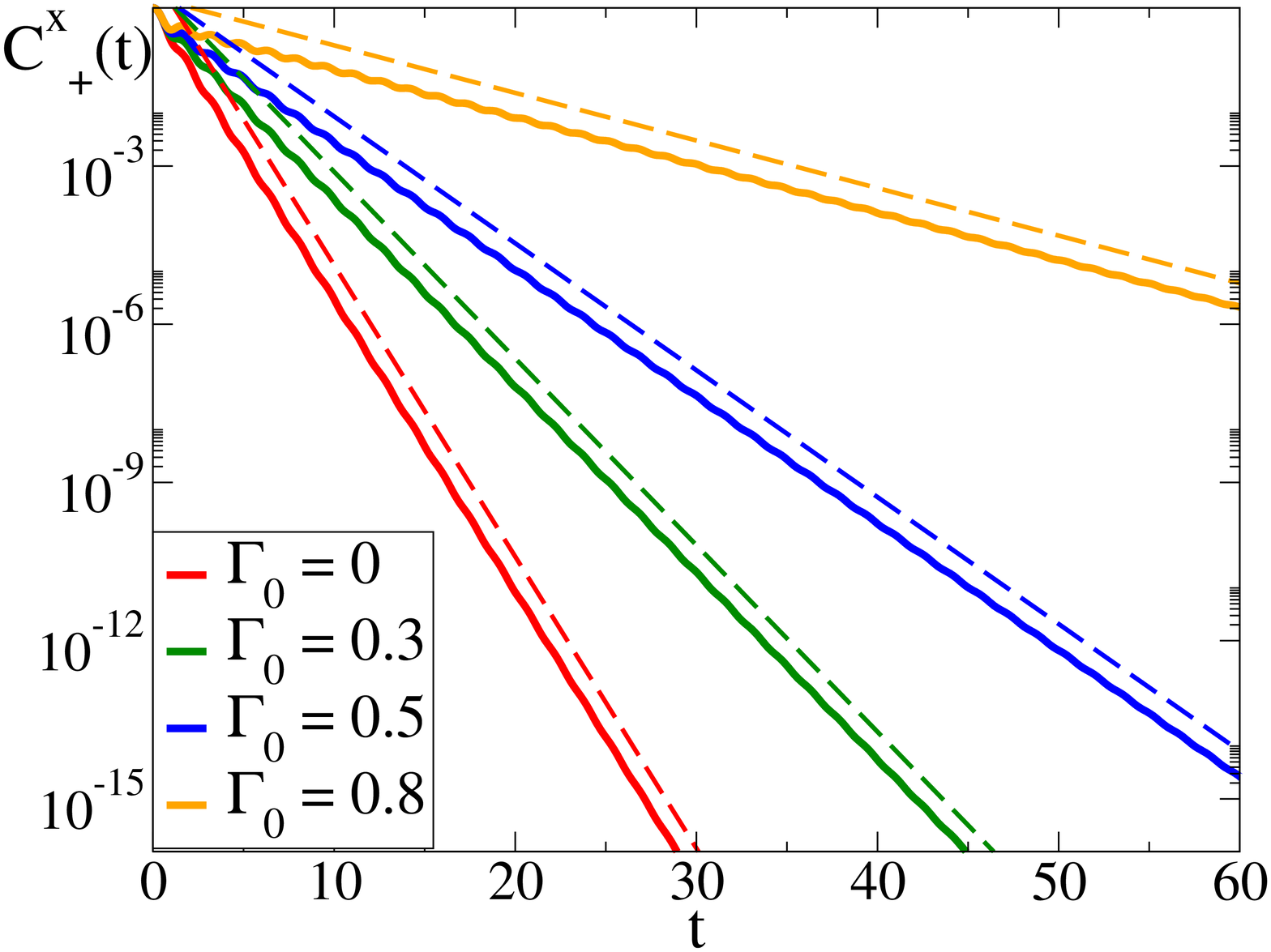}
 \includegraphics[width=0.49\textwidth]{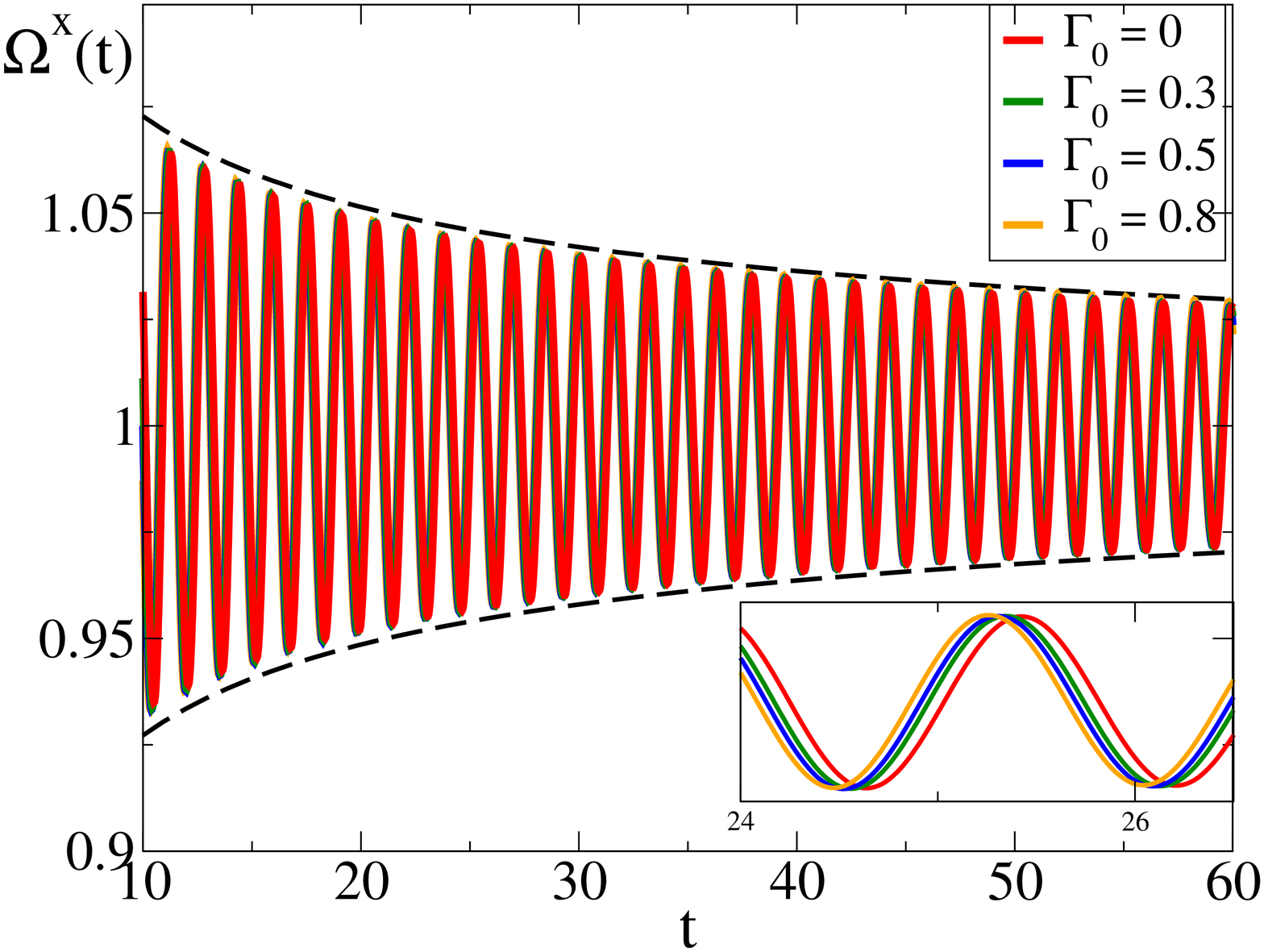}
\caption{Left panel: Time decay of the correlation function $C_{+}^x(t)$ (solid lines)
for several initial values of $\G_0$.  
For each of these values the dashed lines  
correspond to an exponential decay with the rate given by Eq.~\reff{Tau}, which clearly capture the long-time behavior of $C^x_+(t)$. The 
amplitude $A_C$ of this exponential decay (see Eq.~\reff{eq:Cx-fit}) is estimated by the best fit. 
Right panel: Oscillatory correction to
the leading exponential decay of the correlation $C_+^x(t)$. 
Solid lines indicate $\Omega^x(t) = C_+^x(t)\rme^{t/\tau}/A_C$ [see Eq.~(\ref{eq:Cx-fit})],
with $\tau$ determined from Eq.~\reff{Tau} and $A_C$ as indicated above.
The dashed line is an envelope $1\pm0.23/\sqrt{t}$ that shows the amplitude
of the damped  oscillations. The inset zooms in the time interval $t\in[24,27]$ and shows the 
slight phase shift among the different curves.
In both panels red, green, blue and yellow lines correspond
 to $\Gamma_0=0,0.3,0.5$, and 0.8, respectively. For the sake of simplicity we do not show
here $R^x(t)$, but we found the same characteristic time and 
a similar oscillatory behavior.
}
\label{fig:CX_tau}
\end{figure}

Figure~\ref{fig:RCz} presents $C^x_+(t)$ (black) and $R^x(t)$ (green) as functions of time $t$,
for $\Gamma_0=0.3$.  The upper inset shows a zoom of the main plot for large times 
and compares it with the leading exponential decay (yellow dashed curve) with the rate 
$\tau^{-1}$ given by Eq.~\reff{Tau}. The lower inset shows the correction to the leading decay, 
 $(\rme^{t/\tau}C_+^x/A_C -1)/a_C$ (black) and $(\rme^{t/\tau}R^x/A_R-1)/a_R$  (green), 
  estimated 
in Eqs.~\reff{eq:Cx-fit} and \reff{eq:Rx-fit} and
the yellow 
dashed line represents the envelope $t ^{-1/2}$. 
Although several fitting parameters are involved 
in Eqs.~\reff{eq:Cx-fit} and \reff{eq:Rx-fit},
we tested these expressions for several values of  
 $\Gamma_0 \in \{0,0.1,0.2,\dots,0.9\}$ 
and they turned out to be always 
remarkably accurate. 
The exponential decay sets in very soon, usually
already after the first oscillation of $C_+^x$ and $R^x$. The next-to leading order
oscillatory correction involving a power-law decay is more difficult to determine, but still
we found a good agreement quite soon,
with $a_C\simeq 0.21$ and $a_R \simeq 0.52$ rather independent of the value of $\G_0$. 
While the parameters $A_{R,C}$ depend on $\Gamma_0$, 
their ratio 
\beq
A_C/A_R=1.210(5)
\label{eq:ratA}
\eeq 
does not within our numerical accuracy and the range considered.  
In fact, for the fixed initial time $t_0=10$ that we have primarily considered in our computations, 
 only the ratio corresponding to $\Gamma_0=0.9$ is slightly larger than the others. 
 We argue that this is due to non-stationary contributions and indeed, upon increasing $t_0$, 
 the corresponding value of the ratio decreases towards the value obtained for smaller $\G_0$. 
 This observation agrees with the fact that for the correlation functions $C^x_{\pm}$ to
be  stationary after the quench, one has generically to take $t_0 \gg \tau$, with the coherence time $\tau$ given in Eq.~\reff{Tau}.  
 As this coherence time diverges for $\G_0\to 1$, the investigation of shallow quenches with $\G_0\simeq \G = 1$ necessarily requires studying increasingly large values of $t_0$. In addition, a large coherence time makes the numerical determination of the effective temperature
in the $\omega\to0$ limit more difficult as it requires the integration of the correlation function 
over a longer time interval. This is the reason why we have limited our analysis to sufficiently deep quenches with $\tau \lesssim t_0=10$, \ie, from Eq.~\reff{Tau}, $\G_0 \lesssim 0.9$.
In the left panel of Fig.~\ref{fig:CX_tau} we show the exponential
decay of $C_{+}^x(t)$ for several initial conditions $\Gamma_0$ and its  
comparison with the rate in Eq.~\reff{Tau}. The exponential decay of $R^x(t)$ is precisely the same.
In the right panel we show the damped oscillations displayed by 
$C_{+}^x(t) ~ \rme^{\frac{t}{\tau}} / A_C$ 
[where the value of $A_C$ is determined by the best fit and $\tau$ from Eq.~\reff{Tau}]
and with a dashed 
line the envelope $\propto t^{-1/2}$. 
The inset of this panel highlights the phase shift $\phi$ among the various curves. 
Note that --- apart from this tiny shift --- the curves almost overlap, 
suggesting that the oscillatory corrections in Eqs.~\reff{eq:Cx-fit} and \reff{eq:Rx-fit} might be actually independent of $\G_0$ and that 
these observed differences are finite-size or numerical effects.
%

Since the correlation and response functions are computed on a chain of finite length $L$,
they display the $L$-independent behavior described by  Eqs.~\reff{eq:Cx-fit} and \reff{eq:Rx-fit} only for times $t$ smaller than a 
certain cross-over time set by $L$, after which finite-size effects dominate. Accordingly, in comparing numerical data with these 
asymptotic expressions we restricted to suitable long times within the "early" regime, being the detailed discussion of finite-size effects beyond the scope of our study.

The discussion in Secs.~\ref{Sec:sigmaz} and \ref{Sec:Mz} reveals  that the qualitative 
behavior of the various effective temperatures for the transverse magnetization can be affected by the spatial structure of the quantity 
under study, as indeed local and global quantities display different features.  
In order to understand  the extent up to which the spatial structure of the correlation and response functions influences 
such effective temperatures, we investigated this issue numerically  by computing the 
two-time and two-point functions of the local order parameter:
\beq
C^x_{\pm}(r,t_0+t,t_0) = \frac{1}{2} \langle \Big[ \hat{\sigma}^x_{i+r}(t_0+t) ,
\hat{\sigma}^x_i(t_0) \Big]_{\pm} \rangle,
\label{eq:defCpm-mt}
\eeq
which provide the symmetric correlation function $C_+^x$ and the linear response function
$R^x(r,t_0+t,t_0) = 2 i C^x_{-}(r,t_0+t,t_0)$, for $t>0$~\cite{Kubo}.
The computation of this
quantity is still based on the same determinant equation as Eq.~\reff{eq:toeplitz} where the 
indices $j_1$ and $l_1$ run from $\frac{L}{2}+1$ to $L-r$,
$j_2$ and $l_2$ from $\frac{L}{2}+2$ to $L-r+1$,
$j_3$ and $l_3$ from $1$ to $\frac{L}{2}-r$, and 
$j_4$ and $l_4$ from $2$ to $\frac{L}{2}-r+1$.
%
\begin{figure}[h]
\centering
 \includegraphics[width=0.49\textwidth]{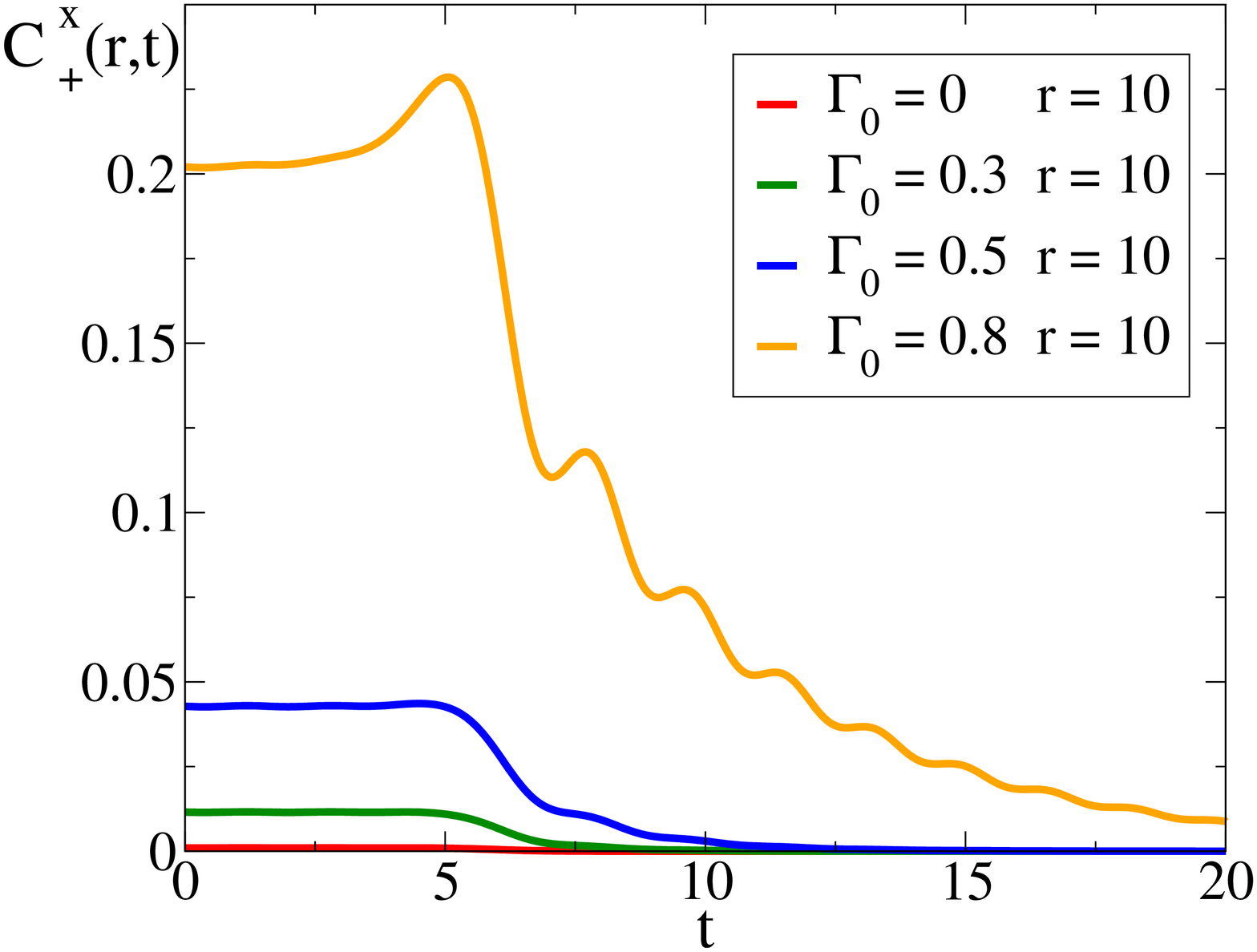}
 \includegraphics[width=0.49\textwidth]{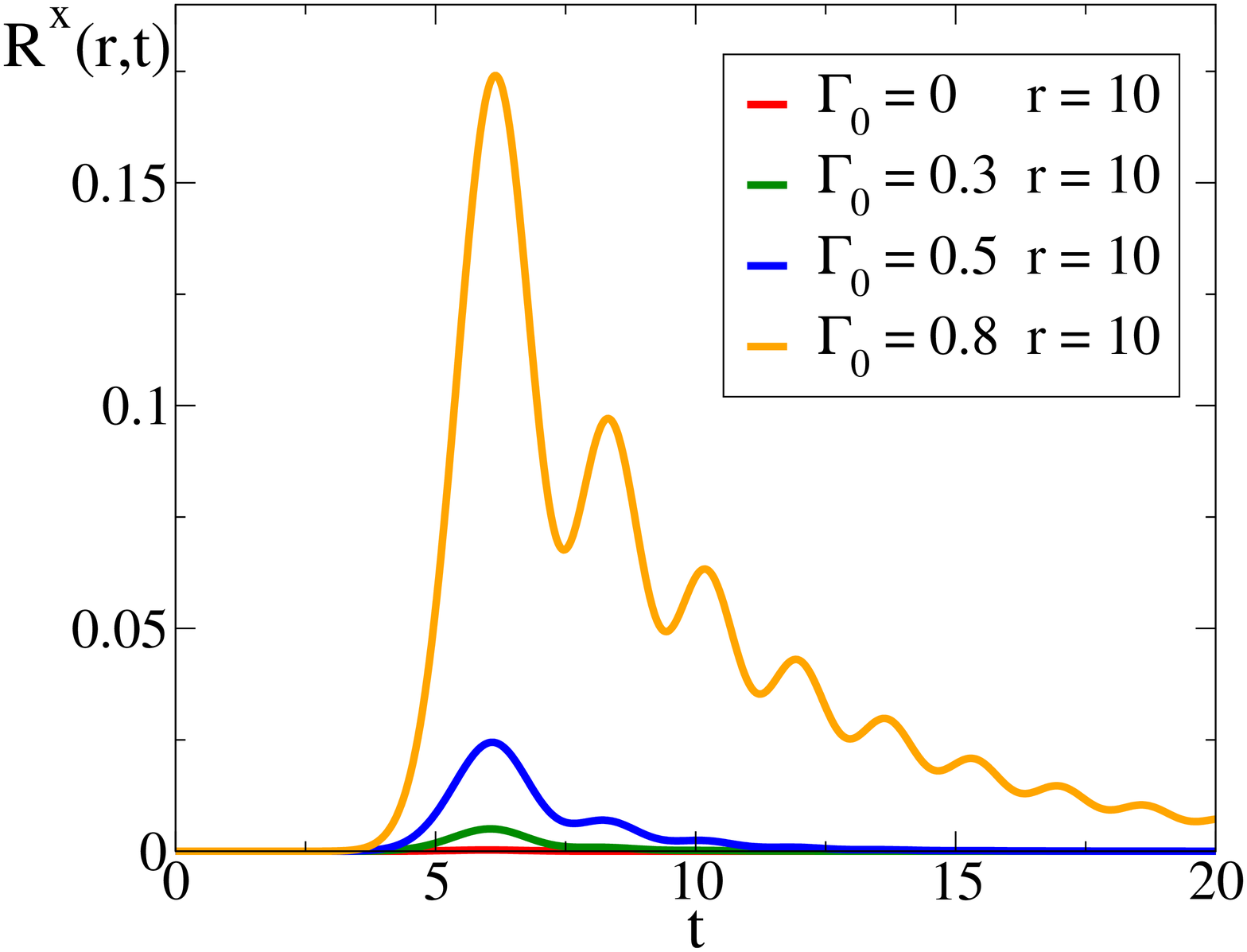}
\caption{Time-dependence of the stationary correlation and linear response function 
of two $\hat\sigma^x$ spins (order parameter) separated by a distance $r=10$ in 
units of the lattice spacing, after a quench to the critical point $\G=1$. The correlation function
$C_{+}^x(r=10,t)$ is reported in the left panel, whereas the response 
function $R^x(r=10,t)$ in the right one. From bottom to top, the various curves 
refer to the different initial conditions $\G_0 = 0$ (red), 0.3 (green), 0.5 (blue) and 0.8 (yellow). 
Both $C_{+}^x(r=10,t)$ and $R^x(r=10,t)$ display clear light-cone effects, which are discussed in the main text 
and summarized in Fig.~\ref{fig:lightcone}.
}
\label{fig:CRXtime_R10}
\end{figure}
%
%
We fixed the length of the chain $L=1000$
and the waiting time $t_0=10$  and we studied the dependence on $t$ after quenches to the critical point $\G=1$, starting from an initial 
condition which is the ground state of the Hamiltonian corresponding to $\Gamma_0 \in \{0,0.1,0.2,\dots,0.9\}$, exactly as done before for the 
case $r=0$.

The results of the numerical calculation of $C^x_+(r=10,t)$ and $R^x(r=10,t)$ are presented in Fig.~\ref{fig:CRXtime_R10} (left and 
right panel, respectively) as functions of the time separation $t$ at which the two observables are measured, see Eq.~\reff{eq:defCpm-mt}. 
%
%
%
\begin{figure}[h]
\centering
 \includegraphics[width=0.49\textwidth]{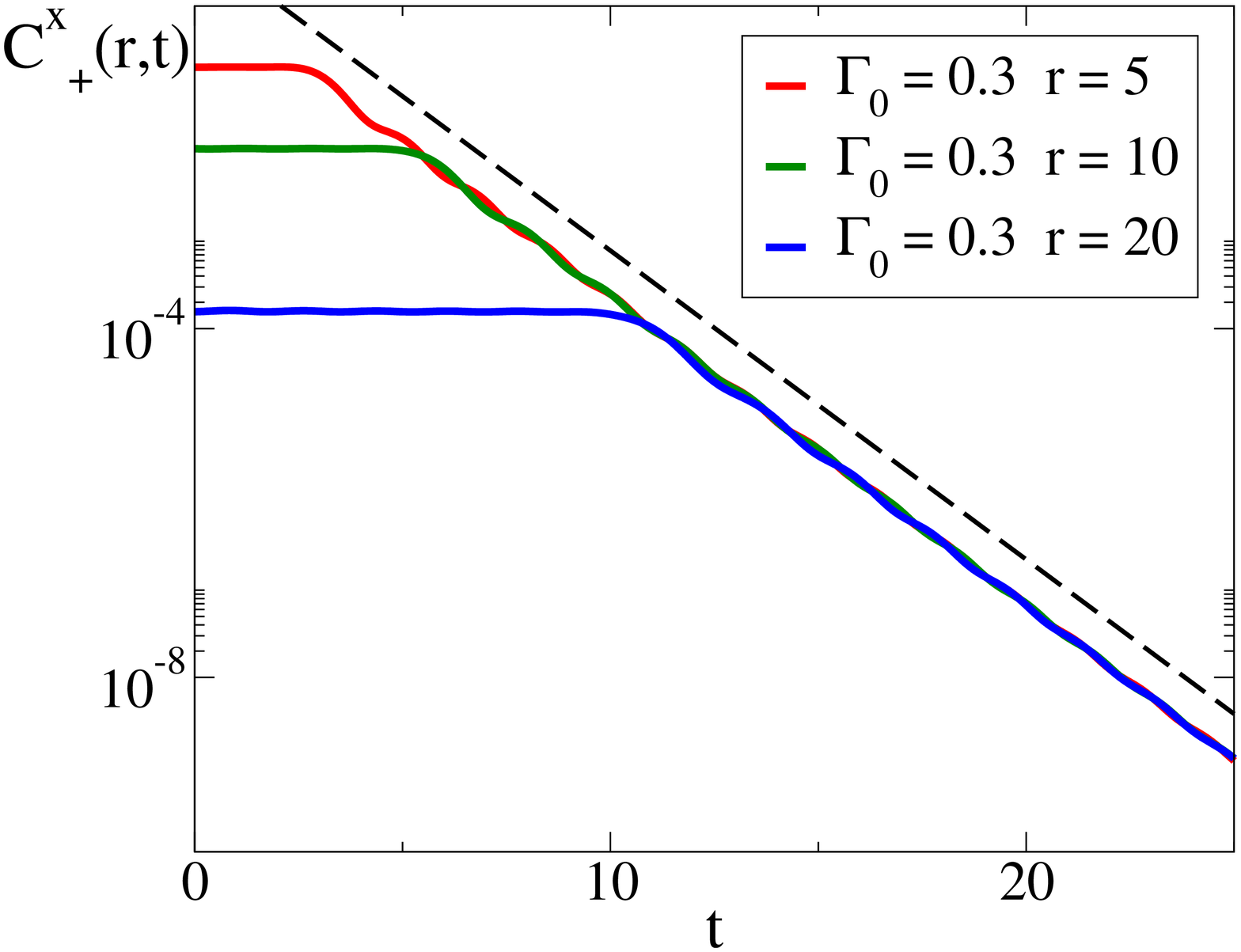}
 \includegraphics[width=0.49\textwidth]{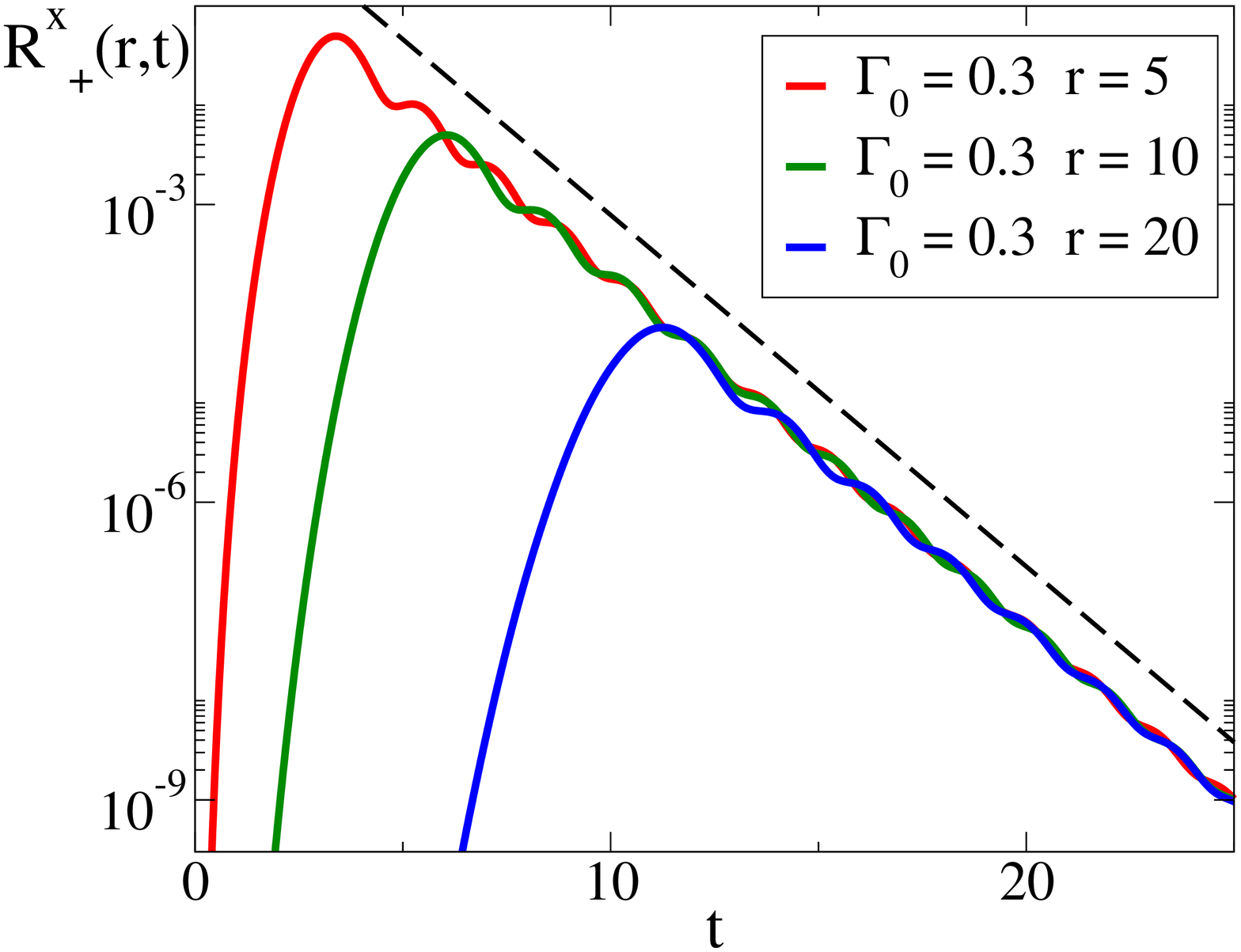}
\caption{Time-dependence of the correlation and response functions 
$C_{+}^x(r,t)$ (left panel) and $R^x(r,t)$ (right panel), respectively, 
for $\Gamma_0=0.3$ and, from top to bottom, various values of $r = 5$ (red curve), 10 (green) and 20 (blue). 
The behavior of both functions at short times is compatible with a light-cone effect with characteristic time 
$r/v_{\rm m} = r/2$ 
for the present case $\Gamma=1$ (see the main text).  Before this characteristic time, the correlation function is almost constant whereas the response function is negligible. 
The eventual exponential decay (highlighted by the choice of the logarithmic scale) is clearly independent 
of $r$ and the dashed lines correspond to a decay rate given by Eq.~\reff{Tau}, which we conjecture to capture the behavior for $r=0$. 
The dependence on $r$ of the correlation function $C^x_+(r,t\simeq 0)$ at small times is compatible with a spatial exponential decay with correlation length $\xi\simeq 2.3$ given by Eq.~\reff{xi}.
}
\label{fig:CRXtime_R_G0_03}
\end{figure}
%
%
Both correlation and response functions are enhanced for small values of $|1-\Gamma_0|$
and, as expected, they vanish in the limit of large time separations. However, 
differently from the case $r=0$ shown in Fig.~\ref{fig:RCz}, these correlations  with $r\neq 0$
display a light-cone effect due to the finite maximum speed 
of the quasi-particles of the model which move ballistically,
with a maximal speed $v_{\rm m}$.
This effect manifests itself in the fact that both the (connected) correlation and the response functions in 
Fig.~\ref{fig:CRXtime_R10} remain almost constant up to 
 times $t\simeq r/2$. After that, the correlation function decays oscillating
 towards its asymptotic vanishing value, whereas the response function first abruptly takes non-negligible values and then decays 
 as well.
This feature is highlighted in Fig.~\ref{fig:CRXtime_R_G0_03}, which shows on a logarithmic scale
the functions $C_{+}^x(r,t)$ (left panel) and $R^x(r,t)$ (right panel) with
fixed $\Gamma_0=0.3$ and, from top to bottom, various values of $r=5$ (red), 10 (green), 20 (blue). 
The behavior can be understood by extending to dynamical quantities in the stationary case the 
qualitative picture of Refs.~\cite{Calabrese06,Calabrese11,Igloi11,rieger2011}, as depicted and summarized 
in Fig.~\ref{fig:lightcone}.
Indeed, one expects the response function $R^x(r,t>0)$ to be negligible
(vanishing in the scaling limit in analogy with what happens for the correlation function \cite{CEF12}) and the correlation function $C_+^x(r,t)$
to be almost constant as long as the point located at $(r,t+t_0)$ does not belong to the forward "light cone" of the point $(r=0,t_0)$. 
Taking into account that the forward light cone expands with a speed 
$v_{\rm m} \equiv \mbox{max}_{k}\, |\rmd\epsilon_k(\Gamma)/\rmd k| = 2\mbox{min}\{\Gamma,1\}$ around the vertex 
point, this is compatible with the horizon effect that appears for $\Gamma=1$ at $t \simeq r/v_{\rm m}= r/2$. 
Note that the emergence of a finite speed at which correlations and responses propagate
is also present at equilibrium~\cite{SacYou97}. 
\begin{figure}[h]
\centering\includegraphics[width=0.5\textwidth]{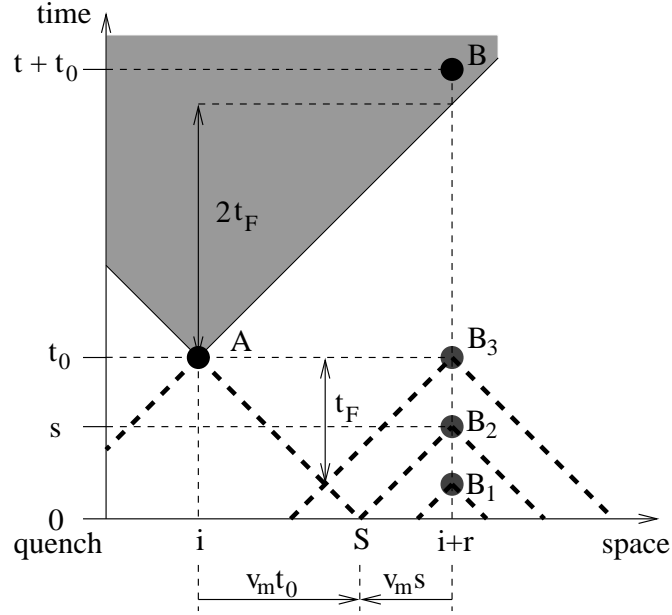} 
\caption{Schematic plot of the various light-cone effects after a quench occurring at $t=0$, in the ``space-time" diagram. Correlations 
between two points A and B separated by a spatial distance $r$ change because of the quench only if their backward light-cones, 
indicated in the figure by dashed lines originating from the points, include spatial points which were significantly correlated in the initial 
state. This is not the case for A and B$_1$ (assuming the spatial extent $\xi_0$ of the initial correlations to be negligible compared to the 
distance between the cones) but it occurs for A and B$_2$, as there is an overlap in the point S, which extends into a wider region at later 
times \cite{Calabrese06}. Indicating by $t_0$ and $s$ the times of A and B$_2$, respectively, the existence of an overlap requires 
$v_{\rm m}(t_0 + s)> r$, where $v_{\rm m}$ is the speed with which the light cone extends around a point. 
For an equal-time correlation function such as the one between A and  B$_3$ (with $s=t_0$), this occurs only if a time larger than the 
``Fermi time" $t_F = r/(2 v_{\rm m})$  \cite{CEF12} has elapsed since the quench, while equal-time connected correlation functions are 
approximately constant and equal to their equilibrium values at zero temperature for $t_0 < t_F$. Similarly, the 
symmetric correlation function  is expected to be constant and the response function 
negligible if the point $B$ does not belong to the forward light cone (indicated by the shaded area) of the 
point A at which the perturbation is applied. This occurs if $t \simeq r/v_{\rm m} = 2 \times t_F$, as it can be seen
graphically from the figure and as discussed in the main text.}
\label{fig:lightcone}
\end{figure}

The eventual exponential time-decay of both $C_{+}^x(r,t)$ and $R^x(r,t)$ in Fig.~\ref{fig:CRXtime_R_G0_03} 
is actually independent of $r$ and occurs with the rate in Eq.~\reff{Tau} which we conjectured to describe the exponential 
decay of the correlation and response functions with $r=0$ [see Eqs.~\reff{eq:Cx-fit}, \reff{eq:Rx-fit} and 
Table~\ref{tau_fitted}], indicated by the dashed lines in Fig.~\ref{fig:CRXtime_R_G0_03}.
Focusing on the behavior of the correlation function $C_+$ at short times, it is possible to study the dependence of its plateau 
values for $t \lesssim r/v_{\rm m} = r/2$ on the distance $r$ and in particular how they
decay as functions of $r$. This issue has been addressed in Refs.~\cite{Calabrese11,CEF12} for 
$\langle \hat{\sigma}^x_{i+r}(t_0)\hat{\sigma}^x_{i}(t_0) \rangle$ [which corresponds to the case $t=0$ of the 
dynamic correlation $C^x_+(r,t)$ studied here]  and, as expected, 
the spatial de-correlation of these plateau values in the present case follows an exponential decay with  
the correlation length $\xi$ found in Refs.~\cite{Calabrese11,CEF12}: 
\beq
\label{xi}
\xi^{-1} = - \int_0^\pi \frac{\rmd k}{\pi} \; \ln\cos\Delta_k(1,\Gamma_0) =  \ln \left( 1 + \frac{1}{\sqrt{\Y}}\right) .
\eeq

\subsubsection{Effective temperatures}

\label{sec:ET}

\begin{figure}[h]
\centering
  \includegraphics[width=0.49\textwidth]{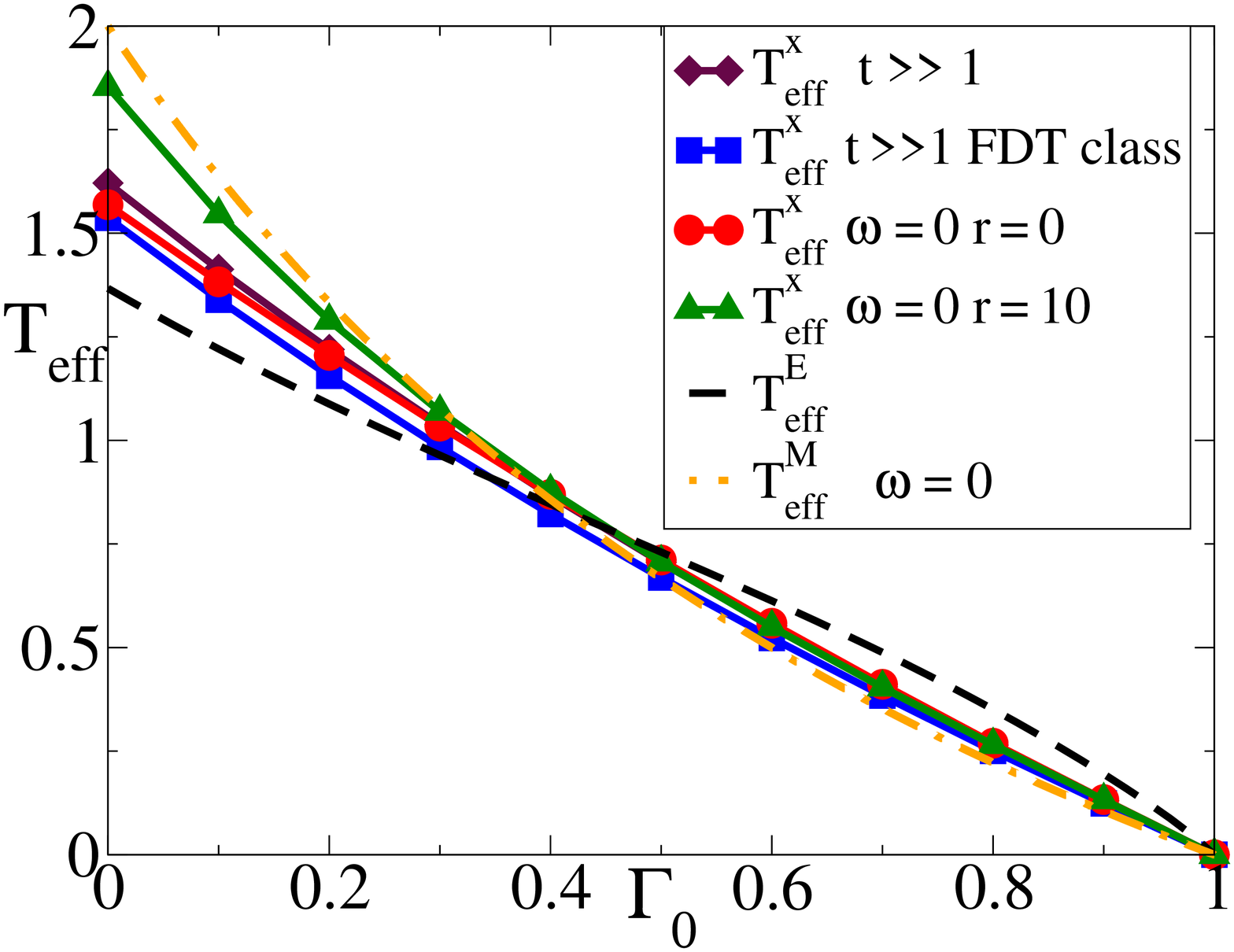} 
   \includegraphics[width=0.49\textwidth]{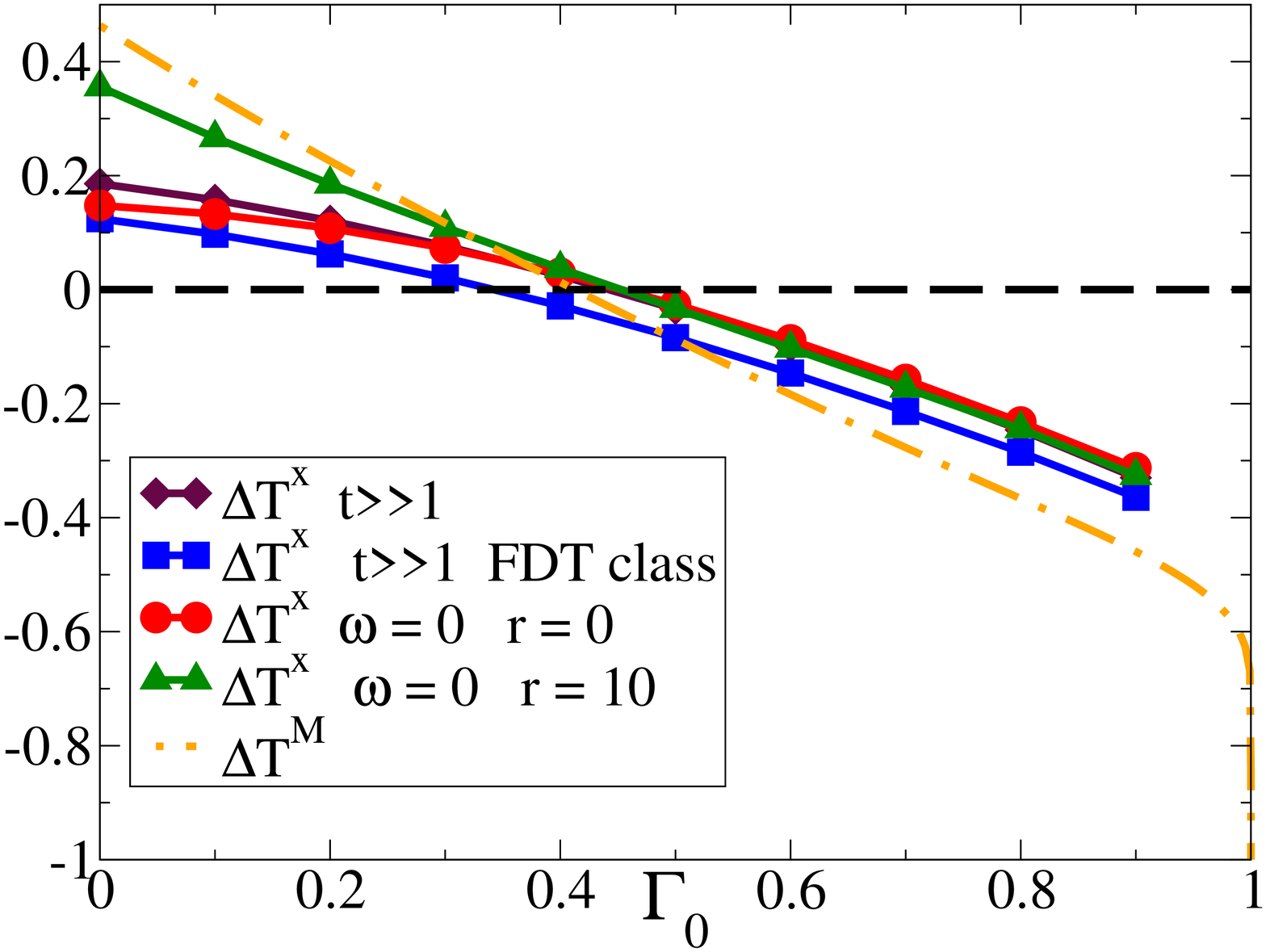}
\caption{Left panel: various parameters $T^x_{\rm eff}$ (symbols and solid lines, see the key) as a  function of $\G_0$, 
compared with the effective temperature $T^E_{\rm eff}$ defined from the energy [dashed line, see Eq.~\reff{Teff_energy}]. 
The solid lines, from bottom to top, indicate the values determined 
on the basis of the classical limit of the FDR in the time domain (squares), 
of the limit $\omega\to 0$ of the frequency-domain FDR (circles),
of Eq.~\reff{Teff_x_time} (diamonds), and of the limit $\omega\to 0$ of the frequency-domain FDR
but for spins separated by a distance $r=10$ (triangles). 
The orange dashed-dotted line shows the effective temperature $T^M_{\rm eff}(\omega\to0)$ 
obtained for the total transverse magnetization in the limit of $\omega\to0$.
Right panel: Relative differences $\Delta T^{\alpha} = (T^{\alpha}_{\rm eff} -T^E_{\rm eff})/T^E_{\rm eff}$
between the various effective temperatures $T^{\alpha}_{\rm eff}$ reported on the left panel and $T^E_{\rm eff}$ taken as a reference, as functions of $\G_0$.  
Colors and symbols are the same as those of the left panel.
}
\label{fig:Teff-Gamma0}
\end{figure}

Here we focus on the effective temperatures associated with the stationary dynamics of the order
parameter $\hat{\sigma}^x$ both in the time and in the frequency domain at small 
frequencies $\omega\to0$, starting from the latter.
Due to the numerical nature of the calculation of $C^x_{\pm}(t)$ as a function of time, 
it is very difficult to access reliably the full frequency dependence of $\tilde{C}^x_{\pm}(\omega)$.
Still, the numerical results obtained for $\omega\to0$ are robust, also against 
the fact that a finite time interval for the integration has necessarily to be taken.
In fact, we have studied the small-frequency regime
in a time interval $t \in [0,t_{\rm max}]$ for different values of $t_{\rm max}$
and  we found that it stabilizes rapidly with increasing $t_{\rm max}$, $t_{\rm max}$
being an increasing function of $\tau$. Small frequencies capture
the exponential decay and they are less sensitive to other details. For these reasons,
their extrapolation in a sufficiently large, but finite, time interval  is expected
to provide reliable results. Even if a complete characterization in the full frequency spectrum 
of the correlation functions and the associated effective temperatures 
is definitely an interesting issue, the region of small $\omega$ is for us the most interesting because 
we aim at comparing $\beta^x_{\rm eff}(\omega\to0)$ with $\beta^{\ast}_{\rm eff}$ 
obtained from Eq.~\reff{eq:FDT-t-Teff} in the long-time limit.  
The zero-frequency limit of Eq.~\reff{eq:FDT-omega-Teff} yields:
\beq\label{Teff_omega0}
\displaystyle
\beta_{\eff}^{x}(\omega \to 0) = 
\frac{\int_0^{\infty}\!\rmd t\, t R^x(t)}{\int_0^{\infty}\!\rmd t\,C^x_{+}(t)},
\eeq
that we take as our working definition of $\beta_{\eff}^x(\omega \to 0)$.
The corresponding temperature $T^x_{\rm eff}(\omega=0) = 1/\beta^x_{\rm eff}(\omega=0)$ 
is indicated by circles in Fig.~\ref{fig:Teff-Gamma0} as a function of $\G_0<1$.
Alternatively, as discussed after Eq.~\reff{Eq-beff-timedomain}, we can enforce a constant value 
$\beta_{\eff}^{\ast}$ in the (generalized) FDT (\ref{eq:FDT-t-Teff}) 
in the time domain.
This operation allows one to interpret the r.h.s.~of that equation as a  
series of time derivatives of $C^x_+(t)$ and, after the substitution into this series of the long-time behaviors in Eqs.~\reff{eq:Cx-fit}
 and \reff{eq:Rx-fit}, it yields
\begin{equation}
\label{Teff_x_time}
\hbar A_R/(2A_C) =  \tan(\hbar\beta_{\eff}^x/2 \tau)
\end{equation}
for $t\to\infty$, in which one neglects the oscillatory 
corrections in Eqs.~\reff{eq:Cx-fit} and \reff{eq:Rx-fit}. 
[Note that the oscillatory terms could not be discarded in the analogous analysis presented for $\hat M$ after Eq.~\reff{eq:Teff0M} 
because they were actually providing the leading contributions.] 
Note that $\hbar$ has been reinstated in Eq.~\reff{Teff_x_time} for completeness, but we shall set again $\hbar=1$ in what follows. 
As we pointed out in Eq.~\reff{eq:ratA}, $A_R/A_C$ --- and therefore the l.h.s.~of Eq.~\reff{Teff_x_time} --- does not depend 
significantly on $\G_0$ within the range of values of $\G_0$ investigated here;
accordingly, $\beta^x_{\rm eff}$ on the r.h.s.~of Eq.~\reff{Teff_x_time} can be expressed in terms of $\tau$ as 
$\beta^x_{\rm eff} \simeq 0.78 \tau$ 
and inherits its dependence on $\G_0$ given by Eq.~\reff{Tau}.
The corresponding temperature $T^x_{\rm eff} = 1/\beta^x_{\rm eff}$ is referred to as ``$T^x_{\rm eff} \ t \gg 1$" in 
Fig.~\ref{fig:Teff-Gamma0}, where it is plotted as a function of $\G_0$ and the corresponding data points are denoted by diamonds.
In this case we find that the effective temperatures determined in 
frequency  (circles) and time domain (diamonds) are almost indistinguishable, especially for $\G_0\simeq 1$.
For $\hbar\beta_{\rm eff}^x/2\tau \ll 1$ in Eq.~(\ref{Teff_x_time}) one recovers 
the classical limit (\ref{FDT_classical})
$\beta^x_{\eff} \simeq -R^x(t)/[\rmd C^x(t)/\rmd t]
\simeq \tau A_R/A_C \simeq 0.83\tau$. 
The value of the corresponding temperature, referred to as ``$T^x_{\rm eff}\ t \gg 1$ FDT class." is indicated by squares in Fig.~\ref{fig:Teff-Gamma0}.
All the three determinations of $T_{\rm eff}^x$ discussed so far are compared in
Fig.~\ref{fig:Teff-Gamma0} with the effective temperature 
$T^E_{\rm eff}$ (dashed line) determined on the basis of Eq.~(\ref{Teff_energy})~\cite{Rossini10_short}.
In the same figure we also show (triangles)
the zero-frequency effective temperature
obtained from Eq.~\reff{Teff_omega0} on the basis of the two-point functions 
$C^x_{+}(r,t)$ and $R^x(r,t)$ with $r=10$. 

%
In order to facilitate the quantitative comparison between the various effective temperatures $T^{\alpha}_{\rm eff}$ (which we label generically by $\alpha$) reported in Fig.~\ref{fig:Teff-Gamma0}, its right panel shows their relative difference $\Delta T^{\alpha} = (T^{\alpha}_{\rm eff}-T^E_{\rm eff})/T^E_{\rm eff}$ as a function of $\G_0$, 
where the effective temperature $T^E_{\rm eff}$ (dashed lines in both panels of the figure) obtained from the energy balance is taken as the reference.
As we discuss in App.~\ref{app:TeffEnergyShallow},
\beq
T^E_{\rm eff} \simeq \frac{2}{\sqrt{\Y} }\frac{\sqrt{6~ ( \log \Y  + 4 \log 2 - 3)}}{\pi} 
\eeq
for $\Y \to \infty$, \ie, in the limit of shallow quenches $\G_0\to \G = 1$ (see Eq.~\reff{eq:defY}).
Accordingly, taking into account Eq.~\reff{eq:Teff0M}, the ratio $\Delta T^M = (T^M_{\rm eff} - T^E_{\rm eff})/T^E_{\rm eff}$ tends to $ - 1$ for $\Gamma_0\to1$, with the logarithmic approach clearly displayed by the dash-dotted curve on the right panel of Fig.~\ref{fig:Teff-Gamma0}.
Both panels of this figure demonstrate that, in general,  the temperatures defined from the FDR associated with the stationary dynamics of the various quantities do not coincide with the ``static" temperature $T^E_{\rm eff}$ defined on the basis of the energy of the system. 
However, in the neighborhood of $\Gamma_0\simeq0.4$, the relative discrepancies $\Delta T^\alpha_{\rm eff}$ are less pronounced (see right panel), which can be heuristically traced back to the fact that the temperature $T^k_{\rm eff}$
of the various modes (see Eq.~\reff{eq:TGGE} and the right panel of Fig.~\ref{fig:TeffM-omega}) 
is practically constant and equal to $T^{k=0}_{\rm eff}$ within a rather wide interval of momenta $k$. 
Upon moving away from $\G_0\simeq 0.4$, such an interval shrinks, as it is clearly shown by the various curves on the right panel of Fig.~\ref{fig:TeffM-omega}, which correspond to different values of $\G_0$.
On the other hand, differently from the cases of $\hat M$ and $\hat\sigma^z$, the frequency-dependent dynamic effective temperature for the order parameter takes a non-vanishing value for $\omega=0$ (circles in Fig.~\ref{fig:Teff-Gamma0}), which can be recovered with great accuracy from 
the long-time limit of the (classical and especially quantum) FDR in the time domain (squares and diamonds in Fig.~\ref{fig:Teff-Gamma0}). Whereas these values 
are almost indistinguishable for $\G_0\simeq 1$, slight discrepancies emerge away from the critical point. However, these might be due to the numerical accuracy of the calculation.  
More consistent discrepancies, instead, emerge in comparison with the case 
$r=10$ (and with the temperature $T^E_{\rm eff}$), but they become anyhow negligible for $0.5 \lesssim \G_0 \lesssim 2$, \ie, in a rather wide neighborhood of the critical point. 

\section{Summary and conclusions}
\label{sec:conclusions}

In this Section we first summarize our findings and we then discuss the
perspectives of this work, which was partly anticipated by the
brief account in Ref.~\cite{Foini11}.

Aiming at understanding whether and how thermalization to the canonical
ensemble described by a Gibbs distribution
may arise after a quench of an isolated quantum system, we focused on the
fluctuation-dissipation
relations (FDRs) between
a set of dynamic quantities. By comparing these FDRs to the canonical FDT
holding in equilibrium, we
extracted parameters, actually functions, that in any (even partial)
equilibrium situation should be equal for all
choices of observables and, moreover,  constant. With a certain abuse of
terminology
we called these functions effective temperatures and we examined
whether they satisfy, and under which circumstances they do, the constancy
property just mentioned.

For illustration purposes we pursued this approach by investigating
the dynamics of the quantum Ising chain in a transverse field $\Gamma$
(see Sec.~\ref{sec:Ising}), which is
initially prepared in the ground state of the Hamiltonian $\hat H(\Gamma)$
with $\G=\G_0$ and then quenched
to the critical point $\G=1$. We
computed  correlation and (linear) response functions of the local
transverse magnetization $\hat{\sigma}^z$, the global
magnetization $\hat{M}^z = (\sum_i  \hat{\sigma}^z_i)/L$ and the
local order parameter $\hat{\sigma}^x$, extracting the associated
effective temperatures.
We chose to work with this model knowing that sound evidence has
accumulated  over the
last years for the fact that its stationary properties, as well as those
of more general integrable systems,
can be consistently described by the so-called generalized Gibbs ensemble
(GGE)~\cite{Rigol,CEF12,CIC-12}.
This implies that the effective temperatures defined from the canonical FDRs
are not expected {\it a priori} to take a common value. Nonetheless, they
may anyhow signal regimes of
partial equilibration and thus capture important
features of the dynamics of the system, as it happens in a number of
classical non-equilibrium
cases in which the Gibbs stationary ensemble cannot be reached because of
the slow relaxation~\cite{Cugliandolo-review}.

As a first observation let us note that the FDRs can be used
to identify effective temperatures independently of the functional form
of the correlation and (linear) response functions involved in the FDRs.
This is of great practical advantage since it
allows us to investigate also quantities with power-law (or any other kind
of) decays, as opposed to the
strategies that base the comparison between the quenched (non-equilibrium)
and the thermal (equilibrium) dynamics
on the analysis of the characteristic time- or length-scales primarily
associated with exponential decays.
Moreover, it is well-known from the analysis
of classical~\cite{Cugliandolo97} (and to a certain extent also
quantum~\cite{Culo,Caso}) glassy and coarsening
dissipative models that the functional form of the decay of the
correlation functions can be highly non-trivial
(non-stationary, non-exponential) and yet, in some circumstances, an
effective
thermal behavior can develop asymptotically in certain time regimes. It is
therefore highly desirable not to confuse the qualitative features of time
decays with the possible emergence of a thermal behavior.
The investigation of the  system's properties
in terms of FDRs provides the possibility to analyze separately its behavior
within different time (or frequency) regimes.

In order to illustrate this general approach, we calculated the (self) FDR
for three observables of the Ising model
that are local ($\hat{\sigma}_i^{x,z}$) or non-local ($\hat{M}$) in space
and local
($\hat{\sigma}_i^z$, $\hat{M}$) or non-local ($\hat{\sigma}_i^x$) in the
quasi-particles.
By inspection of the FDR and of the frequency-dependent $T_{\rm eff}^z$
parameter that
we extract from it, we concluded that the dynamics of $\hat{\sigma}_i^z$
is not compatible with Gibbs thermal equilibrium
at any effective temperature, in spite of the fact that thermal-like
behavior --- compatible with a Gibbs distribution at an effective
temperature $T_{\rm eff}^E$ set by the initial energy --- is observed
for the stationary expectation value $ \langle \hat{\sigma}^z
\rangle_{\mathcal Q} =  \langle \hat{\sigma}^z \rangle_{T=T_{\rm eff}^E}$.
Note that this instance clearly demonstrates the importance of the
alternative approach to the issue of thermalization we are currently
proposing
beyond the specific model investigated here: indeed, the analysis of
specific time-independent quantities (in this case,
$ \langle \hat{\sigma}^z \rangle_{\mathcal Q}$) would suggest a picture of
thermalization which can be easily disproved by focussing on
dynamical quantities (in this case, the two-time response and correlation
functions $\hat{\sigma}^z$).

The FDR  in the frequency-domain yields for $\hat{M}$ a finite $T^M_{\rm
eff}(\Gamma_0)$
in the limit $\omega \to 0$ (see Fig.~\ref{fig:TeffM-omega}, right panel
and the
dot-dashed line in Fig.~\ref{fig:Teff-M}), which is actually connected
to the fact that low-energy modes have a
finite mode-dependent effective temperature $T^k_{\rm eff}$ [see
Eq.~\reff{Teff_GGE}]
for $k \to 0$, in relation to the GGE [see Eq.~\reff{eq:TGGE}].
This feature is peculiar of quenches to (and from)
the critical point and it does not carry over to the other cases.
However, contrary to the heuristic expectation, we did not find a way to
recover the value
$T^M_{\rm eff}(\Gamma_0)$ directly from the long-time limit of the FDR in
the time domain.
In fact, the effective temperature $T_{\eff}^{\ast}$
of the long-time dynamics of $C_{\pm}^M(t)$ (and derivatives)
is somehow controlled by the behavior at high frequencies, as discussed in
Sec.~\ref{Sec:Mz}.
This is a natural consequence of
the presence of oscillating terms at the leading order
in $C_{\pm}^M(t)$ with a frequency equal to the frequency $\omega_{\rm
max}$ above which the Fourier transform of $C_\pm^M(t)$ vanishes
identically.  This structure highlights that not only the slow power-law
decay of $C_{\pm}^{M,z}(t)$ is important,
but also the oscillations are a constitutive ingredient. 
This structure,
as well as the presence of the maximum
frequency $\omega_{\rm max}$,  is the combined result of the presence of the lattice cut-off that
bounds the dispersion
relation and of the particular quadratic form of the observable with
respect to the elementary excitations of the model.

The dynamics of the operator $\hat{\sigma}_i^x$ shows, instead, a very
different behavior.
Both $C^x_+(t)$ and $R^x(t)$ decay exponentially (see Figs.~\ref{fig:RCz},
\ref{fig:CX_tau}, \ref{fig:CRXtime_R_G0_03} and the
discussion further below) with a characteristic time
given by Eq.~\reff{Tau}, which decreases upon increasing the energy
injected into the system.
Differently from the correlations of $\hat\sigma^z$,
the time decay of $C^x_+(t)$ and $R^x(t)$,
for quenches at the critical point $\Gamma=1$,  does not display
oscillations at the leading order for large times.
The oscillatory part, which intervenes only in the leading corrections, is
therefore less relevant
and one would expect that the long-time limit of the corresponding
$T_{\eff}^{\ast}$ could be regularly
recovered from a small-frequency expansion, as it is actually the case
(compare circles and diamonds in Fig.~\ref{fig:Teff-Gamma0}).
$T_{\rm eff}^M$ and $T_{\rm eff}^x$ have the same qualitative dependence
on $\Gamma_0$ but they still present some differences (and they also
differ from $T_{\rm eff}^E$),
thus excluding a single temperature effective Gibbs description.
Finally, we have shown that, by choosing appropriate
observables, one can recover from the study of FDRs
 the effective temperatures $T^k_{\rm eff}$ associated with the modes,
which in turn characterize the GGE~\cite{Rigol} [see Eq.~\reff{eq:TGGE}].

While the primary focus of this work was on the application of
fluctuation-dissipation relations for addressing the issue of the
possible thermalization of isolated quantum systems long after a quench,
our study of two-time quantities (connected
correlation and response functions)  lead to the following observations as
far as their behavior is concerned:
(i) For a quench to the critical point $\G=1$, the dynamical two-time
quantities after the quench are invariant under
$\G_0 \leftrightarrow \G_0^{-1}$ in the \emph{stationary regime}. Such an
invariance is broken when at least one of the two times
approaches the moment of the quench. This statement is supported by
analytical calculations in the case of the local and global
transverse magnetizations $\hat \sigma^z_i$ (see App.~\ref{app:sz}) and
$\hat M^z$, whereas it is based on
numerical observations in the case of $\hat \sigma^x_i$.
Moreover, the expectation value of the energy [see Eq.~\reff{eq:EnQ}] and
of the transverse magnetization [see Eq.~\reff{eq:Qavsz}] for
$\G=1$ possess the same invariance as functions of $\G_0$.
We conjecture that this should be not only a generic property of the
dynamics in the \emph{stationary state} but it should also be
largely independent of the specific quantity under study, being somehow
connected to the duality of the equilibrium model.
To us, a deeper understanding of the nature and the limits of such a
symmetry --- which is broken by non-stationary contributions ---
is definitely an interesting issue.
(ii) In the stationary state, the correlation and response functions of
the order parameter $\hat\sigma^x_i$ decay exponentially for
large time separations, as a function of it [see Eqs.~\reff{eq:Cx-fit} and
\reff{eq:Rx-fit}]. The decoherence time $\tau$ of this decay
[see Eq.~\reff{Tau}] is
(numerically) the same as the one which characterizes the relaxation of
the average of the order parameter
 $ \langle\hat{\sigma}^x_i(t) \rangle \propto \exp(-t/\tau)$
after the quench from the ferromagnetic phase
and of the equal-time correlation
$\langle \hat{\sigma}^x_i(t)\hat{\sigma}^x_j(t)  \rangle \propto \exp(- 2
t/\tau)$ for $|i-j|\gg 4t$,
 calculated in Refs.~\cite{Calabrese11,CEF12}.
Even though we only have  numerical evidence for this fact (for $\G=1$),
we conjecture that this should be true beyond this case, \eg, for quenches
towards non-critical states (see Sec.~\ref{Sec:sigmax} for additional
arguments).
An analytical proof of these features [and particularly of
Eqs.~\reff{eq:Cx-fit} and \reff{eq:Rx-fit}] should be possible within the
approaches
discussed in Refs.~\cite{Calabrese11,CEF12,ES12}. 
(iii) The two-point dynamical
correlation functions of the order parameter in the stationary regime
exhibit ---
similarly
to the relaxation dynamics of the two-points and equal-time
correlations~\cite{Igloi11,rieger2011,Calabrese11,CEF12}
 --- light-cone effects which are compatible with the expectations
 based on the results for correlation functions of CFTs (\ie, in the
scaling limit) \cite{Calabrese06}.
This picture turns out to extend also to the \emph{response} function, which
is negligible up to times $t\simeq r/v_{max}=r/2$ (see
Fig.~\ref{fig:lightcone} for a comprehensive schematic summary of these
effects).

In conclusion, concerning the issue of thermalization in quantum quenches
which primarily motivated our study, we emphasize that a \emph{bona fide}
thermal
behavior should always be accompanied by the validity of suitable FDRs,
also in the context of isolated systems.
In this respect it would be desirable to compute FDRs in the cases in which
thermalization to a Gibbs ensemble is eventually expected, such as
non-integrable systems.
However, due to the general lack of analytic solutions even for the
simplest models,
this would require the use of numerical methods, for which it is still
rather difficult to access the long-time stationary regime.
In spite of this difficulty, it is worth mentioning that the approach we
propose here for
 probing the possible thermalization of a
quantum system involves only basic quantities --- correlation and response
functions --- which, at least in principle, can be naturally
accessed both in numerical and experimental investigations.
Drawing an analogy with the case of non-equilibrium \emph{classical} systems,
the effective temperatures defined above might still
provide a useful description of some coarser aspects of the physics of the
system or of its
thermodynamic, such as, \eg, its energy
exchange with or response to a device used as a thermometer.
Clearly, this question is of interest especially  when there is no
eventual thermalization to a
Gibbs state, \eg, in the case of the
Ising model considered in the present work or more generally for
integrable systems.

\acknowledgments
We gratefully acknowledge useful discussions with P.~Calabrese, F.~Essler, M.~Fagotti, J.~P.~Garrahan, and 
I.~Lesanovsky.
AG is supported by MIUR within "Incentivazione alla mobilit\`a di
studiosi stranieri e italiani residenti all'estero."
AG thanks the Galileo Galilei Institute for Theoretical Physics in Florence for the hospitality during the workshop "New quantum states of 
matter in and out of equilibrium". LFC and LF thank financial support from  ANR-BLAN-0346
(FAMOUS).

\appendix

\section{Transverse magnetization}
\label{app:sz}

This Appendix provides the details of some analytical calculations concerning the transverse magnetization 
whose results have been presented in the main text. As explained in Sec.~\ref{sec:model},
we consider a chain of finite size $L$ and then we take the
thermodynamic limit, which formally amounts to the substitution 
$1/L \sum_k \to  \int_{-\pi}^{\pi} \rmd k/(2\pi)$~\cite{Rossini10_long}.
In the following we focus on the (connected) symmetrized and antisymmetrized  autocorrelation
functions $C_+$ and $C_-$, respectively:
\beq
C_{+}^{z}(t + t_0,t_0) = \frac12 \bra{\psi_0}  \{ \hat{\sigma}^z_{i} (t + t_0) \, , \,  \hat{\sigma}^z_i (t_0) \}  \ket{\psi_0}  - 
 \bra{\psi_0}  \hat{\sigma}^z_{i} (t + t_0)  \ket{\psi_0} \bra{\psi_0}   \hat{\sigma}^z_i (t_0)  \ket{\psi_0} 
\eeq
and
\beq
C_{-}^{z}(t + t_0,t_0) = \frac{1}{2} \bra{\psi_0} [  \hat{\sigma}^z_{i} (t + t_0) \, , \,  \hat{\sigma}^z_i (t_0) ]  \ket{\psi_0},
\eeq
where the expectation value is taken over the ground state $\ket{\psi_0}=\ket{0}_{\Gamma_0}$ 
of the Hamiltonian $\hat H(\Gamma_0)$ before the quench.
We consider the connected correlations since in general the expectation value $\bra{\psi_0}   \hat{\sigma}^z_i (t)  \ket{\psi_0}$ does not vanish.
The Kubo formula [see Eq.~\reff{Kubo}] allows one to express the linear response function $R^z$ in terms of $C_-$, whereas Eq.~\reff{eq:ReC} connects $C_+$ and $C_-$ to the real and imaginary part, respectively, of the correlation function $C^z(t+t_0,t_0) \equiv \bra{\psi_0}\hat{\sigma}^z_{i} (t + t_0) \hat{\sigma}^z_i (t_0) \ket{\psi_0}$. 
As outlined in Sec.~\ref{Sec:dyn_quench} 
this expectation value is calculated by first expressing $\hat{\sigma}^z_{i}(t)$ in terms of $\hat c_i(t)$ [see Eq.~\reff{sz-c}] and then $\hat c_i(t)$ in terms of the quasi-particles 
$\{ \hat \gamma_k^{\Gamma_0}\}$ of $\hat H(\Gamma_0)$ via
the matrix elements defined in Eq.~\reff{Eq:Dynamics_quench}.  
Introducing the simplified notation
$v_k(t)= v_k^{\Gamma,\Gamma_0}(t)$ and $ u_k(t)= u_k^{\Gamma,\Gamma_0}(t)$,  
the symmetric and antisymmetric autocorrelations for the transverse magnetization are given by the real and the imaginary part (see Eqs.~\reff{eq:ReC} and \reff{eq:ImC}) of 
\beq
C^z(t+t_0,t_0) = \frac{4}{L^2}  \sum_{k,l} \Big[ v_k(t + t_0) \, v_k^*(t_0) \, u_l(t + t_0) \, u_l^*(t_0) 
 + v_k(t + t_0) \, v_l^*(t_0) \, u_l(t + t_0) \, u_k^*(t_0) \Big].
 \label{eq:Czraw}
\eeq
Note that, according to Eqs.~\reff{eq:ukvk} and \reff{eq:tanth}, 
\beq
u_{-k} = u_k \quad \mbox{whereas} \quad v_{-k} = - v_{k}. 
\label{eq:ukvk-sym}
\eeq
This implies that the second term in the sums is odd 
with respect to $k$ and therefore it does not contribute to the total (symmetric) sum.
Then, expressing the first term via the angles of the Bogoliubov rotation introduced in Eqs.~\reff{eq:ukvk}  and 
\reff{Eq:Dynamics_quench}, one eventually finds ($\epsilon_k \equiv \epsilon_k(\Gamma)$, see Eq.~\reff{eq:energy})
\beq
\label{eq:Cpzraw}
\begin{split}
 C_{+}^{z}(t + t_0,t_0)     = & \frac{4}{L^2} \sum_{k} \left[
  \cos( \epsilon_k  t ) \left( \sin^2 \theta_k \cos^2 \delta_k
   + \cos^2 \theta_k \sin^2 \delta_k  \right)
   - 2 \cos(\epsilon_k (t + 2 t_0))  \sin \theta_k  \cos \delta_k \cos \theta_k \sin \delta_k
   \right] \\
& \times \sum_{l} \left[
\cos(\epsilon_l t) \left(  \cos^2 \theta_l \cos^2 \delta_l
+  \sin^2 \theta_l \sin^2 \delta_l \right) 
+ 2 \cos(\epsilon_l (t + 2 t_0))  \sin \theta_l  \cos \delta_l \cos \theta_l \sin \delta_l
 \right] \\
 & -  \frac{4}{L^2}\sum_{k} \left[
  \sin( \epsilon_k t)  \left( \sin^2 \theta_k  \cos^2 \delta_k
   -  \cos^2 \theta_k \sin^2 \delta_k  \right)
   \right] 
\sum_{l} \left[\sin( \epsilon_l t ) \left(  \cos^2 \theta_l \cos^2 \delta_l 
-  \sin^2 \theta_l \sin^2 \delta_l \right)
 \right].
\end{split}
\eeq
In this expression (and in the analogous one for $C^z_-$, which can be obtained from the imaginary part of Eq.~\reff{eq:Czraw}) one can easily recognize a stationary part $C_+^z(t) \equiv \lim_{t_0\to\infty}C^z_+(t+t_0,t_0)$ which depends only on $t$ and a non-stationary contribution $C_+^{z,{\rm ns}}(t+t_0,t_0)\equiv C_+^z(t+t_0,t_0) - C^z_+(t)$ which depend on both times $t$ and $t_0$. 
We anticipate here that the long-time limit $t_0\to\infty$ of the terms in Eq.~\reff{eq:Cpzraw}  which involve 
$\cos(\epsilon_k(t+2 t_0))$ vanishes, as it is clear both from a numerical study of the corresponding sums and from the analysis of the long-time behavior of similar expressions in the thermodynamic limit, done further below in App.~\ref{app:asympt}. (The same applies to the identical terms which appear in the analogous equation for $C_-^z(t+t_0,t_0)$, which, however, we do not report here.) Accordingly, the stationary part of $C^z_+$ is given by 
\beq
\begin{split}
 C_{+}^{z}(t) = &  \frac{4}{L^2} \sum_{k} \Big[
  \cos( \epsilon_k t) \Big( \sin^2 \theta_k  \cos^2 \delta_k   
  + \cos^2 \theta_k \sin^2 \delta_k  \Big)   \Big]
\sum_{l} \Big[
\cos(\epsilon_l t) \Big(  \cos^2 \theta_l \cos^2 \delta_l
+  \sin^2 \theta_l \sin^2 \delta_l \Big)  \Big] \\
 & - \frac{4}{L^2}\sum_{k} \Big[
  \sin( \epsilon_k t)  \Big( \sin^2 \theta_k  \cos^2 \delta_k
   -  \cos^2 \theta_k \sin^2 \delta_k  \Big)
   \Big]
\sum_{l} \Big[\sin( \epsilon_l t ) \Big(  \cos^2 \theta_l \cos^2 \delta_l 
-  \sin^2 \theta_l \sin^2 \delta_l \Big)
 \Big] . 
\end{split}
\label{eq:Cp-app1}
\eeq
For later convenience we report here also the expression of the non-stationary contribution 
$C_+^{z,{\rm ns}}(t+t_0,t_0)$, which takes the form:
\beq\label{non-stationary}
\begin{split}
 C_{+}^{z, {\rm ns}}(t + t_0,t_0)     = &  \frac{4}{L^2} \sum_{l , k} \Big[
  \cos( \epsilon_k  t )  (\sin^2 \theta_k - \cos^2 \theta_k) ( \cos^2 \delta_k - \sin^2\delta_k)
 2 \cos(\epsilon_l (t + 2 t_0))  \sin \theta_l  \cos \delta_l \cos \theta_l \sin \delta_l
 \Big] \\
 & 
 - \frac{4}{L^2} \Big[\sum_{l }2 \cos(\epsilon_l (t + 2 t_0))  \sin \theta_l  \cos \delta_l \cos \theta_l \sin \delta_l \Big]^2 \\
  & = - \frac{1}{L} \sum_{k} \Big[
  \cos( \epsilon_k  t )  \cos (2 \theta_k) \cos\Delta_k \Big] W_+(t,t_0)
 - W_+^2(t,t_0)
\end{split}
\eeq
where we introduced $\Delta_k \equiv  2\delta_k$,
\beq
\label{eq:Wp}
W_+(t,t_0) = \frac{2}{L} \sum_{l } \cos(\epsilon_l (t + 2 t_0))  \sin (2 \theta_l)\sin\Delta_l,
\eeq
and 
\beq
\label{eq:sinDk}
\sin\Delta_k(\Gamma,\Gamma_0) =\frac{(\G_0-\G)\sin k }{\sqrt{1 + \G_0^2 - 2 \G_0 \cos k}\sqrt{1 + \G^2 - 2 \G \cos k}} \ .
\eeq
As expected from their very definitions, $W_+(t,t_0)$ and $C_{+}^{z, {\rm ns}}$ decay to zero for $t_0\to\infty$.
The stationary part $C_-^z(t) \equiv \lim_{t_0\to\infty} C_-^z(t+t_0,t_0)$ of $C_-^z(t+t_0,t_0)$, instead, is given by 
\beq
\label{eq:Cm-app1}
\begin{split}
 C_-^{z}(t) 
 =  &  \frac{4}{L^2}    \sum_{k} \Big[
 \sin( \epsilon_k t)  ( - \sin^2 \theta_k  
   +   \sin^2 \delta_k  )
   \Big]  \sum_{l} \Big\{
\cos( \epsilon_l t)  \Big[ \cos^2 \theta_l 
-  \cos^2 (2\theta_l) \sin^2 \delta_l\Big]
 \Big\} \\  
  & + \frac{4}{L^2}     \sum_{k} \Big\{
  \cos( \epsilon_k t) \Big[ \sin^2 \theta_k  
   + \cos^2 (2\theta_k) \sin^2 \delta_k\Big]
    \Big\}
\sum_{l} \Big[
\sin( \epsilon_l t)   ( - \cos \theta_l^2 
+  \sin \delta_l ^2 )
 \Big]  .
 \end{split}
\eeq
Due to the fact that the terms in the sums of Eqs.~\reff{eq:Cp-app1} and \reff{eq:Cm-app1} are all even functions of $k$, in what follows we will replace $\sum_k$ with $2 \sum_{k>0}$. Note that the expressions reported above apply to an arbitrary quench.

\subsection{Quenches to the critical point $\Gamma=1$}\label{Appendix_sz_critical}

The primary interest of the present study is in the case  of a quench to the critical point $\Gamma=1$, for which we can further specialize the equations reported above.
In particular, the r.h.s.~of Eq.~\reff{eq:tanth}
is always positive, and therefore this relation for $k>0$ can be inverted with $0 \le 2\theta_k \le \pi/2$:
\beq
\cos (2\theta_k) = \frac{1}{\sqrt{1+\tan^2(2\theta_k)}} = \sin (k/2).
\label{eq:costhcrit}
\eeq
In view of this equation and of Eq.~\reff{eq:tanth} for $\cos \Delta_k$ 
one can conveniently express all the trigonometric 
functions of $\delta_k$ and $\theta_k$ in Eqs.~\reff{eq:Cp-app1} and \reff{eq:Cm-app1} in 
terms of $\cos (2\delta_k)$ and $\cos(2\theta_k)$. Indeed, $C^z_\pm(t)$ in the stationary state depend on $\delta_k=\Delta_k/2$ and $\theta_k$ via $\sin^2$ and $\cos^2$ and therefore they can unambiguously be expressed in terms of the $\cos$ of the corresponding double angles, \ie, $\Delta_k$ and $2\theta_k$, and in particular in terms of $\cos \Delta_k(\G,\G_0)$, which encodes the dependence on the initial state.
However, this is not the case for the non-stationary contribution, which indeed requires also the introduction of the $\sin$ of these angles, as in Eqs.~\reff{non-stationary} and \reff{eq:Wp}, and in particular of 
$\sin \Delta_k(\G,\G_0)$ (see Eq.~\reff{eq:sinDk}).
As a consequence, the stationary and non-stationary parts of $C^z_\pm$ at the critical point $\G=1$ display different behaviors under the mapping $\G_0 \to \G_0^{-1}$ of the initial condition. In particular, 
as we discuss after Eq.~\reff{eq:defY} in Sec.~\ref{Sec:sigmaz}, $\Delta_k(\G=1,\G_0) = - \Delta_k(\G=1,\G_0^{-1})$ and therefore $\cos \Delta_k(\Gamma=1,\G_0)$ is invariant under this mapping, whereas $\sin \Delta_k(\Gamma=1,\G_0)$ changes sign, as it can be easily verified also by a direct analysis of Eq.~\reff{eq:sinDk}. Accordingly, the stationary parts $C_\pm^z(t)$ of the dynamics in Eqs.~\reff{eq:Cp-app1} and \reff{eq:Cm-app1} for $\G=1$ are invariant for $\G_0 \to \G_0^{-1}$ --- a fact which will be apparent from the expressions discussed further below --- while $W(t,t_0)$ changes sign, making $C_{+}^{z, ns}(t + t_0,t_0)$ not invariant. According to the general strategy outline in Sec.~\ref{sec:Teff_dynamic} we will focus below only on the stationary parts of the correlation and response function, given by Eq.~\reff{eq:Cp-app1} and related to Eq.~\reff{eq:Cm-app1}, respectively. In particular, 
%
%
%
%
%
\begin{figure}[h]
\centering
 \includegraphics[width=0.43\textwidth]{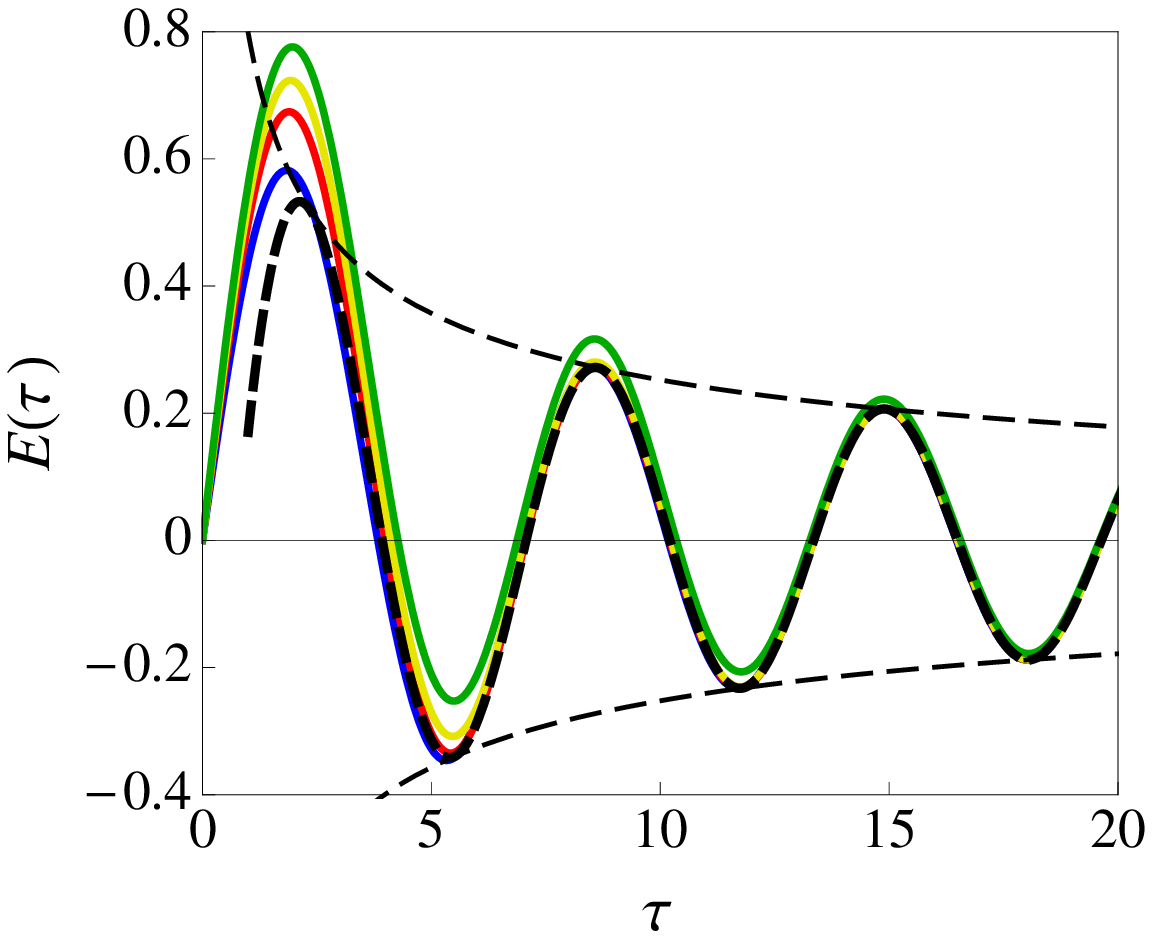} 
 \includegraphics[width=0.43\textwidth]{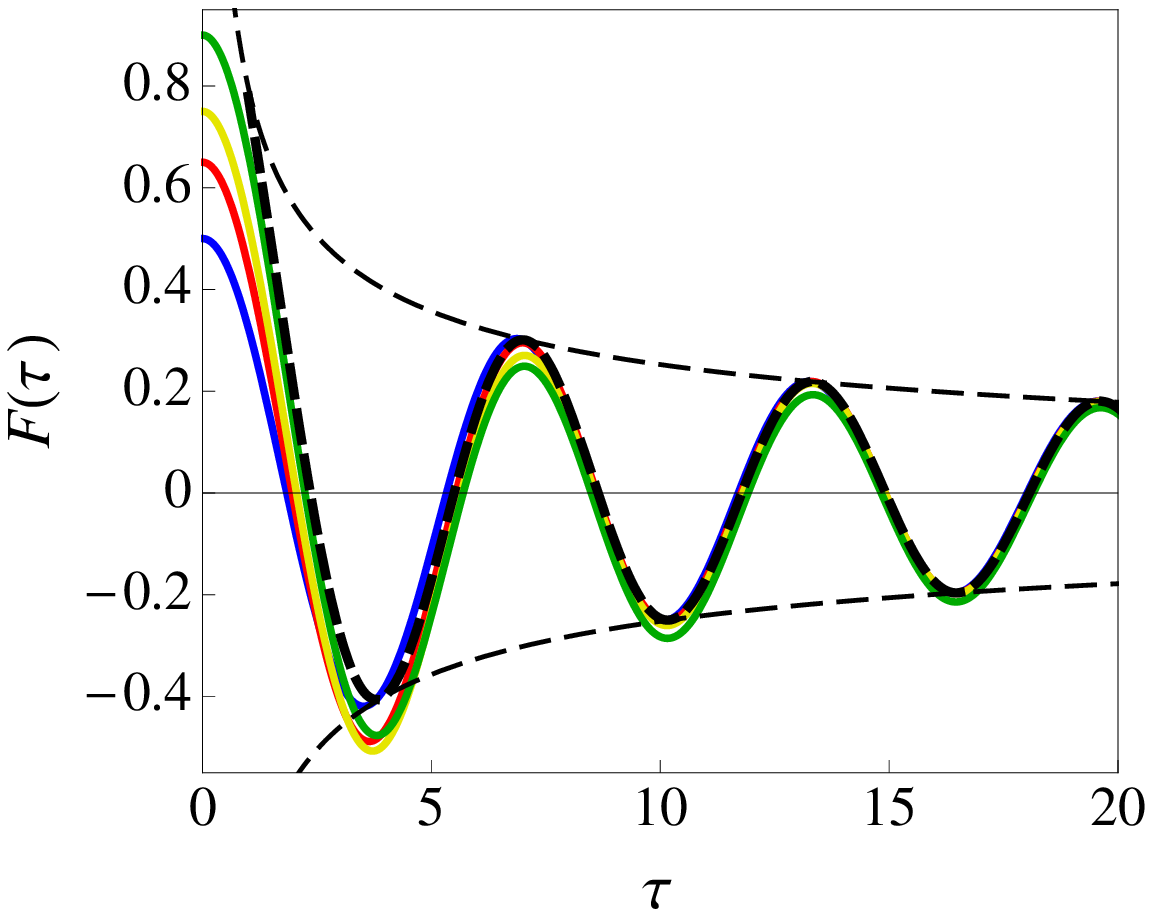} 
\caption{Dependence of the functions $E(\tau)$ and $F(\tau)$ (see Eqs.~\reff{eq:app-defE}
and \reff{eq:app-defF}, respectively) on the time $\tau$ 
for various values of $\Gamma_0$. The blue, red, yellow and green solid lines refer
respectively to $\Gamma_0=0$, 0.3, 0.5, and 0.8. The curves corresponding to $\G_0>1$ can be obtained from those with $\G_0<1$ taking into account that they are both functions of $\Y$ (see Eq.~\reff{eq:defYapp}) 
and therefore invariant for $\G_0 \to \G_0^{-1}$.  
The dashed lines show
the corresponding leading asymptotic behaviors for large $\tau$ (see Eqs.~\reff{eq:app-asyE} and \reff{eq:Fasy}, respectively), which are actually attained rather early in the evolution. The thin dashed lines highlight the
asymptotic leading algebraic time decay of the envelope $\sim 2/\sqrt{2 \pi \tau}$. 
}
\label{fig:E-F-app}
\end{figure}
%
%
for quenches ending at the critical point $\Gamma=1$, one has
$\epsilon_k \equiv \epsilon_k(\Gamma=1)= 4 |\sin (k/2)|$ [see Eq.~\reff{eq:energy}],
\beq
\cos \Delta_k(\Gamma=1,\Gamma_0) = \frac{2(\Gamma_0+1)}{\epsilon_k(\Gamma_0)} |\sin(k/2)| 
=  \frac{ \sqrt{\Y} \, | \sin (k/2) | }{\sqrt{1 + (\Y-1) \sin^2 (k/2)}}
\label{eq:cosDcrit}
\eeq
[see Eq.~\reff{eq:Delta_k}], where $\Delta_k = 2 \delta_k$ and
\beq
\Y = \left(\frac{1+\Gamma_0}{1-\Gamma_0}\right)^2 > 1.
\label{eq:defYapp}
\eeq 
[Note that $\Y$ and therefore $\cos \Delta_k(\Gamma=1,\Gamma_0)$ are invariant under the transformation 
$\Gamma_0 \mapsto \Gamma_0^{-1}$.]
Accordingly, Eq.~\reff{eq:Cp-app1} 
can be simplified as
\beq
\label{eq:Cp-app-crit}
\begin{split}
C_+^z(t) = & \left[ \int_0^{\pi} \frac{\rmd k}{\pi}
  \cos( \epsilon_k t) \right]^2
-  \left[\int_0^{\pi} \frac{\rmd k}{\pi}
  \cos( \epsilon_k t) \cos \Dt_k  \sin(k/2) \right]^2 \\
& + \left[ \int_0^{\pi} \frac{\rmd k}{\pi}
  \sin( \epsilon_k t)   \sin (k/2)  \right]^2
-  \left[\int_0^{\pi} \frac{\rmd k}{\pi} \sin( \epsilon_k t)   \cos \Dt_k \right]^2 ,
\end{split}
\eeq
while the stationary response takes the form [see Eqs.~\reff{Kubo} and~\reff{eq:Cm-app1}]:
\beq
\label{eq:Rz-app-crit}
\begin{split}
R^z(t) =& 4\,\theta(t)\left[   \int_0^{\pi} \frac{\rmd k}{\pi}
 \sin( \epsilon_k t)  \cos \Dt_k   \int_0^{\pi} \frac{\rmd l}{\pi} \cos( \epsilon_l t)\right.  \\
   & \left.\quad\quad\quad\quad -  \int_0^{\pi} \frac{\rmd k}{\pi}
 \sin( \epsilon_k t)   \sin(k/2)  \int_0^{\pi} \frac{\rmd l}{\pi}
\cos( \epsilon_l t)  \sin (l/2) \cos \Dt_l  \right] .
\end{split}
\eeq 
By direct inspection of the previous equations, one realizes that $C_+^z$ and $R^z$ can be conveniently  expressed in terms of the following integrals
\begin{align}
 \int_0^{\pi} \frac{\rmd k}{\pi}
  \cos( \epsilon_k t) &=  J_0(4t), \label{eq:appJ0}\\
 \int_0^{\pi} \frac{\rmd k}{\pi}
  \sin( \epsilon_k t) \sin(k/2) &= -\frac{1}{4}\frac{\rmd }{\rmd t}  J_0(4t) = J_1(4 t), \label{eq:appJ1}\\
\int_0^\pi  \frac{\rmd k}{\pi} \sin( \epsilon_k t)   \cos \Dt_k  &\equiv E(4 t) \label{eq:app-defE},\\
\int_0^\pi \frac{\rmd k}{\pi}  \cos( \epsilon_k t) \cos \Dt_k  \sin(k/2) &=\frac{1}{4}\frac{\rmd }{\rmd t} E(4t) = E'(4 t),
\label{eq:app-defEp}
\end{align}
(see, \eg, 10.9.1 and 10.6.2 in Ref.~\cite{HMF}), where $J_{\alpha}$ is the Bessel function of the first kind and order $\alpha$, as
\begin{eqnarray}
C_+^z(t) &=& J_0^2(4t) - [E'(4 t)]^2 + J_1^2(4t) - E^2(4t), \quad \mbox{and} \label{eq:Cp-app-crit-2}\\[2mm]
R^z(t) &=& 4 \,\theta(t) [E(4t)J_0(4t) -J_1(4t) E'(4 t) ].\label{eq:Rz-app-crit-2}
\end{eqnarray}
The function $E$ introduced in Eq.~\reff{eq:app-defE} can be explicitly written as
\beq
E(\tau) =  \int_0^{\pi}  \frac{\rmd k}{\pi} \sin( \tau \sin (k/2) )     \frac{ \sqrt{\Y} \, \sin (k/2) }{\sqrt{1 + (\Y-1) \sin^2 (k/2)}}
\label{eq:app-defE-2},\
\eeq
where $\Y$ is given in Eq.~\reff{eq:defYapp}. 
The left panel of Fig.~\ref{fig:E-F-app} shows $E(\tau)$ as a function of the time 
$\tau$ for various values of $\G_0$. The leading asymptotic behavior of the function for $\tau \gg 1$ --- which is discussed further below in App.~\ref{app:asympt} --- is indicated by dashed lines. The right panel of the figure, instead, presents for comparison the behavior of the function $F$ which will be introduced in App.~\ref{app:gtM} [see Eq.~\reff{eq:app-defF}].
Later on in the analysis of the effective temperatures, we will need $E'(0)$, which can be readily expressed as
\beq
E'(0) = \frac{2\sqrt{\Y}}{\pi} \frac{ E_e(1-\Y)   - K_e(1-\Y)}{\Y-1},
\label{eq:Ep0}
\eeq
where $K_e(k)$,  $E_e(k)$ are the complete elliptic integrals of the first and second kind, respectively (see, \eg, chapter 19 in Ref.~\cite{HMF}). For later convenience we report here their definitions:
\begin{align}
K_e(k) &= \int_0^{1}\frac{\rmd \varepsilon}{\sqrt{1-\varepsilon^2}} \frac{1}{\sqrt{1 - k \varepsilon^2}}, \label{eq:EI-i}\\
E_e(k) &= \int_0^{1}\frac{\rmd \varepsilon}{\sqrt{1-\varepsilon^2}} \sqrt{1 - k \varepsilon^2}. \label{eq:EI-ii}
\end{align}
Compared to the standard notation for these functions (see, \eg, Ref.~\cite{HMF}), we have added a subscript $e$ to  the corresponding symbols in order to avoid confusion between $E$ and $E_e$. In addition, adopting the same convention as {\tt Wolfram Mathematica},  we indicate by $k$ the square of the moduli of the elliptic functions, which are usually defined in the mathematical literature as in Eqs.~\reff{eq:EI-i} and \reff{eq:EI-ii} with $k \mapsto k^2$; see, \eg, chapter 19 in Ref.~\cite{HMF}.

\subsection{Correlation and response functions for critical quenches with initial conditions $\Gamma_0=0$, $1$, $\infty$}
\label{app:crit-polarized}

For $\Gamma_0=0$ or $\Gamma_0=\infty$ and $\Gamma=1$, the parameter $\Y$ in Eq.~\reff{eq:cosDcrit}  takes the value $\Y=1$ 
[see Eq.~\reff{eq:defYapp}] and therefore $\cos \Dt_k = |\sin (k/2)|$. Accordingly, the function $E$ [see Eq.~\reff{eq:app-defE}] which 
appears in Eqs.~\reff{eq:Cp-app-crit-2} and \reff{eq:Rz-app-crit-2} for $C_+$ and $R^z$, respectively, can be expressed as a Bessel 
function (compare with Eq.~\reff{eq:appJ1} and see, \eg, 10.6.1 in Ref.~\cite{HMF}):
\beq
E(\tau ) = J_1(\tau) \quad \mbox{and therefore}\quad E'(\tau) = J_1'(\tau)=  [J_0(\tau) - J_2(\tau)]/2.
\label{eq:EEp}
\eeq
Accordingly, 
\begin{align}
C^z_+(t) &= J^2_0(4 t) - \frac{1}{4} [ J_0(4t) - J_2(4 t) ]^2 ,\\[2mm]
 R^z(t) &= 2 \,\theta(t) J_1(4t) [ J_0(4 t) + J_2(4 t) ].
\end{align}

The case of the critical quench with $\Gamma_0=1 (= \Gamma)$, \ie, $\Y \to \infty$ (see Eq.~\reff{eq:defYapp}) actually corresponds 
to the equilibrium situation at $T=0$, for which one expects the correlation and response functions to be stationary immediately after 
the ``quench" at $t=0$ because, effectively, no quench takes place.
The corresponding expressions for $C^z_+$ and $R^z$ are readily derived from Eqs.~\reff{eq:Cp-app-crit-2} and 
\reff{eq:Rz-app-crit-2}, by taking into account that in this case $\cos \Delta_k(\Gamma=1,\Gamma_0=1)=1$ 
[see Eq.~\reff{eq:cosDcrit}] and therefore 
\beq
E(\tau) = H_0(\tau), \quad \mbox{with} \quad E'(\tau) = H_{-1}(\tau),
\label{eq:EH0}
\eeq
where $H_\alpha$ are the so-called Struve functions (see, \eg, 11.5.1 and 11.4.27 in Ref.~\cite{HMF}).
Accordingly, the correlation and response functions are
\begin{align}
C_+^z(t) &= J_0^2(4t) - H_{-1}^2(4 t) + J_1^2(4t) - H_0^2(4t), \quad \mbox{and} \label{eq:Cp-app-crit-eq}\\[2mm]
R^z(t) &= 4 \,\theta(t)[H_0(4t)J_0(4t) -J_1(4t) H_{-1}(4 t) ].\label{eq:Rz-app-crit-eq}
\end{align}
The expressions \reff{eq:Cp-app-crit-eq} and \reff{eq:Rz-app-crit-eq}
agree with those for the symmetric correlation derived in Ref.~\cite{Igloi00}
and for the response
function after quenches originating from the fully polarized state $\Gamma_0=\infty$ derived in Ref.~\cite{Karevski}.

\subsection{Asymptotic expansions for large times}
\label{app:asympt}

For quenches at the critical point $\Gamma=1$ and generic value of $\Gamma_0$, we are interested in the asymptotic expansion for $t\to\infty$ 
of the correlation function $C^z_+$ and of the response function $R^z$ in Eqs.~\reff{eq:Cp-app-crit-2} and \reff{eq:Rz-app-crit-2}, respectively. 
These expressions involve the function $E(\tau)$ introduced in Eq.~\reff{eq:app-defE} and the Bessel functions  $J_{0,1}$. The asymptotic 
expansion of the latter is well-known (see, \eg, 10.17.3 in Ref.~\cite{HMF})
%
%
and therefore we need only to determine the asymptotic behavior of $E(\tau)$, defined in Eqs.~\reff{eq:app-defE} 
and \reff{eq:app-defE-2}.
Introducing the integration variable $y = 1- \sin (k/2)$, the latter expression becomes (the procedure below is analogous to the 
one adopted in Ref.~\cite{Rossini_kinks}, in particular see App.~B.2 therein):
\beq
E(\tau) = 
   \int_0^{1} \rmd y 
  \frac{\sin( (1-y) \tau)}{\sqrt{y}} \times  \frac{2\sqrt{\Y}}{\pi} 
  \frac{(1-y)}{ \sqrt{(2-y)[1 + (\Y-1)  (1-y)^2] }} .
  \label{eq:appEint}
\eeq
The asymptotic behavior of this integral can be inferred from the one of 
\beq
e(\tau; f) = \int_0^1\rmd y \frac{\rme^{i y \tau}}{\sqrt{y}} f(y)
\label{eq:app-defe}
\eeq
where $f(y)$ is a generic function assumed to have a regular expansion for $y=0$, \ie, $f(y) = \sum_{n=0}^\infty f_n y^n$, such that $e(\tau;f) = 
\sum_{n=0}^\infty f_n (-i \rmd/\rmd \tau)^n e(\tau;1)$, with $e(\tau;1) = \sqrt{2\pi/\tau} \times [C(\sqrt{2 \tau/\pi}) + i S(\sqrt{2\tau/\pi})]$. Here $C$ 
and $S$ are the so-called Fresnel integrals: see, \eg, 7.2.7 and 7.2.8 in Ref.~\cite{HMF} for their definitions and 7.5.3, 7.5.4, 7.12.2 and 7.12.3 for the corresponding asymptotic expansions.
%
%
The asymptotic behavior of Eq.~\reff{eq:app-defe} for large $\tau$ can be calculated on the basis of these expansions 
and turns out to be:
\beq
\begin{split}
e(\tau; f)  = &\sqrt{\pi}\, \rme^{i \pi/4} \left[ \frac{f(0)}{\tau^{1/2}} + \frac{i f'(0)}{2 \tau^{3/2}} - \frac{3 f''(0)}{8 \tau^{5/2}} + {\cal O}(\tau^{-7/2})\right] \\
& +  \left(\frac{1}{i\tau} - \frac{1}{2\tau^2} \right) \rme^{i\tau} f(1) + \frac{\rme^{i\tau}}{\tau^2} f'(1) + {\cal O}(\tau^{-3}).
\end{split}
\eeq
Note that, alternatively, the first line on the r.h.s.~of this equation could have been inferred from the stationary-phase approximation \cite{Appel}. One can take advantage of the previous asymptotic expansion in order to calculate the one of the integral 
\beq
\begin{split}
 \int_0^{1} \rmd y 
  \frac{\rme^{i(1-y) \tau}}{\sqrt{y}}  f(y) = &\ \rme^{i\tau}e^*(\tau,f) \\
  = &\ \sqrt{\pi}  \left[ f(0) \frac{\sin(\tau+\pi/4)}{\tau^{1/2}}  -  \frac{f'(0)}{2} \frac{\cos (\tau+\pi/4)}{\tau^{3/2}}\right]\\
  & + i \sqrt{\pi}  \left[ - f(0) \frac{\cos(\tau+\pi/4)}{\tau^{1/2}} + \frac{f(1)}{\sqrt{\pi}} \frac{1}{\tau} -  \frac{f'(0)}{2} \frac{\sin (\tau+\pi/4)}{\tau^{3/2}}\right] + {\cal O}(\tau^{-2}),
  \end{split}
\label{eq:app-int-int}
\eeq
the imaginary part of which has the same structure as $E$ in Eq.~\reff{eq:appEint}. 
Taking into account the specific expression of $f$ in Eq.~\reff{eq:appEint} one eventually finds
\beq
E(\tau) = - \frac{1}{\sqrt{2\pi}} \frac{2}{\tau^{1/2}} \cos(\tau+\frac{\pi}{4}) +\frac{4 \Y^{-1}-1}{32\sqrt{2\pi}}\frac{8}{\tau^{3/2}}\sin(\tau+\frac{\pi}{4}) + {\cal O}(\tau^{-5/2}),
\label{eq:app-asyE}
\eeq
where the term $\propto 1/\tau$ in Eq.~\reff{eq:app-int-int} --- which emerges upon resumming the expansion of $f(u)$ --- drops out because $f(1)=0$ in the case of a quench with $\Y^{-1}\neq 0$ 
(which excludes the equilibrium case at zero temperature $\G_0=1 (=\G)$ discussed in Sec.~\ref{app:crit-polarized} and further below, corresponding to $\Y^{-1} = 0$). By using Eqs.~\reff{eq:app-asyE} and the asymptotic expansions of $J_{0,1}$ (see, \eg, 10.17.3 in Ref.~\cite{HMF}) 
in Eqs.~\reff{eq:Cp-app-crit-2} and \reff{eq:Rz-app-crit-2}, we finally find:
\begin{align}
C_+^z(t) &= - \frac{1}{8\pi t^2} \cos(8 t)  + {\cal O}(t^{-3}) , \label{eq:app-asy-Cz}  \\ 
R^z(t) &= \frac{1}{4\pi t^2} \Big[\Y^{-1} - \sin(8 t) \Big] + {\cal O}(t^{-3}), \label{eq:app-asy-Rz}
\end{align}
for $t\gg 1$.
In addition, in the special case $\Y=1$ --- see Sec.~\ref{app:crit-polarized} --- $E(\tau) = J_1(\tau)$  and indeed one recovers from Eq.~\reff{eq:app-asyE} the well-known expansion of the Bessel function $J_1$ (see, \eg, 10.17.3 in Ref.~\cite{HMF}). 

In the equilibrium case at zero temperature $\Y^{-1} = 0$, $f(1) = 2/\pi$ and Eq.~\reff{eq:app-asyE} with the additional 
term $2/(\pi \tau)$ stemming from $f(1)\neq 0$ reproduces the known large-$\tau$ expansion of the Struve function $H_0(\tau)$ (see 11.6 in Ref.~\cite{HMF}), 
which $E$ reduces to in this case [see Eq.~\reff{eq:EH0}]. Accordingly, one finds
\begin{align}
C_+^z(t) & = \frac{1}{\sqrt{2} (\pi t)^{3/2}} \cos(4 t+\frac{\pi}{4}) - \frac{1}{8 \pi t^2}\left[ \frac{2}{\pi} + \cos (8t)\right] + {\cal O}(t^{-5/2}), \label{eq:app-asy-Cz-eq}\\
R^z(t) & =  \frac{\sqrt{2}}{(\pi t)^{3/2}} \sin(4 t+\frac{\pi}{4}) - \frac{1}{4\pi t^2}\sin(8 t) +  {\cal O}(t^{-5/2}),\label{eq:app-asy-Rz-eq}
\end{align}
which display a different leading asymptotic behavior compared to the non-equilibrium case after the quench in Eqs.~\reff{eq:app-asy-Cz} and \reff{eq:app-asy-Rz}.

\section{Global transverse magnetization}
\label{app:gtM}

By adopting the same approach and notation as those used for the calculation of the (self-)correlation and (self-)response function of $\hat{\sigma}_i^z(t)$ summarized in App.~\ref{app:sz},
we calculate the correlation and linear response functions of the total magnetization:
\beq
 \hat{M}^z(t) = \displaystyle \frac1L \sum_i   \hat{\sigma}_i^z(t).
\eeq
Its connected correlation function is
\beq
\label{Corr_M}
\langle  \hat{M}^z(t_2)  \hat{M}^z(t_1)\rangle - \langle  \hat{M}^z(t_2)\rangle 
\langle  \hat{M}^z(t_1)\rangle 
=  \displaystyle \frac{8}{L^2} \sum_{k} v_k^{\ast}(t_1)v_k(t_2)u_k^{\ast}(t_1)u_k(t_2) .
\eeq
After a multiplication of this correlation by a factor $L$ --- which is required in order for the fluctuations not to vanish in the thermodynamic limit --- one obtains the symmetric and antisymmetric correlation for $L\to\infty$:
\beq
\begin{array}{l}\label{CS_CA_M}
C^M_+(t_2,t_1) = \displaystyle  L\left[\frac12 \langle \{ \hat{M}^z(t_2) ,  \hat{M}^z(t_1) \} \rangle - 
\langle  \hat{M}^z(t_2)\rangle\langle  \hat{M}^z(t_1)\rangle\right] \stackrel{L\to\infty}{\longrightarrow}  8
 \int_0^{\pi} \frac{\rmd k}{\pi}~ \RE\Big[ v_k^{\ast}(t_1)v_k(t_2)u_k^{\ast}(t_1)u_k(t_2) \Big],\\[4mm]
C^M_-(t_2,t_1) = \displaystyle \frac{L}{2} \langle [  \hat{M}^z(t_2) ,  \hat{M}^z(t_1) ] \rangle  
\stackrel{L\to\infty}{\longrightarrow} 8 i  \int_0^{\pi} \frac{\rmd k}{\pi}~ \IM\Big[ v_k^{\ast}(t_1)v_k(t_2)u_k^{\ast}(t_1)u_k(t_2) \Big] .
\end{array} 
\eeq
As we are primarily concerned with the behavior in the long-time stationary regime, we focus on the stationary functions $C^M_\pm(t) \equiv \lim_{t_1\to\infty} C^M_\pm(t_2= t+t_1,t_1)$.  
The response function $R^M$ is related to $C_-^M$ via Eq.~\reff{Kubo}, \ie,
$R^M(t) = 2 i \theta(t) \, C_-^M(t)$.
Inserting the expression for the matrix elements in terms of  
the Bogoliubov angles, the stationary parts of the correlations in Eqs.~\reff{CS_CA_M} become:
\begin{align}
C^{M}_+(t)  &=   2 \int_{0}^{\pi} \frac{\rmd k}{\pi} \left\{
 (1-\cos^2\Dt_k) \cos^2 (2\theta_k)   +  
 [1-\cos^2 (2\theta_k)] \frac{\cos^2\Dt_k+1}{2} \cos(2 \epsilon_k t) 
 \right\}, \label{eq:CMp-1} \\[2mm]
R^M(t)  &=  4 \, \theta(t)\int_{0}^{\pi} \frac{\rmd k}{\pi} 
[1-\cos^2 (2\theta_k)]   \sin(2 \epsilon_k t) \cos\Dt_k .  \label{eq:CMm-1}
\end{align}
 While these expressions are valid for a generic quench
we will focus on the critical case, for which $\epsilon_k =\epsilon^{\Gamma=1}_k = 4 \sin (k/2)$ [Eq.~\reff{eq:energy}], $\cos(2\theta_k) = \sin(k/2)$ [Eq.~\reff{eq:costhcrit}] and $\cos \Delta_k$ can be 
expressed as in Eq.~\reff{eq:cosDcrit}. In particular, substituting into Eqs.~\reff{eq:CMp-1} and \reff{eq:CMm-1} this expression for $\cos (2\theta_k)$, renders Eqs.~\reff{CM} and \reff{RM} which, in turn, 
can be expressed in terms of the integrals in Eqs.~\reff{eq:appJ0}, \reff{eq:appJ1}, \reff{eq:app-defE} and of the following ones:
\begin{eqnarray}
\int_0^\pi\frac{\rmd k}{\pi} \sin^2(k/2) \cos(2\epsilon_k t) &=& \frac{1}{8}\frac{\rmd }{\rmd t} J_1(8t) = \frac{J_0(8t)-J_2(8t)}{2}, \label{eq:int1}\\
\int_0^\pi\frac{\rmd k}{\pi} \cos^2\Delta_k \cos(2\epsilon_k t) &\equiv& F(8t), \label{eq:app-defF} \\
\int_0^\pi\frac{\rmd k}{\pi} \sin^2(k/2) \cos^2\Delta_k \cos(2\epsilon_k t) &=& - F''(8t), \label{eq:app-Fs}\\
\int_0^\pi\frac{\rmd k}{\pi} \sin^2(k/2)  \sin(2\epsilon_k t) \cos \Delta_k &=&  - E''(8t).
\end{eqnarray}
We note that the connected symmetric correlation in Eq.~\reff{eq:CMp-1} contains a constant term, which is attained for $t \to \infty$ and which can also be expressed in terms of the previous functions 
[see Eqs.~\reff{eq:int1} and \reff{eq:app-Fs}] as
\beq
C = 2 \int_{0}^{\pi} \frac{\rmd k}{\pi} \Big[(1-\cos^2\Dt_k) \cos^2( 2\theta_k)\Big]  = 1 + 2 F''(0)
\label{eq:defC-app}
\eeq
(where we used the fact that $J_0(0)=1$ and $J_2(0)=0$). Accordingly, one has
\begin{align}
C^{M}_+(t)  &= C +  \frac{J_0(8t)+J_2(8t)}{2}  + F(8t)  + F''(8t), \label{eq:CMp-fin}\\
R^{M}(t)  &= 4\,\theta(t)[E(8t) + E''(8t)],  \label{eq:CMm-fin}
\end{align}
with [see Eq.~\reff{eq:app-defF}]
\beq
F(\tau) = \int_0^\pi\frac{\rmd k}{\pi} \cos(\tau \sin(k/2)) \frac{\Y \sin^2(k/2)}{1 + (\Y-1)\sin^2(k/2)}.
\label{eq:appF-exp}
\eeq
The right panel of Fig.~\ref{fig:E-F-app} shows $F(\tau)$ as a function of the time 
$\tau$ for various values of $\G_0$. The leading asymptotic behavior of the function for $\tau \gg 1$ --- which is discussed further below in App.~\ref{app:M-long-time}, see Eq.~\reff{eq:Fasy} --- is indicated by dashed lines. 
A straightforward calculation of $F''(0)$ from Eq.~\reff{eq:appF-exp} gives [see Eq.~\reff{eq:defC-app}]
\beq
C = \frac{1}{(1+ \sqrt{\Y})^2},
\label{eq:valC}
\eeq
whereas
\beq
F(0) = \frac{\sqrt{\Y}}{1 + \sqrt{\Y}}.
\label{eq:F-0}
\eeq
In order to define an effective temperature according to Eq.~\reff{eq:Teff-cl-st}, we will be interested in
\beq
C^{M}_+(0) - C^{M}_+(t=-\infty) = \frac{1}{2} + F(0) + F''(0) = \frac{1}{2(1+\sqrt{\Y})^2} + \frac{\sqrt{\Y}}{1+\sqrt{\Y}},
\label{eq:appCM-0-inf}
\eeq
which follows from Eqs.~\reff{eq:CMp-fin}, \reff{eq:defC-app}, \reff{eq:valC} and the fact that $J_{0,2}$ and $F$ vanish for large 
arguments (as we will discuss in App.~\ref{app:M-long-time} below), whereas $J_0(0)=1$ and $J_2(0)=0$.
Note that in the fully polarized case $\Gamma_0 = 0$ or $\Gamma_0 = \infty$, both corresponding to $\Y=1$ (see Eqs.~\reff{eq:defYapp} and \reff{eq:cosDcrit}), the integrals on the 
l.h.s.~of Eqs.~\reff{eq:app-defF} and \reff{eq:int1} are identical and therefore 
\beq
F(\tau) = [J_0(\tau)-J_2(\tau)]/2, \quad \mbox{with} \quad F''(\tau) = - (3/8) J_0(\tau) +(1/2) J_2(\tau) - (1/8) J_4(\tau) = (1-3/\tau^2) J_2(\tau). 
\eeq
Taking into account this relation,  Eqs.~\reff{eq:valC}, \reff{eq:EEp}, \reff{eq:CMp-fin}, \reff{eq:CMm-fin} and the properties of the Bessel functions (see, \eg, Ref.~\cite{HMF}), one finds
\begin{align}
C^{M}_+(t)  &= \frac{1}{4} +  \frac{5}{8}J_0(8t)+ \frac{1}{2}J_2(8t) - \frac{1}{8}J_4(8t), \label{eq:CMp-pol}\\[2mm]
R^{M}(t)  &= \theta(t)[J_1(8t)+J_3(8t)] =  \theta(t) \frac{J_2(8t)}{2 t},  \label{eq:CMm-pol}
\end{align}
for a critical quench starting from a fully polarized case $\G_0=0$ (or, equivalently, $\G_0=\infty$).

In the equilibrium case $\Gamma_0 = 1 (= \Gamma)$, $\Y \to \infty$ and [see Eq.~\reff{eq:cosDcrit}] 
$\cos \Delta_k (\G=1,\G_0=1) = 1$. Accordingly, the integral on the l.h.s.~of 
Eq.~\reff{eq:app-defF}  becomes identical (up to a trivial rescaling of $t$) to the one on the l.h.s.~of Eq.~\reff{eq:appJ0}, so that
\beq
F(\tau) = J_0(\tau), \quad \mbox{with} \quad F''(\tau) = - [J_0(\tau) - J_2(\tau)]/2. 
\eeq
Taking into account Eq.~\reff{eq:EH0}, one finds $E''(\tau) = H_{-1}'(\tau) = H_{-2}(\tau) + \tau^{-1} H_{-1}(\tau)$ (see, \eg, 11.4.27 in 
Ref.~\cite{HMF}) and therefore 
[see Eqs.~\reff{eq:CMp-fin}, \reff{eq:CMm-fin}, \reff{eq:valC} and 11.4.23 in Ref.~\cite{HMF}]
\begin{align}
C^{M}_+(t)  &= J_0(8t) + J_2(8t), \label{eq:CMp-eq}\\[2mm]
R^{M}(t)  &= \theta(t)  \left[ \frac{1}{\pi t} - \frac{H_{-1}(8t)}{2t}\right],  \label{eq:CMm-eq}
\end{align}
in the equilibrium critical case at zero temperature (no quench). 

\subsection{Long-time behavior}
\label{app:M-long-time}

In order to determine the long-time behavior of $C^M_+$ and $R^M$ in Eq.~\reff{eq:CMp-fin} and \reff{eq:CMm-fin}, one has first to determine the long-time behavior of the function $F$ introduced in 
Eqs.~\reff{eq:app-defF} and \reff{eq:appF-exp}.
After the change of variable $y = 1- \sin (k/2)$, $F$ can be cast in the form
\beq
F(\tau) =  \int_0^1\rmd y \frac{\cos((1-y)\tau)}{\sqrt{y}} \times \frac{2\Y}{\pi} \frac{ (1-y)^2}{\sqrt{2-y}[1 + (\Y-1)(1-y)^2]},
\eeq
which has the same form as the real part of the integral analyzed in Eq.~\reff{eq:app-int-int}, with $f(0) = \sqrt{2}/\pi$ and
$f'(0) = (1-8 \Y^{-1})/(2\sqrt{2}\pi)$.
Accordingly, the asymptotic expansion of $F(\tau)$ is given by
\beq
F(\tau)= \frac{1}{\sqrt{2\pi}}  \frac{2}{\tau^{1/2}} \sin(\tau+\frac{\pi}{4})  + \frac{8 \Y^{-1}-1}{4\sqrt{2\pi}} \frac{1}{\tau^{3/2}}\cos (\tau+\frac{\pi}{4}) + {\cal O}(\tau^{-2}).
\label{eq:Fasy}
\eeq
Taking into account Eqs.~\reff{eq:Fasy}, \reff{eq:app-asyE}, 
and the standard expansion of $J_{0,2}(\tau)$ (see, \eg, 10.17.3 in Ref.~\cite{HMF}), one eventually finds 
from Eqs.~\reff{eq:CMp-fin} and \reff{eq:CMm-fin}:
\begin{align}
C^{M}_+(t)  &= C  - \frac{1}{8\sqrt{\pi}}\sin(\frac{\pi}{4}-8t) \frac{1}{t^{3/2}} + {\cal O}(t^{-5/2}),
\label{eq:CMp-asy} \\
R^{M}(t) 
&= - \frac{1}{4 \sqrt{\pi}}\cos(\frac{\pi}{4}-8t) \frac{1}{t^{3/2}}  + {\cal O}(t^{-5/3}) . \label{eq:CMm-asy}
\end{align}
Note that the case of equilibrium dynamics at zero temperature can be recovered from these expressions by setting 
$\Gamma_0 = \Gamma = 1$, \ie,  for $\Y^{-1} = 0$. As we discussed after Eq.~\reff{eq:app-asyE}, the asymptotic behavior of 
$E(\tau)$ changes due to the additional contribution of a term $\propto \tau^{-1}$ which renders the leading decay of $R^M(t)$ 
slower and $\propto t^{-1}$, whereas the one of $C^M_+$ is not altered compared to Eq.~\reff{eq:CMp-asy} and --- apart 
from the specific value of $C$ --- equal to the non-equilibrium one.

\subsection{Fourier transform of the response and correlation functions}
\label{app:sub-M-FT}

In order to determine the frequency-dependent effective temperature defined in Eq.~\reff{eq:FDT-omega-Teff} we calculate here the 
Fourier transforms of $C_+^M$ and $R^M$ in 
Eqs.~\reff{eq:CMp-fin} and \reff{eq:CMm-fin} according to the convention~\reff{eq:FT}.  Consider first $C_+$: the functions $J_0$, 
$J_2$ [see Eq.~\reff{eq:int1}] and $F$ 
[Eqs.~\reff{eq:app-defF} and \reff{eq:appF-exp}] in terms of which $C_+$ is expressed involve all an integration over the 
wave-vector $k$ of $\cos (2 \epsilon_k t)$. As 
expected from the fact that $C_+^M(t) = C_+^M(-t)$, this function is even under time reversal. Accordingly, the Fourier 
transform $\tilde C_+^M(\omega) = 
\int_{-\infty}^{+\infty} \!\rmd t\, \rme^{i \omega t} C_+^M(t)$ is real and involves integrals over momenta of  
$\int_{-\infty}^{+\infty}\! \rmd t \,\rme^{i \omega t} \cos (2 \epsilon_k t)  = 
\pi [\delta(\omega+2\epsilon_k) + \delta(\omega-2\epsilon_k)]$,
which can be calculated straightforwardly and give Eq.~\reff{eq:CMomega}. 
The response function $R^M$, instead, is given in Eq.~\reff{eq:CMm-fin} in terms of
the function  
$E$ 
defined in Eqs.~\reff{eq:app-defE} and \reff{eq:app-defE-2}. Note that its Fourier transform $\tilde R^M(\omega) = 
\int_0^\infty\!\rmd t\, \rme^{i\omega t} R^M(t)$ 
can be also expressed as
\beq
\tilde R^M(\omega) = \frac{1-(\omega/8)^2}{2} \tilde E_>(\omega/8) - \frac{E'(0)}{2} \quad \mbox{where} \quad
  \tilde E_>(\omega) \equiv \int_0^\infty\rmd\tau \, \rme^{i \omega\tau} E(\tau) 
\label{app:FT-R-E}
\eeq
is the Fourier transform of $\theta(\tau)E(\tau)$ and $E'(0)$ is given by Eq.~\reff{eq:Ep0}. Taking into account that $E(\tau)$ involves 
an integral over momenta of 
$\sin(\varepsilon_k \tau)$ [see Eq.~\reff{eq:app-defE-2}], with $\varepsilon_k \equiv \sin(k/2)$, 
$\tilde E_>(\omega)$ involves an integral over momenta of
\beq
\begin{split}
\int_0^\infty\!\rmd \tau\, \rme^{i\omega \tau} \sin(\varepsilon_k \tau) & = \int_0^\infty\!\rmd \tau\,  
\left\{ \frac{\sin((\omega+\varepsilon_k)\tau) - \sin((\omega-\varepsilon_k)\tau)}{2} 
+
 i \frac{\cos((\omega-\varepsilon_k)\tau) - \cos((\omega+\varepsilon_k)\tau)}{2} \right\} \\[2mm]
& =  \frac{1}{2}{\rm pv} \frac{1}{\omega+\varepsilon_k} - \frac{1}{2}{\rm pv} \frac{1}{\omega-\varepsilon_k} + i \frac{\pi}{2} 
[\delta(\omega-\varepsilon_k) - \delta(\omega+\varepsilon_k)],
\end{split}
\eeq
(see, \eg, Ref.~\cite{Appel}) where ``pv" indicates that the principal value of the subsequent integral over momenta has to 
be considered. 
Accordingly, from Eq.~\reff{eq:app-defE-2}, 
\beq
\begin{split}
\tilde E_>(\omega) = & \frac{\sqrt{\Y}}{\pi} \int_0^1\frac{\rmd \varepsilon}{\sqrt{1-\varepsilon^2}} \frac{\varepsilon}{\sqrt{1 + (\Y-1)\varepsilon^2}} \left( {\rm pv}
\frac{1}{\omega+\varepsilon} - {\rm pv} \frac{1}{\omega-\varepsilon}
\right)\\
 & + i \frac{\omega}{\sqrt{1-\omega^2}} \frac{\sqrt{\Y}}{\sqrt{1 + (\Y-1)\omega^2}}  \theta(1-|\omega|).
\end{split}
\label{app:FT-E}
\eeq
As expected, due to the fact that $\tilde E_>(\omega)$ is the Fourier transform of the "causal" function $\theta(\tau)E(\tau)$ which vanishes for $\tau<0$, the real and imaginary parts of $\tilde E_>(\omega)$ are connected by a Kramers-Kronig relation~\cite{Appel}. In addition, for this specific case, the real part of $\tilde E_>(\omega)$
can be cast in the form 
\beq
{\rm Re\;} \tilde E_>(\omega)   = \frac{2 \sqrt{\Y}}{\pi} \int_0^1\frac{\rmd \varepsilon}{\sqrt{1-\varepsilon^2}} \frac{1}{\sqrt{1 + (\Y-1)\varepsilon^2}} 
\left( 1 -  \omega^2\; {\rm pv} \frac{1}{\omega^2 - \varepsilon^2}
 \right),
\label{eq:ReEw}
\eeq
in which one recognizes the definitions of  the complete elliptic integrals of the first and third kind, $K_e(k)$ and $\Pi(\alpha,k)$, respectively (see 19.2.4, 19.2.7, and 19.2.8 in Ref.~\cite{HMF}), which we report in Eq.~\reff{eq:EI-i} and here for convenience:
\beq
\Pi(\alpha,k) = \int_0^{1}\frac{\rmd \varepsilon}{\sqrt{1-\varepsilon^2}} \frac{1}{\sqrt{1 - k \varepsilon^2}} \; {\rm pv}\frac{1}{1-\alpha \varepsilon^2}.
\label{eq:EI-iii}
\eeq
(See also the remark after Eq.~\reff{eq:EI-ii} about our usage of $k$.)
Note that  for $\alpha> 1$ the integrand of $\Pi$ is singular within the domain of integration and the principal value has to be considered as a part of the definition. In particular, one finds 
$\Pi(\alpha,k) = K_e(k) - \Pi(k/\alpha,k)$ (see 19.6.5 of Ref.~\cite{HMF}), which connects the behavior for $\alpha >1$ to the one for $\alpha < 1$ (assuming $k<1$).
Additional properties which will be useful later are: $\Pi(0,k) = K_e(k)$, $\Pi(\alpha,0) = \pi/(2\sqrt{1-\alpha^2})$, and $K_e(0)=\pi/2$.
According to Eqs.~\reff{eq:EI-i} and \reff{eq:EI-iii}, one can express Eq.~\reff{eq:ReEw} as
\beq
{\rm Re\;} \tilde E_>(\omega)  = \frac{2\sqrt{\Y}}{\pi} 
\begin{cases}
K_e(1-\Y) -  \Pi(\omega^{-2},1-\Y) \quad &\mbox{for} \quad |\omega| >1, \\
 \Pi((1-\Y)\omega^2,1-\Y) \quad &\mbox{for} \quad |\omega| <1, 
\end{cases}
\label{eq:ReEwPi}
\eeq
in terms of the elliptic integral $\Pi(\alpha,k)$ with $\alpha < 1$. 

The imaginary part of $\tilde R^M(\omega)$ --- which enters into the definition of the effective temperature via Eq.~\reff{eq:FDT-omega-Teff} --- can be calculated 
straightforwardly from the imaginary part of $\tilde E_>(\omega)$ above and gives Eq.~\reff{eq:RMomega}. 

In order to define the effective temperature according to Eq.~\reff{eq:Teff-cl-st} one needs to determine $\tilde R^M(\omega=0)$, which is related to $ \tilde E_>(\omega=0) = \RE \, \tilde E_>(\omega=0)$ in Eq.~\reff{app:FT-R-E} via Eqs.~\reff{app:FT-E} and \reff{eq:ReEwPi}:
\beq
\tilde E_>(0) = \frac{2 \sqrt{\Y}}{\pi} \Pi(0,1-\Y)
= \frac{2 \sqrt{\Y}}{\pi}  K_e(1-\Y),
\label{app:FT-E0} 
\eeq
where $K_e(k)$ is 
defined in Eq.~\reff{eq:EI-i} (see, \eg, chapter 19 in Ref.~\cite{HMF} and the remarks after  Eq.~\reff{eq:EI-ii}). Accordingly, taking into account also Eqs.~\reff{app:FT-R-E} and \reff{eq:Ep0}, one eventually finds:
\beq
\tilde R^M(\omega=0) = \frac{\sqrt{\Y}}{\pi} \frac{\Y K_e(1-\Y) - E_e(1-\Y)}{\Y-1}.
\label{app:FT-R0}
\eeq

\section{Short-time expansion of the order parameter correlations}
\label{Appendix_order_parameter}

We aim at calculating $C^{x}(t+t_0,t_0) = \langle \sigma_j^x(t+t_0)  \sigma_j^x(t_0) \rangle = 
 {}_{\Gamma_0}\!\!\bra{0} \hat{\sigma}_j^x(t+t_0)  \hat{\sigma}_j^x(t_0) \ket{0}_{\Gamma_0}$ for $t \ll \tau \ll t_0$ and $\tau$
 given in Eq.~(\ref{Tau}). 
For the purpose of the present discussion we consider the Schr\"odinger picture of the dynamics and we indicate by 
$\ket{\psi(t_0)} \equiv \rme^{- i \hat{H} t_0}\ket{0}_{\Gamma_0}$ the state obtained by evolving the ground state of $\hat{H}(\Gamma_0)$ with $\hat H \equiv \hat H(\G)$ up to the time $t_0$.
Expanding the evolution operator for small $t$ one obtains:
\beq
\begin{split}
C^{x}(t+t_0,t_0)  
&= {}_{\Gamma_0}\!\!\bra{0} \rme^{i \hat{H} (t+t_0)} \hat{\sigma}_j^x \rme^{- i \hat{H}t} \hat{\sigma}_j^x \rme^{- i \hat{H} t_0}\ket{0}_{\Gamma_0}
= \bra{\psi(t_0)} \rme^{i \hat{H} t} \hat{\sigma}_j^x \rme^{- i \hat{H} t} \hat{\sigma}_j^x \ket{\psi(t_0)} \\[3mm]
&=  \bra{\psi(t_0)} \left(\hat{1} + i t [\hat{H},  \hat{\sigma}_j^x ] \hat{\sigma}_j^x   
  + \frac{1}{2} t^2 \left[ [ \hat{H} , \hat{\sigma}_j^x ] , \hat{H} \right] \hat{\sigma}_j^x   + \OO(t^3)\right) \ket{\psi(t_0)},
   \end{split}
\eeq
where $\hat{1}$ is the identity matrix. The commutators in the previous expression can be calculated by taking into account that Pauli matrices at different sites commute, whereas  
$\hat \sigma^a_j \hat\sigma^b_j = \delta_{ab} 1 + i \varepsilon^{a b c } \hat\sigma_j^c$, with 
$\varepsilon^{abc}$ the completely antisymmetric tensor with $\varepsilon^{xyz}=1$. Accordingly, 
setting $J=1$ in the definition of $\hat{H}$ in Eq.~(\ref{H_Ising}), the following
commutation relations hold: 
$[\hat H, \hat\sigma_j^x] \hat \sigma_j^x = - 2 \Gamma \hat\sigma_j^z$ and $\left[[\hat H, \hat\sigma_j^x],\hat H\right] \hat \sigma_j^x =  - 4 \Gamma^2 + 4 \Gamma \, \hat\sigma_j^z \hat\sigma_j^x (\hat\sigma_{j-1}^x + \hat\sigma_{j+1}^x)$.
We note that the last term of the latter expression is an anti-hermitean operator and therefore its expectation value is imaginary, contributing only with a subleading, $\OO(t^2)$ term to the response function $\propto \IM\, C^x$, which receives a contribution $\OO(t)$ from the first commutator calculated above.  In addition, one can actually verify that such a term vanishes in the stationary regime. In fact, by using Eq.~\reff{JW_fermions}  one finds $\hat\sigma^x_j\hat\sigma^x_{j+1} = \hat c^\dagger_j \hat c^\dagger_{j+1} + \hat c^\dagger_j \hat c_{j+1} + \mbox{h.c.} = (\hat c^\dagger_j - \hat c_j) (\hat c^\dagger_{j+1} + \hat c_{j+1})$ and from Eqs.~\reff{sz-c} and \reff{Fourier_Transform}
\beq
\label{zxx1}
\begin{split}
& - \hat \sigma_j^z \hat\sigma_j^x  \hat\sigma_{j-1}^x = 
( \hat c_{j-1}^{\dag} - \hat c_{j-1} )  ( \hat c_{j}^{\dag} - \hat c_{j}) \\[3mm]
&= \displaystyle \frac{1}{L} \sum_{k,k'}  \left\{ \rme^{-i (k+k') j + i k} \hat c_{k}^{\dag} \hat c_{k'}^{\dag}  -  \rme^{-i (k-k') j + i k}  \hat c_k^{\dag} \hat c_{k'} 
-  \rme^{i (k-k') j - i k}  \hat c_{k} \hat c_{k'}^{\dag}   + \rme^{i (k+k') j - i k} \hat c_{k} \hat c_{k'}  \right\} 
 \end{split}
 \eeq
and analogously
\beq\label{zxx2}
\begin{split}
& - \hat \sigma_j^z \hat\sigma_j^x \hat \sigma_{j+1}^x  =  ( \hat c_j^{\dag} + \hat c_j )  ( \hat c_{j+1}^{\dag} + \hat c_{j+1} ) 
\\[3mm]
&= \displaystyle \frac{1}{L} \sum_{k,k'} \left\{ \rme^{-i (k+k') j - i k}   \hat c_{k'}^{\dag}  \hat c_{k}^{\dag}  + \rme^{i (k-k') j + i k}  \hat c_{k'}^{\dag}  \hat c_{k}  
+ \rme^{-i (k-k') j - i k}  \hat c_{k'}  \hat c_{k}^{\dag}   + \rme^{i (k+k') j + i k}  \hat c_{k'}  \hat c_{k}  \right\}.
 \end{split}
 \eeq
Upon adding Eqs.~\reff{zxx1} and \reff{zxx2} and rearranging the terms by using the canonical anticommutation relations of the fermions one obtains:
\beq
\label{zxxs}
\begin{split}
 &-   \hat \sigma_j^z \hat\sigma_j^x (\hat\sigma_{j-1}^x + \hat \sigma_{j+1}^x) \\[3mm]
 & = \frac{2}{L}
 \sum_{k,k'}\left\{ (i \sin k) \rme^{-i (k+k') j} \hat c_{k}^{\dag} \hat c_{k'}^{\dag} 
  +\rme^{ -i (k+k') j} (\cos k' - \cos k) \hat c_{k}^{\dag} \hat c_{k'} 
  + \rme^{i (k+k')j} (- i \sin k) \hat c_{k} \hat c_{k'}\right\}.
\end{split}
\eeq
Its expectation value is therefore given by 
\beq
-\bra{\psi(t_0)}  \hat \sigma_j^z \hat\sigma_j^x (\hat\sigma_{j-1}^x + \hat \sigma_{j+1}^x)  \ket{\psi(t_0)} = 
\frac{8i}{L} \sum_{k>0} (\sin k) \RE\, [v_k(t_0) u^*_k(t_0)] = \frac{4 i}{L} \sum_{k>0} \sin k \sin \Delta_k \sin (2 \epsilon_k t_0),
\label{eq:comm2-st}
\eeq
where we used the fact that 
$\bra{\psi(t_0)}  \hat c^\dag_{k}  \hat c^\dag_{k'} \ket{\psi(t_0)} = 
{}_{\Gamma_0}\!\!\bra{0} \hat c^\dag_{k} (t_0) \hat c^\dag_{k'} (t_0) \ket{0}_{\Gamma_0}
 =  - v_k(t_0) u^*_k(t_0) \delta_{k',-k}$, $\bra{\psi(t_0)}  \hat c^\dag_{k}  \hat c_{k'} \ket{\psi(t_0)} = |v_k(t_0)|^2 \delta_{k',k}$, and $\bra{\psi(t_0)}  \hat c_{k}  \hat c_{k'} \ket{\psi(t_0)} = u_k(t_0) v_k^*(t_0) \delta_{k,-k'}$, which follow from Eqs.~\reff{Eq:Dynamics_quench}, \reff{eq:ukvk} and  \reff{eq:ukvk-sym} (in which we omit the superscripts of $u_k$, $v_k$), with $\epsilon_k \equiv \epsilon_k^\Gamma$ and $\Delta_k \equiv 2 (\theta^\Gamma_k-\theta_k^{\Gamma_0})$.
The expression on the r.h.s.~of Eq.~\reff{eq:comm2-st} can be shown to vanish in the stationary regime $t_0\to \infty$. 
Therefore,  the expressions above give for $t$, $t_0>0$: 
\beq
\begin{split}
C^x_+(t+t_0,t_0) &= \mbox{Re\,} C^x(t+t_0,t_0) = 1- \frac{t^2}{2} 4\Gamma^2 + \OO(t^3), \\[2mm]
R^x(t+t_0,t_0) &= - 2\, \mbox{Im\,} C^x(t+t_0,t_0) = 4 \Gamma t   \bra{\psi(t_0)}\hat\sigma_i^z \ket{\psi(t_0)} + \OO(t^2),
\end{split}
\eeq
which, in the stationary limit $t_0\to\infty$, yield Eqs.~\reff{eq:Cx-st} and \reff{eq:Rx-st}
where, in indicating the corrections, we used the fact that one expects the correlation function in the stationary regime to be an even function of 
time.

\section{Energy-based effective temperature $T^E_{\rm eff}(\G,\G_0)$ in the limit of shallow critical quenches}
\label{app:TeffEnergyShallow}

The effective temperature $T^E_{\rm eff}$ based on the energy of the system is obtained according to Eq.~\reff{Teff_energy}, \ie,  by equating the thermal average of the energy at a certain temperature $T^E_{\rm eff}$, see Eq.~\reff{eq:thexp}, with the 
expectation value in Eq.~\reff{eq:EnQ} of the energy after the quench from $\G_0$ to $\G$. 
Here we are interested in a quench of the Ising 
model in a transverse field $\G$ with critical final value $\Gamma=1$.  As discussed
in Sec.~\ref{Sec:Teff_Ising}, the expectation value of the 
energy after the quench is given by Eq.~\reff{eq:thexp}:
\beq
\begin{split}
{}_{\Gamma_0}\!\langle 0 |\hat{H}(\Gamma) | 0 \rangle_{\Gamma_0} &= - \int_0^\pi\frac{\rmd k}{2\pi} \epsilon_k(\Gamma=1) \cos \Delta_k(\Gamma=1,\Gamma_0) \\
& = - 2 E'(0) = - \frac{4\sqrt{\Y}}{\pi} \frac{E_e(1-\Y)-K_e(1-\Y)}{\Y-1},
\end{split}
\eeq
where we used the fact that, at the critical point, $\epsilon_k(\G=1) = 4 |\sin (k/2)|$ and we took into account 
the definitions of the functions $\cos \Delta_k(\Gamma=1,\Gamma_0)$ and 
$E'(0)$ given in Eqs.~\reff{eq:cosDcrit}, \reff{eq:app-defE-2}, and \reff{eq:Ep0} in order to express the result as a function of $\Y$, see Eq.~\reff{eq:defY}.
For $\G_0\to 1$, $\Y  \to \infty$ and therefore one can expand the expression above:
\beq
{}_{\Gamma_0}\!\langle 0|\hat H(\Gamma=1)|0\rangle_{\Gamma_0}   
= -\frac{4}{\pi} + \frac{-3 + 4 \ln 2 + \ln \Y}{\pi \Y} + {\cal O}(\Y^{-2}, \Y^{-2}\ln \Y) .
\label{eq:Te1}
\eeq
This expansion has to be compared with the thermal average of $\hat H(\G=1)$ at a temperature $\beta^{-1}=T^E_{\rm eff}$, which is expected to vanish for $\G_0 \to 1$. 
The thermal average given by Eq.~\reff{eq:thexp} can be conveniently cast in the form
\beq
\begin{split}
\langle \hat H(\G=1)\rangle_{T=\beta^{-1}} & =  - \int_0^\pi\frac{\rmd k}{2\pi} \epsilon_k(\G=1) \tanh\left(\beta\epsilon_k(\G=1)/2 \right)\\
& = -2 \frac{\rmd }{\rmd \beta} \int_0^\pi \frac{\rmd k}{2\pi} \ln \cosh\left(2\beta \sin (k/2) \right) \\
& = - \frac{4}{\pi} - 2  \frac{\rmd }{\rmd \beta} \int_0^\pi \frac{\rmd k}{2\pi} \ln \left(1 + \rme^{-4\beta\sin(k/2)}\right).
\end{split}
\eeq
In the limit $\beta\to\infty$ we are interested in, the last integral is 
dominated by small values of $k$, and therefore one can approximate
$\sin(k/2)\simeq k/2$ in the integrand:
\beq
 \int_0^\pi \frac{\rmd k}{2\pi} \ln \left(1 + \rme^{-4\beta\sin(k/2)}\right) 
 \stackrel{\beta\to\infty}{\simeq}  \int_0^\infty \frac{\rmd k}{2\pi} \ln \left(1 + \rme^{-2\beta k}\right) 
= \frac{\pi}{48\beta},
\eeq
where the extension of the integral from $[0,\pi]$ to $[0,\infty]$ introduces only exponentially small corrections.
Accordingly,
\beq
\langle \hat H(\G=1)\rangle_{T=\beta^{-1}}  \stackrel{\beta\to\infty}{\simeq} -\frac{4}{\pi} + \frac{\pi}{24\beta^2},
\label{eq:Te2}
\eeq
and, by comparing Eqs.~\reff{eq:Te1} and \reff{eq:Te2}, one finds:
\beq
\beta^E_{\rm eff}(\Y\to\infty) = \frac{1}{T^E_{\rm eff}(\G=1,\G_0\to 1)} \simeq  \frac{\sqrt{\Y}}{2}  \frac{\pi}{\sqrt{6 (\ln \Y + 4 \ln 2 - 3)}}.
\eeq
This asymptotic expression is used in Fig.~\ref{fig:Teff-Gamma0} in order to plot the behavior of $T^E_{\rm eff}(\G=1,\G_0)$ for $\G_0\gtrsim 0.9$.
In addition, this result can be compared with the inverse temperature obtained by studying the zero-frequency limit of the FDT of the global transverse magnetization $\hat M$ (see Sec.~\ref{Sec:Mz} and in particular Eq.~\reff{eq:Teff0M})
\beq
\beta^M_{\rm eff}= \frac{1}{T^M_{\rm eff}} = \frac{\sqrt{\Y}}{2},
\eeq
which shows that $T^M_{\rm eff}/T^E_\eff \to 0$ for $\Gamma_0\to\Gamma=1$.


\begin{thebibliography}{99}
\bibitem{footnote-supercooled}
Super-cooled liquids are systems in equilibrium (metastable with 
respect to the crystal) with the thermal bath in which the decay of the correlation functions is rather complex, 
certainly non-exponential
and for which the identification of a single coherence time is delicate.
\bibitem{footnote1}
For some particular values of $\Gamma$ and $\Gamma_0$, we verified numerically that 
the the long-time exponential decay rate of $C^x_+(t)$ and $R^x(t)$  in the stationary regime 
is well-fitted by Eq.~\reff{Tau} also for non-critical quenches $\G\neq 1$. This suggest that 
the conclusion we draw here for $\Gamma=1$ might extend also to the case $\Gamma\neq 1$.
\end{thebibliography}
\end{document}